\documentclass[12pt]{article}
\usepackage{amsmath,amsthm,latexsym,amssymb,amsfonts,epsfig}

\usepackage[all,arc,dvips,arrow,curve,cmactex]{xy}
\usepackage{amsfonts,amsmath,amssymb,graphics,psfrag}

\newcommand\Sets{{\bf Sets}}
\newcommand\op{{\rm op}}
\newcommand\ps[1]{\underline{#1}}
\newcommand\down[1]{\downarrow\!{#1}}

\makeatletter \@addtoreset{equation}{section} \makeatother

\newtheorem{Theorem}{Theorem}[section]
\newtheorem{Definition}{Definition}[section]
\newtheorem{Lemma}{Lemma}[section]
\newtheorem{Corollary}{Corollary}[section]

\newtheorem{Conjecture}{Conjecture}[section]

\newtheorem{Proof}{Proof}[section]

\def\be{\begin{equation}}
\def\ee{\end{equation}}
\def\ba{\begin{eqnarray}}
\def\ea{\end{eqnarray}}
\def\mv{\mathcal{V}}
\def\mc{\mathcal{C}}
\def\md{\mathcal{D}}
\def\mh{\mathcal{H}}

\def\us{\underline{\Sigma}}

\def\ug{{\underline{G}}}
\def\uom{\underline{\Omega}}
\def\Nl{{\mathchoice
{\setbox0=\hbox{$\displaystyle\rm N$}\hbox{\hbox to0pt
{\kern0.4\wd0\vrule height0.9\ht0\hss}\box0}}
{\setbox0=\hbox{$\textstyle\rm N$}\hbox{\hbox to0pt
{\kern0.4\wd0\vrule height0.9\ht0\hss}\box0}}
{\setbox0=\hbox{$\scriptstyle\rm N$}\hbox{\hbox to0pt
{\kern0.4\wd0\vrule height0.9\ht0\hss}\box0}}
{\setbox0=\hbox{$\scriptscriptstyle\rm N$}\hbox{\hbox to0pt
{\kern0.4\wd0\vrule height0.9\ht0\hss}\box0}}}}
\def\Zl{{\mathchoice
{\setbox0=\hbox{$\displaystyle\rm Z$}\hbox{\hbox to0pt
{\kern0.4\wd0\vrule height0.9\ht0\hss}\box0}}
{\setbox0=\hbox{$\textstyle\rm Z$}\hbox{\hbox to0pt
{\kern0.4\wd0\vrule height0.9\ht0\hss}\box0}}
{\setbox0=\hbox{$\scriptstyle\rm Z$}\hbox{\hbox to0pt
{\kern0.4\wd0\vrule height0.9\ht0\hss}\box0}}
{\setbox0=\hbox{$\scriptscriptstyle\rm Z$}\hbox{\hbox to0pt
{\kern0.4\wd0\vrule height0.9\ht0\hss}\box0}}}}
\def\Ql{{\mathchoice
{\setbox0=\hbox{$\displaystyle\rm Q$}\hbox{\hbox to0pt
{\kern0.4\wd0\vrule height0.9\ht0\hss}\box0}}
{\setbox0=\hbox{$\textstyle\rm Q$}\hbox{\hbox to0pt
{\kern0.4\wd0\vrule height0.9\ht0\hss}\box0}}
{\setbox0=\hbox{$\scriptstyle\rm Q$}\hbox{\hbox to0pt
{\kern0.4\wd0\vrule height0.9\ht0\hss}\box0}}
{\setbox0=\hbox{$\scriptscriptstyle\rm Q$}\hbox{\hbox to0pt
{\kern0.4\wd0\vrule height0.9\ht0\hss}\box0}}}}
\def\Rl{{\mathchoice
{\setbox0=\hbox{$\displaystyle\rm R$}\hbox{\hbox to0pt
{\kern0.4\wd0\vrule height0.9\ht0\hss}\box0}}
{\setbox0=\hbox{$\textstyle\rm R$}\hbox{\hbox to0pt
{\kern0.4\wd0\vrule height0.9\ht0\hss}\box0}}
{\setbox0=\hbox{$\scriptstyle\rm R$}\hbox{\hbox to0pt
{\kern0.4\wd0\vrule height0.9\ht0\hss}\box0}}
{\setbox0=\hbox{$\scriptscriptstyle\rm R$}\hbox{\hbox to0pt
{\kern0.4\wd0\vrule height0.9\ht0\hss}\box0}}}}
\def\Cl{{\mathchoice
{\setbox0=\hbox{$\displaystyle\rm C$}\hbox{\hbox to0pt
{\kern0.4\wd0\vrule height0.9\ht0\hss}\box0}}
{\setbox0=\hbox{$\textstyle\rm C$}\hbox{\hbox to0pt
{\kern0.4\wd0\vrule height0.9\ht0\hss}\box0}}
{\setbox0=\hbox{$\scriptstyle\rm C$}\hbox{\hbox to0pt
{\kern0.4\wd0\vrule height0.9\ht0\hss}\box0}}
{\setbox0=\hbox{$\scriptscriptstyle\rm C$}\hbox{\hbox to0pt
{\kern0.4\wd0\vrule height0.9\ht0\hss}\box0}}}}
\def\Hl{{\mathchoice
{\setbox0=\hbox{$\displaystyle\rm H$}\hbox{\hbox to0pt
{\kern0.4\wd0\vrule height0.9\ht0\hss}\box0}}
{\setbox0=\hbox{$\textstyle\rm H$}\hbox{\hbox to0pt
{\kern0.4\wd0\vrule height0.9\ht0\hss}\box0}}
{\setbox0=\hbox{$\scriptstyle\rm H$}\hbox{\hbox to0pt
{\kern0.4\wd0\vrule height0.9\ht0\hss}\box0}}
{\setbox0=\hbox{$\scriptscriptstyle\rm H$}\hbox{\hbox to0pt
{\kern0.4\wd0\vrule height0.9\ht0\hss}\box0}}}}
\def\Ol{{\mathchoice
{\setbox0=\hbox{$\displaystyle\rm O$}\hbox{\hbox to0pt
{\kern0.4\wd0\vrule height0.9\ht0\hss}\box0}}
{\setbox0=\hbox{$\textstyle\rm O$}\hbox{\hbox to0pt
{\kern0.4\wd0\vrule height0.9\ht0\hss}\box0}}
{\setbox0=\hbox{$\scriptstyle\rm O$}\hbox{\hbox to0pt
{\kern0.4\wd0\vrule height0.9\ht0\hss}\box0}}
{\setbox0=\hbox{$\scriptscriptstyle\rm O$}\hbox{\hbox to0pt
{\kern0.4\wd0\vrule height0.9\ht0\hss}\box0}}}}

\title{{\sf Group Action in Topos Quantum Physics}\\
}

\author{{\sf C. Flori$^1$}\thanks{{\sf cflori@perimeterinstitute.ca}}\\
\\
{\sf $^1$ Perimeter Institute for Theoretical Physics,}\\
{\sf 31 Caroline Street N, Waterloo, ON N2L 2Y5, Canada}}
\date{{\small\sf }}

\begin{document}

\maketitle
\begin{abstract}
Topos theory has been suggested first by Isham and Butterfield, and
then by  Isham and D\"oring, as an alternative mathematical structure
within which to formulate physical theories. In particular it has
been used to reformulate standard quantum mechanics in such a way
that a novel type of logic is used to represent propositions. In
this paper we extend this formulation to include the notion of a
group and group transformation in such a way that we overcome the
problem of twisted presheaves. In order to implement this we need
to change the type of topos involved, so as to render the notion
of continuity of the group action meaningful.
\end{abstract}
\vspace{.5in}
\emph{I would like to dedicate this paper to my uncle Luciano Romani and his son Libero Romani}
\newpage\section{Introduction}
In recent years Isham and D\"oring   have developed a new formulation of quantum theory based on the novel mathematical structure of topos theory first suggested by Isham and Butterfield, [4], \cite{isham1},  \cite{isham2},  \cite{isham3},  \cite{isham4},  \cite{andreas5}. The aim of this new formulation is to overcome the Copenhagen interpretation (instrumentalist interpretation) of quantum theory and replace it with an observer-independent, non-instrumentalist interpretation.

The strategy adopted to attain such a new formulation is to
re-express quantum theory as a type of `classical theory' in a
particular topos.
The notion of classicality in this setting is defined in terms of the notion of \emph{context} or \emph{classical snapshots}. In particular, in this framework, quantum theory is seen as a collection of local `classical snapshots', where the quantum information is determined by the relation of these local classical snapshots.

Mathematically, each classical snapshot is represented by an
abelian von-Neumann sub-algebra $V$ of the algebra
$\mathcal{B}(\mh)$ of bounded operators on a Hilbert space. The
collection of all these contexts forms a category $\mv(\mh)$,
which is actually a poset by inclusion.
As one goes to smaller sub-algebras $V^{'}\subseteq V$ one obtains a coarse-grained classical perspective on the theory.

The fact that the collection of all such classical snapshots forms a category, in particular a poset, means that the quantum information can be retrieved by the relations of such snapshots, i.e. by the categorical structure.

A topos that allows for such a classical local description is the
topos of presheaves over the category $\mv(\mh)$. This is denoted
as $\Sets^{\mv(\mh)^{\op}}$. By utilising the topos $\Sets^{\mv(\mh)^{\op}}$ to
reformulate quantum theory, it was possible to define pure quantum
states, quantum propositions and truth values of the latter
without any reference to external observer, measurement or any
other notion implied by the instrumentalist interpretation. In
particular, for pure quantum states, probabilities are replaced by
truth values, which derive from the internal structure of the
topos itself.
These truth values are lower sets in the poset $\mv(\mh)$, thus they are interpreted as the collection of all classical snapshots for which the proposition is true. Of course being true in one context implies that it will be true in any coarse graining of the latter.

However, this formalism lacked the ability to consider mixed
states in a similar manner as pure states, in particular it lacked
the ability to interpret truth values for mixed states as
probabilities. This problem was solved in \cite{probabilities} by
enlarging the topos $\Sets^{\mv(\mh)^{\op}}$ and considering instead the
topos of sheaves over the category $\mv(\mh)\times (0,1)_L$, i.e.
$Sh( \mv(\mh)\times(0,1)_L)$. Here $(0,1)_L$ is the category
whose open sets are the intervals $(0,r)$ for $0\leq r\leq 1$.
Within such a topos it is possible to define a logical
reformulation of probabilities also for mixed states. In this way
probabilities are derived internally from the logical structure of
the topos itself and not as an external concept related to
measurement and experiment. Probabilities thus gain a more
objective status which induces an interpretation in terms of
propensity rather than relative frequencies.

Moreover it was also shown in \cite{probabilities} that all that was done for the topos $Set^{\mv(\mh)^{\op}}$ can be translated to the topos $Sh( \mv(\mh)\times(0,1)_L)$.

Although much of the quantum formalism has been re-expressed in
the topos framework there are still many open questions and
unsolved issues. Of particular importance is the role of unitary
operators and the associated concept of group transformations. In
\cite{andreas1}, \cite{andreas2}, \cite{andreas3},
\cite{andreas4}, \cite{andreas5} the role of unitary operators in
the topos $\Sets^{\mv(\mh)^{\op}}$ was discussed and it was shown that
generalised truth values of propositions transform `covariantly'.
However, the situation is not ideal since `twisted' presheaves had
to be introduced. More precisely, given the spectral presheaf $\us$ \cite{andreas5} ( the topos analogue 
of the state space), for a group element $g\in G$ we have an arrow
$l_g:\us\rightarrow \us^{\hat{U}_g}$ but not an arrow
$l_g:\us\rightarrow \us$ as would be the case in classical physics. The object $\us^{\hat{U}_g}\in \Sets^{\mv(\mh)^{op}}$ is the twisted presheaf referred to $\us$. Essentially, what it does is to assign to each context $V\in \mv(\mh)$ not its spectrum, but the spectrum of the transformed algebra $l_{\hat{U}_g}(V):= \{\hat{U}_g\hat{A}\hat{U}_g^{-1}|\hat{A}\in V \}$.\\
This situation is (i) inelegant; (ii) not clear how it can be
generalise to an arbitrary topos; and (iii) it does not give a
clear indication of the
potential role of the geometry of the Lie group G and its orbits on $\mv(\mh)$.

Our aim in this paper is to give a precise definition of what a group and associated group transformation is in the topos representation of quantum theory, in such a way that we will not have the problem of twisted presheaves. In order to do this we will have to slightly change the topos we are working with. In particular, similarly as was done to account for probabilities,  we will have to consider the topos of sheaves over an appropriate category rather than the topos of presheaves. The reason for this shift is because we would like the group action to be continuous, thus we require a notion of topology. Such a notion is lacking in the definition of presheaves but it is present in the notion of sheaves.

It is interesting to note that the formulation of quantum theory, that we obtain by introducing the notion of a group and of group transformations, opens the door to a formulation of quantum theory in which we take into account all possible quantisations related by unitary operators.

In such a schema it would be possible to implement the notion of Dirac's covariance which states that, if we consider a physical state $|\psi\rangle\in\mh$ and a physical observable (self adjoint operator) $\hat{A}$ acting on $\mh$, then we would obtain the same physical predictions if we were to consider the state $\hat{U}|\psi\rangle$ and the physical observable $\hat{U}\hat{A}\hat{U}^{-1}$. Here $\hat{U}$ is any unitary operator. 

This means that in the canonical formulation of quantum theory,
the mathematical representatives of physical quantities are
defined only up to arbitrary transformations of the type
above. 

As a consequence, in non-relativistic quantum theory we obtain: 
\noindent
a) the canonical commutation relations; b) the angular-momentum
commutator algebra; c) and the unitary time displacement operator.
Similarly in relativistic
quantum theory we have the Poincar\'e group.

Although this is a topic of future work \cite{cecilia} it is however promising that the formalism obtained by introducing the notion of a group and group transformations sheds light on how to tackle a new formulation of quantum theory, in which all covariant representations are considered at the same time, i.e., a formalism in which all possible quantisations related by group transformations are considered at the same time.

If such a formulation is possible we would be able to precisely
understand/define the concept of quantisation in a topos. Work in
this direction has been done in \cite{quantization}.

\section{Presheaf over $\mv(\mh)$ as Sheaves over $\mv(\mh)$}
In this section we will introduce the notion of a sheaf and we will define the topos formed by the collection of all sheaves over a specific topological space. In such a topos the notion of a group and respective group action will be defined. 

The reason we decided to work with sheaves instead if presheaves, as has been done so far in the literature, stems from the fact that we are now interested in defining a group action, in particular a continuous group action. However, in order to define the notion of continuity we need the notion of topology, but presheaves do not retain any topological information neither of the base space nor of the `stalk space'. What is needed, instead, is a construction that takes into account the topological information of both the base space and the `stalk space'. This is precisely what a sheave achieves. 

A sheaf can essentially be thought of as a bundle with some extra
topological properties. In particular,
given a topological space $I$, a sheaf over $I$ is a pair $(A,p)$ consisting of a topological space $A$ and a continuous map $p:A\rightarrow I$, which is a local homeomorphism. By this we mean that for each $a\in A$ there exists an open set $V$ with $a\in V\subset A$, such that $p(V)$ is open in $I$ and $p_{|V}:V\rightarrow p(V)$ is a homeomorphism.

Thus, pictorially, one can imagine that to each point, in each
fibre, one associates an open disk (each of which will have a
different size) thus obtaining a stack of open disks for each
fibre. These different open discs are then glued together by the
topology on $A$.

The above is the more intuitive definition of
what a sheaf is. Now we come to the technical definition which is
the following:
\begin{Definition}
A sheaf of sets $F$ on a topological space $I$ is a functor
$F:\mathcal{O}(X)^{op}\rightarrow Sets$, such that each open
covering $U=\cup_{i}U_i$, $i\in I$ of an open set $U$ of $I$
determines an equaliser
\[\xymatrix{
F(U)\ar@{{>}->}[r]^e & \prod_i F(U_i)\ar@<3pt>[r]^{p}
\ar@<-3pt>[r]_{q} & \prod_{i,j}F(U_i\cap U_j) }\] where for $t\in
F(U)$ we have $e(t)=\{t|_{U_i}|i\in I\}$ and for a family $t_i\in
F(U_i)$ we obtain \be p\{t_i\}=\{t_i|_{U_i\cap
U_j}\},\;\;\;q\{t_i\}=\{t_j|_{U_i\cap U_j}\} \ee
\end{Definition}
The collection of all sheaves over a topological space forms a topos.

In the case at hand, since our base category $\mv(\mh)$ is a poset
we have an interesting result. In particular, each poset $P$ is equipped with an Alexandroff topology whose basis is given by the collection of all lower
sets in the poset $P$, i.e., by sets of the form $\downarrow
p:=\{p^{'}\in P|p^{'}\leq p\}$, $p\in P$\footnote{ 
Note that a function $\alpha
: P_1\rightarrow P_2$ between posets
$P_1$ and $P_2$ is continuous with respect to the Alexandroff topologies on each poset, if and only if it is order preserving.}.

The dual of such a topology is the topology of upper sets, i.e.
the topology generated by the sets $\uparrow p:=\{p^{'}\in P|
p^{'}\leq p\}$. Given such a topology it is a standard result
that, for any poset $P$, \be \Sets^P\simeq Sh(P^+) \ee where $P^+$
denotes the complete Heyting algebra of upper sets, which are the
duals of lower sets. It follows that \be \Sets^{P^{op}}\simeq
Sh((P^{op})^+)\simeq Sh(P^-) \ee where $P^-$ denotes the set of
all lower sets in $P$. In particular, for the poset $\mv(\mh)$ we have 
\be\label{equ:correspondence}
\Sets^{\mv(\mh)^{op}}\simeq Sh(\mv(\mh)^-) \ee
Thus every presheaf in our theory is in fact a sheaf with respect to the topology $\mv(\mh)^-$. We will denote by $\underline{\bar{A}}$ the sheaves over $\mv(\mh)$, while the respective presheaf will denote by $\underline{A}$. Moreover, in order to simplify the notation we will write $Sh(\mv(\mh)^-)$ as just $Sh(\mv(\mh))$.

We shall frequently use the particular class of lower sets in
$\mv(\mh)$ of the form \be \downarrow V:= \{V^{'}|V^{'}\subseteq
V\} \ee where $V\in Ob(\mv(\mh))$. It is easy to see that the set
of all of these is a basis for the topology $\mv(\mh)^-$. Moreover
\be \downarrow V_ 1\cap\downarrow V_2 =\downarrow (V_1\cap V_2)
\ee
i.e., these basis elements are closed under finite intersections.

It should be noted that $\downarrow V$ is the `smallest' open set
containing $V$ , i.e., the intersection of all open neighbourhoods
of $V$ is $\downarrow V$ . The existence of such a smallest open
neighbourhood is typical of an Alexandroff space.

If we were to include the minimal algebra
$\Cl(\hat{1})$ in $\mv(\mh)$ then, for any $V_1$, $V_2$ the
intersection $V_1\cap V_2$ would be non-empty. This would imply
that $\mv(\mh)$ is non Hausdorff.
To avoid this, we will exclude the minimal algebra from $\mv(\mh)$. This means that, when $V_1\cap V_2$ equals $\Cl(\hat{1})$ we will not consider it. 

More precisely the semi-lattice operation $V_1, V_2\rightarrow V_1 \wedge V_2$ becomes a partial operation which is defined as $V_1\cap V_2$ only if $V_1 \cap V_2\neq \Cl(\hat{1})$, otherwise it is zero.

\noindent
This restriction implies that when considering the topology on the
poset $\mv(\mh)-\Cl(\hat{1})$ we obtain \be
\downarrow V_1\cap\downarrow V_2=\begin{cases}\downarrow (V_1\cap V_2)& {\rm if}\hspace{.1in}V_1 \cap V_2\neq \Cl(\hat{1});\\
\emptyset& \text{otherwise}.
\end{cases}
\ee There are a few properties regarding sheaves on a poset worth
mentioning:
\begin{enumerate}
\item In general the sub-objects
of a presheaf form a Heyting algebra but it may not be complete.
However, for sheaves we have the following theorem
\begin{Theorem}
For any sheaf $\underline{\bar{E}}$ on a site $(C, J)$, the lattice $Sub(\underline{\bar{E}})$ of all
sub-sheaves of $\underline{\bar{E}}$ is a complete Heyting algebra
\end{Theorem}
The proof can be found in \cite{topos7}. It follows that for any
$\underline{\bar{A}}\in Sh(\mv(\mh))$,  $Sub(\underline{\bar{A}})$
is a complete Heyting algebra. Of particular importance is the
collection of sub-objects of the spectral sheaf, i.e.
$Sub(\bar{\us})$.
\item When constructing sheaves it suffices to restrict attention to the basis elements of
the form $\downarrow V$ , $V\in Ob(\mv(\mh)$. For a given presheaf
$\underline{A}$, a key relation between its associated sheaf,
$\underline{\bar{A}}$ is simply \be\label{equ:up}
\underline{\bar{A}}(\downarrow V ) := \underline{A}_V \ee where
the left hand side is the sheaf using the topology $\mv(\mh)^{-}$
and the right hand side
is the presheaf on $\mv(\mh)$.\\
Given a presheaf map, there is an associated restriction map for
sheaves. In particular, given $i_{V_1V} : V_1\rightarrow V$ with
associated presheaf map $\underline{A}(i_{V_1V} ) :
\underline{A}_V\rightarrow \underline{A}_{V_1}$ , then the
restriction map $\rho_{V_1V} :\underline{\bar{A}}(\downarrow V
)\rightarrow\underline{\bar{A}}(\downarrow V_1)$ for the sheaf
$\underline{\bar{A}}$ is defined as \be a_{|\downarrow V_1} =
\rho_{V_1V} (a) := \underline{A}(i_{V_1V} )(a) \ee for all $a\in
\underline{\bar{A}}(\downarrow V )\simeq \underline{A}_V$.
\item Given an open set $\mathcal{O}$ in $\mv(\mh)^-$ such a set is covered by the down set $\downarrow V$, $V\in Ob(\mv(\mh))$. Therefore we have (see \cite{topos3})
\be 
\underline{\bar{A}}(\mathcal{O}) = {\lim_{\longleftarrow}}_{ V\subseteq
\mathcal{O}}\underline{\bar{A}}(\downarrow V) = {\lim_{\longleftarrow}}_{ V\subseteq
\mathcal{O}}\underline{A}_V \ee As an example let us consider the open set
$\mathcal{O}:=\downarrow V_1\cup\downarrow V_2$. Applying the
definition of the inverse limit of sets we obtain \ba
\underline{\bar{A}}(\downarrow V_1\cup\downarrow V_2)&=&{\lim_{\longleftarrow}}_{ V\in\{V_1, V_2\}}\underline{A}_V\\
&=&\{\langle\alpha, \beta\rangle \in \underline{A}_{V_1}\times\underline{A}_{V_2}|\alpha_{|V_1\cap V_2}=\beta_{|V_1\cap V_2}\}\\
&=&\{\langle\alpha, \beta\rangle
\in\bar{\underline{A}}(\downarrow
V_1)\times\bar{\underline{A}}(\downarrow V_2)|\alpha_{\downarrow
(V_1\cap V_2)}=\beta_{|\downarrow(V_1\cap V_2)}\} \ea A direct
consequence of the above is that \be
\underline{\bar{A}}(\mathcal{O})=\Gamma\underline{A}_{|\mathcal{O}}
\ee The connection with \ref{equ:up} is given by the fact that
$\Gamma\underline{\bar{A}}_{|\downarrow V}\simeq\underline{A}_V$.

\item The concept of the bundle $\Lambda\bar{\underline{A}}$ of germs of a sheaf $\bar{\underline{A}}$ simplifies for our Alexandroff base spaces as given any point $V\in\mv(\mh)$, there is a unique smallest open set, namely $\downarrow V$, to which $V$ belongs.

Let $\mathcal{O}_1$ and $\mathcal{O}_2$ be open neighbourhoods of $V\in\mv(\mh)$ with $s_1\in\underline{\bar{A}}(\mathcal{O}_1)$ and $s_2\in\underline{\bar{A}}(\mathcal{O}_2)$.
Then $s_1$ and $s_2$ have the same germ at $V$ if there is some open $\mathcal{O}\subseteq \mathcal{O}_1\cap \mathcal{O}_2$ such that
$s_{1|\mathcal{O}} = s_{2|\mathcal{O}}$. Since $\mv(\mh)$ has the Alexandroff topology, we can see at once that $s_1$ and
$s_2$ have the same germ at $V$ iff 
\be
s_{1|\downarrow V} =s_{2|\downarrow V}
\ee
It follows at once that if $V\in \mathcal{O}$, $s\in\underline{\bar{A}}(\mathcal{O})$, then $germ_V s = s_{|\downarrow V}$ . Then
\ba
(\Lambda\underline{\bar{A}})_V&=&\underline{\bar{A}}(\downarrow V )\\
&\simeq&\underline{\bar{A}}_V 
\ea
This is consistent with the general result that
\be
(\Lambda\underline{\bar{A}})_V = {\lim_{\longrightarrow}}_{V\in \mathcal{O}}=\underline{\bar{A}}(\mathcal{O})
\ee

\item For presheaves on partially ordered sets the sub-object classifier $\uom^{\mv(\mh)}$ has some interesting properties. In particular, given the set $\uom^{\mv(\mh)}_V$ of sieves on $V$, there exists a bijection between sieves in $\uom^{\mv(\mh)}_V$ and lower sets of $V$. To understand this let us consider any sieve $S$ on $V$, we can then define the lower set of $V$
\be L_S:=\bigcup_{V_1\in S}\downarrow V_1 \ee Conversely, given a
lower set $L$ of $V$ we can construct a sieve on $V$ \be
S_L:=\{V_2\subseteq V|\downarrow V_2\subseteq L\} \ee However if
$\downarrow V_2\subseteq L_S$ then $\downarrow V_2\subseteq
\bigcup_{V_1\in S}\downarrow V_1$, therefore $V_2\in S$. On the
other hand if $V_2\in S_L$ ($S_L$ sieve on $V$), then
$V_2\subseteq V$ and $\downarrow V_2\subseteq L$, therefore
$V_2\in \bigcup_{V_1\in S}\downarrow V_1$, i.e. $V_2\in L_S$. This
implies that the above operations are inverse of each other.
Therefore \be \bar{\uom}^{\mv(\mh)}(\downarrow
V):=\uom^{\mv(\mh)}_V\simeq \Theta(V) \ee where $\Theta(V)$ is
the collection of lower subsets (i.e. open subsets in $\mv(\mh)$)
of $V$. This is equivalent to the fact that, in a topological
space $X$, we have that $\Omega^X(O)$ is the set of all open
subsets of $O\subseteq X$.
\end{enumerate}

\subsection{Sheaves as Bundles with Topology}
In this section we will analyse how it is possible to give a certain topology to the sheaves we are considering. In fact, when we consider a sheaf as a bundle $p:X\rightarrow Y$, where the map $p$ is continuous, the space $X$ is not always given a topology a priori (it does however have the etal\'e topology given by the fact that it is a sheaf). Of course if we consider the sheaf $\underline{X}$ as an etal\'e bundle $p:X\rightarrow Y$ (whose total space is the bundle, $\Lambda{X}$, of germs of $\underline{X}$) then, the topology of each fibre of such a bundle will be discrete, i.e. the bundle will be equipped with the etal\'e topology. However there will be some sheaves in which this is not the case, i.e. the stalks will not have a discrete topology on them. A prime example is the spectral presheaf $\us$ in which each stalk $\us_V$ for each $V\in \mv(\mh)$ has a priori the spectral topology, which in the case that $\mh$ is infinite dimensions  may not be discrete. Thus the question is how to incorporate the spectral topology in the sheaf structure. 

In particular the objects whose topology we would like to define are (i) spectral sheaf $\us$; (ii) the clopen sub-objects of $\us$; and (iii) the quantity-value sheaf.

\noindent
Let us start with the spectral sheaf $\us$ and define the set $\Sigma:=\coprod_{V\in \mv(\mh)}\us_V:=\bigcup_{V\in\mv}\{V\}\times \us_V$, with associated map $p_{\Sigma}:\Sigma\rightarrow \mv(\mh)$ defined by $p_{\Sigma}(\lambda)=V$ where $V$ is the context, such that $\lambda\in\us_V$. In this context each $\us_V=p_{\Sigma}^{-1}(V)$, i.e. they are the fibres of the map $p_{\Sigma}$. 

Our aim is to give $\Sigma$ a topological structure with the
minimal requirement that the projection map $p_{\Sigma}$ is
continuous.
A possible choice would be the disjoint union topology determined by the spectral topology on the subsets $\us_V$. Such a topology is the strongest topology on $\Sigma$, such that it induces the spectral topology on each fibre.

\noindent
Given such a topology we then have $p_{\Sigma}^{-1}(\downarrow
V):=\coprod_{V^{'}\subseteq V}\us_{V^{'}}$. Since each $\us_V$ is
open in $\Sigma:=\coprod_{V\in \mv(\mh)}\us_V$, the map
$p_{\Sigma}$ is continuous in such topology.

However, we would like to incorporate in the topology of $\Sigma$,
not only the `vertical' topology of each fibre, but also the
`horizontal' Alexandroff topology coming from the base space
$\mv(\mh)$. Thus we are looking for a topology on $\Sigma$ which,
locally, would look like the product topology of the `horizontal'
and fibre topology. Unfortunately, the topology defined above does not
allow for such product topology since at each context (locally) we
have the open sets $\us_V$.

An alternative topology that one could consider is the topology
associated to the clopen sub-objects of $\us$. In particular, to
each clopen sub-object $\underline{S}\subseteq \us$ we associate
the subset $S$ of $\Sigma$ defined as
$S:=\coprod_{V\in\mv(\mh)}\underline{S}_V$. The collection of all
these subsets is algebraically closed under finite unions and
intersections, but not under arbitrary unions. In fact, arbitrary
unions of clopen subset of $\us_V$ are open but not necessarily
clopen. However, De Groote has shown in \cite{Groote} that for any unital abelian
von-Neumann algebra $V$ , the collection of clopen subsets of
$\us_V$ form a base for the
spectral topology on $\us_V$. Thus a possible topology on $\Sigma$ is the topology whose basis sets are the collection of clopen sub-objects of $\Sigma$. We will call this the \emph{spectral topology} on $\Sigma$.

Clearly, the topology induced on each fibre $\us_V$ by the
spectral topology is the original spectral topology on the
Gel'fand spectrum $\us_V$ of $V$. This implies that the spectral
topology on $\Sigma$ is weaker than the canonical topology
(product topology).
This, in turn, implies that the map $p_{\Sigma}:\Sigma\rightarrow \mv(\mh)$ is continuous with respect to the spectral topology on $\Sigma$. In fact $p^{-1}(\downarrow V )=\coprod_{V^{'}\in \downarrow V}\us_{V^{'}}$ represents the clopen sub-object of $\Sigma$ which has value $\us_{V^{'}}$ for each $V^{'}\in \downarrow V$ and $\emptyset $ everywhere else.

Given the above discussion, it follows that the bundle $p_{\Sigma}
:\Sigma\rightarrow \mv(\mh)$ is generally not etal\'e. For example
if $\mh$ is infinite-dimensional the spectral topology on a
spectral space $\us_V$, $V\in\mv(\mh)$ may not be discrete.
However, if we consider the sheaf $\bar{\us}$ associated to the
presheaf $\Sigma$ we know that $\Lambda\bar{\us}_V\simeq\us_V$ for
all $V\in\mv(\mh)$. Thus the fibres of the etal\'e bundle
$p_{\bar{\us}}:\Lambda\bar{\us}\rightarrow \mv(\mh)$ are
isomorphic to
those of the bundle $p_{\Sigma}:\Sigma\rightarrow \mv(\mh)$. 

In particular, if we consider both $\Lambda \bar{\us}$ and $\us$
as sets, there exists a bijective bundle map $i:\Lambda
\bar{\us}\rightarrow \us$. Since each fibre in $\Lambda \bar{\us}$
is discrete the map is obviously continuous. However, in the
infinite-dimensional case $i$ may not be bi-continuous.

The topology on the clopen sub-objects of $\bar{\us}$ is simply the subspace topology. In particular, denoting the
spectral topology on $\Sigma:=\coprod_{V\in \mv(\mh)}\us_V$, by
$\tau$, the topology on any clopen subset
$S:=\coprod_{V\in\mv(\mh)}\underline{S}_V$ is defined as \be
\tau_S:=\{S\cap U|U\in \tau\} \ee
Since both $S$ and $U$ are clopen subsets, their intersection also is a clopen subset. \\

We now want to analyse the topology for the quantity value object.
To this end we recall the definition of the quantity value object.
\begin{Definition}
The quantity valued object is identified with the $\Rl$-valued
presheaf, $\underline{\Rl}^{\leftrightarrow}$ of order-preserving
and order-reversing functions on $\mv(\mh)$ which is defined as
follows:
\begin{itemize}
\item [i)] On objects $V\in Ob(\mv(\mh))$:
\be \underline{\Rl}^{\leftrightarrow}_V:=\{(\mu, \nu)|\mu\in
OP(\downarrow V,\Rl)\;, \nu\in OR(\downarrow V, \Rl),\; \mu\leq
\nu\} \ee The condition $\mu\leq \nu $ implies that for all
$V^{'}\in \downarrow V$, $\mu(V^{'})\leq\nu(V^{'})$.
\item[ii)] On morphisms $i_{V^{'},V} :V^{'}\rightarrow V$, ($V^{'}\subseteq V$) we get
\ba
\underline{\Rl}^{\leftrightarrow}(i_{V^{'},V} ):\underline{\Rl}^{\leftrightarrow}_V&\rightarrow&\underline{\Rl}^{\leftrightarrow}_{V^{'}}\\
(\mu, \nu)&\mapsto&(\mu_{|V^{'}}, \nu_{|V^{'}}) \ea where
$\mu_{|V^{'}}$ denotes the restriction of $\mu$ to $\downarrow
V^{'}\subseteq \downarrow V$ and analogously for $\nu_{|V^{'}} $.
\end{itemize}
\end{Definition}
Given such a presheaf we define the set $\Rl^{\leftrightarrow}:=\coprod_{V\in\mv(\mh)}\underline{\Rl}^{\leftrightarrow}_V$ with associated map $p_{\mathcal{R}}:\Rl^{\leftrightarrow}\rightarrow \mv(\mh)$ such that $p_{\mathcal{R}}(\mu,\nu)=V$ for $(\mu,\nu)\in\underline{\Rl}^{\leftrightarrow}_V$. We would like to define a topology on $\Rl^{\leftrightarrow}$ such that the map $p_{\mathcal{R}}$ be continuous. 

A possibility would be to define the discrete topology on each fibre $p_{\mathcal{R}}^{-1}(V)= \underline{\Rl}^{\leftrightarrow}_V$
which would accommodate for the fact that the bundle is an etal\`e bundle. We could then define the disjoint union topology, but this would not account for the `horizontal' topology on the base category $\mv_f(\mh)$. 

Another possibility would be to consider as a basis for the topology on $\Rl^{\leftrightarrow}$, the collection of all open sub-objects. Thus a basis set would be of the form $\underline{S}=\coprod_{V\in\mv(\mh)}\underline{S}_V$ such that $\underline{S}_V$ is open in $\underline{\Rl}^{\leftrightarrow}_V$. In such a setting the `horizontal' topology would be accounted for by the presheaf maps.

\noindent
Since each $\underline{\Rl}^{\leftrightarrow}_V$ is equipped with the discrete topology, the topology on the entire set $\Rl^{\leftrightarrow}$ would essentially be the discrete topology in which all sub-objects of
$\Rl^{\leftrightarrow}$ are open.

\noindent
Obviously, with respect to such a topology, the bundle map
$p_{\mathcal{R}}$ would be continuous since for each $\downarrow V$, $p^{-1}(\downarrow V)=\coprod_{V^{'}\in\down V}\underline{\Rl}^{\leftrightarrow}_{V^{'}}$ will represent the open sub-object whose value is $\underline{\Rl}^{\leftrightarrow}_{V^{'}}$ for all $V^{'}\in \down V$ and $\emptyset $ everywhere else.


\section{Group Action on the Original Base Category $\mv(\mh)$}

\subsection{Alexandroff Topology}
We will now analyse the group action on our original category $\mv(\mh)$ equipped with the Alexandroff topology. We recall that the Alexandroff topology on $\mv(\mh)$ is the topology whose basis are all the lower sets $\downarrow V$, for all $V\in\mv(\mh)$.\\
Let us consider a unitary operator $\hat{U}$ which acts on
$\mv(\mh)$. Such an action is defined as \ba
l_{\hat{U}}:\mv(\mh)&\rightarrow& \mv(\mh)\\
V&\mapsto&\hat{U}V\hat{U}^{-1} \ea
where $\hat{U}V\hat{U}^{-1}:=\{\hat{U}\hat{A}\hat{U}^{-1}|\hat{A}\in V\}$.\\
It is easy to see that this action is continuous with respect to
the Alexandroff topology since it preserves the partial order on
$\mv(\mh)$. Therefore we obtain a representation of the Lie group $G$ of the form
\ba
g\leadsto l_{g} : \mv(\mh) &\rightarrow &\mv(\mh)\\
V&\mapsto&\hat{U}_gV\hat{U}_{g^{-1}} \ea
where each map $l_g$ for $g\in G$ is continuous. \\
Moreover, the map $G\rightarrow \mathcal{U}(\mh)$, $g\rightarrow \hat{U}_g$, is strongly continuous, i.e., the map $g\rightarrow \hat{U}_g|\psi\rangle$ is a norm-continuous function for all $|\psi\rangle\in \mh$.

However, the definition of a proper representation of the topological group $G$ also requires the following map to be
continuous: \ba\label{equ:alexcont}
\Phi:G\times\mv(\mh)&\rightarrow& \mv(\mh)\\
(g,V)&\mapsto&l_g(V) \ea 
To prove continuity it suffices to consider only the basis open sets only, i.e., the sets 
$\downarrow V, V\in \mv(\mh)$. Thus we consider \ba
\Phi^{-1}(\downarrow V)&=&\{(g,V^{'})|l_{g}V^{'}\in\downarrow V\}\\
&=&\{(g,V^{'})|l_{g}V^{'}\subseteq V\} \ea A necessary condition
for this to be continuous is that, for each $V\in\mv(\mh)$,  the
induced map \ba
f_V : G&\rightarrow&\mv(\mh)\\
g&\mapsto& l_g(V) \ea is continuous. If we consider the open set
$\downarrow V\in\mv(\mh)$ we then have \ba
f^{-1}_V(\downarrow V)&=&\{g\in G|l_gV\in\downarrow V\}\\
&=&\{g\in G|l_gV\subseteq V\}\\
&=&\{g\in G|l_gV= V\}\\
&=:&G_V \ea
where $G_V$ is the stabiliser of $V$. The last equality follows since the group action can not transform an algebra into a proper sub-algebra of itself.

\noindent
Thus in order to show that the action is continuous we need to show that the stability group $G_V$ is open.

We know from the result in the Appendix that if $\mv(\mh)$ is Hausdorff, then $G_V$ is closed, and since a typical Lie group does not have clopen subgroups it follows that the action is not continuous.

\noindent
However $\mv(\mh)$ is not Hausdorff. In fact, given $V_1,
V_2\in\mv(\mh)-\Cl\hat{1}$, with corresponding lower sets
$\downarrow V_1, \downarrow V_2$, the smallest neighbourhood
containing both is \be
\downarrow V_1\cap \downarrow V_2=\begin{cases} \downarrow (V_1\cap V_2)& {\rm if}\hspace{.1in}V_1\cap V_2\neq \Cl\hat{1};\\
\emptyset& \text{otherwise}.
\end{cases}
\ee Obviously the RHS might not be empty therefore
to prove that $G_V$ is closed we will need to use another
strategy.
\begin{Lemma}\label{lem:closedV}
For each $V\in\mv(\mh)$ the stabiliser $G_V$ is a closed subgroup
of the topological group $G$.
\end{Lemma}
\begin{Proof}
Given a unitary representation of $G$ on the Hilbert space $\mh$,
the map $G\rightarrow \mathcal{U}(\mh)$ is strongly continuous, i.e., the
map $g\rightarrow \hat{U}_g |\psi\rangle$ is a norm continuous
function for every $|\psi\rangle\in\mh$. Now let $g_{\nu}$,
$\nu\in I$ (a directed index set), be a net of elements of $G$ in
$G_V$, i.e., \be \hat{U}_{g_{\nu}}V\hat{U}_{g^{-1}_{\nu}}=V \ee
for all $\nu\in I$. In other words, given any self-adjoint
operator $\hat{A}\in V$ we obtain \be
\hat{U}_{g_{\nu}}\hat{A}\hat{U}_{g^{-1}_{\nu}}\in V \ee for all
$\nu\in I$. We assume that the net of group elements converges
with $\lim_{\nu\in I}g_{\nu}=g$. Since the $G$ representation is
strongly continuous then $\hat{U}_{g_{\nu}}$ converges strongly to
$\hat{U}_g$. We will denote strong convergence by $
\hat{U}_{g_{\nu}}\mapsto_{s}\hat{U}_g$. In order to show that
$G_V$ is closed we need to show that \be
\hat{U}_g\hat{A}\hat{U}_{g^{-1}}\in V \ee However, operator
multiplication is such that if $
\hat{U}_{g_{\nu}}\mapsto_{s}\hat{U}_g$ then $
\hat{U}_{g_{\nu}}\hat{A}\mapsto_{s}\hat{U}_g\hat{A}$. Since
$\hat{U}^{\dagger}_{g_{\nu}}\mapsto_{s}\hat{U}^{\dagger}_g$ it
follows that \be
\hat{U}_{g_{\nu}}\hat{A}\hat{U}_{g_{\nu}^{-1}}\mapsto_{w}\hat{U}_g\hat{A}\hat{U}_{g_{\nu}^{-1}}
\ee
where $\mapsto_{w}$ denotes convergence in the weak operator topology.

Von Neumann algebras are weakly closed, thus
$\hat{U}_g\hat{A}\hat{U}_{g_{\nu}^{-1}}$ belongs to $V$ and $G_V$
is closed.
\end{Proof}
It follows that the group action in equation \ref{equ:alexcont} is
not continuous.

\subsection{Vertical Topology on $\mv(\mh)$}
We will now try to construct a different topology on $\mv(\mh)$
which we will call the \emph{vertical topology}. Such a topology
will take into account the usual topology on coset spaces. Before
defining the `vertical' topology for $\mv(\mh)$, we will first list
certain properties of the coset topology. Such properties will be
useful in subsequent sections.

Given a closed subgroup $H$ of a Hausdorff topological group G,
the topology on the coset space $G/H$  is given by the
identification topology whose open sets are $\{U\subseteq
G/H|\text{ iff } p^{-1}(U) \text{ open in } G\}$. Here $p :
G\rightarrow G/H$ is the quotient map.
\begin{Lemma}
The map $p : G\rightarrow G/H$ is open.
\end{Lemma}
\begin{Proof}
Given an open set $O\subseteq G$, we have \be\label{equ:open}
p^{-1}(p(O))=\bigcup_{h\in H}r_h(O) \ee where $r_h : G\rightarrow
G$ is the right translation by $h\in H$. However $r_h$ is a
homeomorphism of
$G$ with itself, therefore $r_h(O)$ is an open subset of G. Since the finite union of open sets is open, equation \ref{equ:open} implies that $p^{-1}(p(O))$ is open.

If $O$ is open in $G$ implies that $p^{-1}(p(O))$ is also open
in $G$ then $p(O)$ is open in $G/H$, therefore $p$ is an open map.
Conversely, if $p$ is an open map then $O$ open in $G$ implies
that $p(O)$ open in $G/H$ therefore $p^{-1}p(O)$ open in $G$.
\end{Proof}
\begin{Lemma}
$G/H$ is Hausdorff.
\end{Lemma}
\begin{Proof}
Let us consider any two distinct elements $p(w_1), p(w_2)$ in $G/H$. We then have that $w_1$ and $w_2$ are not related. Moreover, since $G$ is Hausdorff, it is possible to fine two open sets $V_1$ and $V_2$ such that $w_1\in V_1$, $w_2\in V_2$ and $V_1\cap V_2=0$. Since $p$ is open the sets $p(V_1)$ and $p(V_2)$ are open and $p(w_1)\in p(V_1)$ while $p(w_2)\in p(V_2)$. However since the elements $w_1$ and $w_2$ are not related it follows that one can choose the two opens $V_1$ and $V_2$ such that they belong to $H^c$ (complement of $H$, where $H$ is closed). Thus $p(V_1)\cap p(V_2)=0$

%
\end{Proof}
We will now prove a lemma which will reveal itself very important
in subsequent sections.
\begin{Lemma}\label{lem:vertical}
For all $g\in G$ the map $\Phi:G\times G/H\rightarrow G/H$,
$(g, g_oH)\mapsto gg_0H$ is continuous.
\end{Lemma}
\begin{Proof}
Consider the chain of maps \ba
G\times G&\xrightarrow{id_G\times p}&G\times G/H\xrightarrow{\Phi}G/H\\
( g, g_0) &\mapsto& (g, g_0H)\mapsto gg_0H \ea This
can be combined with the chain \ba
G\times G&\xrightarrow{\mu}&G\xrightarrow{p}G/H\\
( g, g_0) &\mapsto& gg_0\mapsto gg_0H \ea to give a
commutative diagram:

\[\xymatrix{
G\times G\ar[rr]^{id_G\times p}\ar[dd]_{\mu}&&G\times G/H\ar[dd]^{\Phi}\\
&&\\
G\ar[rr]_{p}&&G/H\\
}\]

Here the map $\mu:G\times G\rightarrow G$ represents multiplication.

\noindent
Because of commutativity of the diagram we have that $\Phi\circ
(id_G\times p)=p\circ \mu$ where, by definition $p\circ \mu$ is
continuous. Thus $\Phi\circ (id_G\times p)$ is continuous.

Now we need to show that $id_G\times p$ is:
(i) continuous; (ii) open.

(i) We consider an open set $U$ in $G/H$, which by definition of
the identification topology is open iff $p^{-1}(U) $ is open in
$G$. Thus, without loss of generality we can choose, in the
product topology, of $G\times G/H$ the open $\langle p^{-1}(U) ,U\rangle$. We then have \ba (id_G\times p)^{-1}\langle
p^{-1}(U) ,U\rangle=\langle id_G^{-1}p^{-1}(U)
,p^{-1}(U)\rangle=\langle p^{-1}(U) ,p^{-1}(U)\rangle \ea
This again is open by the definition of the identification topology\footnote{Continuity of $id_G\times p$ also follows from the definition of product map $id_G\times p:=\langle ig_G\circ pr_1, p\circ pr_2\rangle$ and the fact that both $p$ and $id_G$ are continuous. }.

(ii) Let us now consider any open set $\langle A, B\rangle \in G\times G$, then $(id_G\times p)(\langle A, B\rangle)=\langle ig_G\circ pr_1, p\circ pr_2\rangle(\langle A, B\rangle)= \langle A, p(B)\rangle$ is open since $A$ is open and $p$ is an open map. 

Given the above we can now show that $\Phi$ is continuous. To this
end let us consider an open set $U\in G/H$, we then have that \be
\Big(\Phi\circ (id_G\times p)\Big)^{-1}(U)=(id_G\times
p)^{-1}\circ \Phi^{-1}(U) \ee is open. However if
$\Phi^{-1}(U)$ were to be not open, then \be
(id_G\times p)\circ\Big((id_G\times p)^{-1}\circ
\Phi^{-1}(U)\Big)=(id_G\times p)\circ(id_G\times p)^{-1}\circ
\Phi^{-1}(U)=\Phi^{-1}(U) \ee
is closed. The second equality follows for the definition of product maps and the fact that both $id_G$ and $p$ are surjective ($p^{-1}p(U)=U$).

However, we know that $(id_G\times p)$ is open and that
$(id_G\times p)^{-1}\circ \Phi^{-1}(U)$ is an open set. Thus we
get a contradiction, therefore $\Phi $ is continuous.
\end{Proof}
We are now ready to define the `vertical' topology of $\mv(\mh)$.
This is the weak topology associated with the orbits of $G$ on
$\mv(\mh)$. Hence its basis open sets are \be \mathcal{O}(V,
N):=\{l_g(V)|g\in N\subseteq G\} \ee
where $V\in \mv(\mh)$ and $N$ is open in $G$.

Since the action of $G$ is transitive on each orbit by
construction, it suffices to let $N$ be a neighbourhood of the
identity element, $e$, of $G$. The sets $\mathcal{O}(V,N)$ are then a basis
of the neighbourhood filter of $V$. Given this definition we have
the following result: \ba\label{ali:ver}
\mathcal{O}(V,N_1)\cap \mathcal{O}(V,N_2)& = &\{l_g(V) |g\in N_1\subseteq G\}\cap \{l_g(V) | g\in N_2\subseteq G\}\\
&=& \{l_g(V) | g\in N_1\cap N_2\}\\
&=&\mathcal{O}(V,N_1\cap N_2) \ea Because of the `vertical' nature of the
topology, and the fact that $G$ acts continuously on each orbit, intuitively one would guess that the G-action on $\mv(\mh)$ is continuous
in the `vertical' topology; i.e., the map $G\times
\mv(\mh)\rightarrow\mv(\mh)$ is continuous. A formal proof is as
follows.
\begin{Lemma}\label{lem:vcontinuos}
The map $G\times\mv(\mh)\rightarrow \mv(\mh)$ is continuous in the
`vertical' topology.
\end{Lemma}
In the proof of this lemma we will use a standard result in
topology which we will report here for completeness reasons.
\begin{Theorem}\label{the:continuous}
Given a topological space $Y$ and a topological space $X$ whose topology is determined by the
family $\{A_{\alpha}|\alpha\in I\}$ of subsets of $X$, each with
its own topology, then a map $f:X\rightarrow Y$ is continuous iff
each $f|A_{\alpha}:A_{\alpha}\rightarrow Y$ is continuous.
\end{Theorem}
The proof can be found in \cite{dugund}. We will now give the
proof for Lemma \ref{lem:vcontinuos}.
\begin{Proof}
Given the nature of the poset $\mv(\mh)$ it follows that it can be
written as \be \mv(\mh)=\coprod_{w\in\mv(\mh)/G}\mathcal{O}_w \ee
where $\mathcal{O}_w$ is the orbit associated with the coset $w$.
Thus the group action map is now \ba
\Phi: G\times \coprod_{w\in\mv(\mh)/G}\mathcal{O}_w&\rightarrow& \coprod_{w\in\mv(\mh)/G}\mathcal{O}_w\nonumber\\
 \coprod_{w\in\mv(\mh)/G}\mathcal{O}_w\times G&\rightarrow& \coprod_{w\in\mv(\mh)/G}\mathcal{O}_w
\ea
where we have used that $G\times \coprod_{w\in\mv(\mh)/G}\mathcal{O}_w\simeq  \coprod_{w\in\mv(\mh)/G}G \times\mathcal{O}_w $. Since the $G$-action is `vertical', i.e., the G-action acts ÔverticallyÕ on each individual fibre/orbit, then according to theorem \ref{the:continuous} we have that
$\Phi: G\times \coprod_{w\in\mv(\mh)/G}\mathcal{O}_w \rightarrow
\coprod_{w\in\mv(\mh)/G}\mathcal{O}_w$ is continuous iff
$\Phi|\mathcal{O}_w:G\times \mathcal{O}_w\rightarrow
\coprod_{w\in\mv(\mh)/G}\mathcal{O}_w$ is continuous. However the group action on each orbit is continuous by definition
thus $\Phi$ is continuous.
\end{Proof}
We will denote $\mv(\mh)$ with the `vertical' topology as
$\mv(\mh)_{ver}$.

\subsection{Bucket Topology}
From the results of the previous section it is clear that in order
to arrive to the situation where the $G$-action is continuous we
need to change the topology on $\mv(\mh)$. One striking feature of
the Alexandroff topology on $\mv(\mh)$ is that the induced topology
on each orbit of the form $G/G_V$ is discrete, so it is hardly
surprising that things go wrong!

A possible way of defining a topology on $\mv(\mh)$, which renders
the action continuous, is to combine the `vertical' topology with
the Alexandroff topology. To do this we define (following very
useful discussions with Ieke Moerdijk) the basis of the `bucket'
topology as all sets of the form \be\label{equ:bucket}
\downarrow\mathcal{O}(V, N):=\bigcup_{g\in N}\downarrow l_g(V) \ee
where $V\in \mv(\mh)$ and $N\subseteq G$ is an open neighbourhood
of $e\in G$. These `buckets' are a basis
for the neighbourhood filter of $V$ in the bucket topology. 

We note that since $\downarrow (V_1\cap V_2) = \downarrow V_1\cap
\downarrow V_2$, \ref{ali:ver} shows that \be \downarrow
\mathcal{O}(V, N_1)\cap \downarrow \mathcal{O}(V, N_2)=\downarrow
\mathcal{O}(V, N_1\cap N_2) \ee The following lemma will now be
useful; here $GV$ indicates the $G$ orbit through $V$,
i.e. $GV:=\{l_gV|g\in G\}$.
\begin{Lemma}\label{lem:orbit}
if $V_1\subseteq V$ then $\downarrow \mathcal{O}(V, N)\cap G V_1$
is open in the `vertical' topology.
\end{Lemma}
\begin{Proof}
Given an element $V_0\in \downarrow \mathcal{O}(V, N)\cap G V_1$,
then there exists an element $W\in \mathcal{O}(V, N)$ such that
$V_0\subseteq W$. Since $ \mathcal{O}(V, N)$ is open in the
`vertical' topology there exists some $N_0$ such that
$W\in\mathcal{O}(W, N_0)\subseteq \mathcal{O}(V, N)$. We have seen
above that the action of each $g\in G$ on $\mv(\mh)$ is order preserving,
thus $V_0\subseteq W$ implies that $l_gV_0\subseteq  l_gW$ for all
$g\in G$. Therefore we have \be V_0\in \mathcal{O}(V_0,
N_0)\subseteq \downarrow \mathcal{O}(W, N_0)\cap G V_1\subseteq
\downarrow \mathcal{O}(V, N)\cap G V_1 \ee
\end{Proof}
A direct consequence of the above lemma is that the bucket topology induces the `vertical' topology on each $G$-orbit. Moreover it is clear from equation \ref{equ:bucket} that each bucket is the union of Alexandroff open sets. Therefore, every set open in the bucket topology is also open in the Alexandroff topology. The inverse however is not true. If follows that the bucket topology is \emph{weaker} than the Alexandroff topology.
\begin{Lemma}
The bucket topology is not Hausdorff.
\end{Lemma}
\begin{Proof}
Given any two elements $V_1, V_2\in\mv(\mh)$ the smallest neighbourhoods of each are respectively $\downarrow\mathcal{O}(V_1,N_1)$ and $\downarrow\mathcal{O}(V_2,N_2)$. However
\be
\downarrow\mathcal{O}(V_1,N_1)\cap
\downarrow\mathcal{O}(V_2,N_2):=\downarrow\mathcal{O}(V_1\cap
V_2,N_1\cap N_2) \ee and, we know that \be
 V_1\cap\downarrow V_2=\begin{cases}\downarrow (V_1\cap V_2)& {\rm if}\hspace{.1in}V_1 \cap V_2\neq \Cl\hat{1};\\
\emptyset& \text{otherwise}.
\end{cases}
\ee is not always empty. Therefore $\downarrow\mathcal{O}(V_1,N_1)\cap
\downarrow\mathcal{O}(V_2,N_2)$ is not always empty.
\end{Proof}

\section{Sheaves on $\mv(\mh)$ with Respect to the Bucket Topology}
The main reason for introducing the bucket topology on $\mv(\mh)$
is to render the group action of G continuous. However, the bucket
topology is strictly weaker than the Alexandroff topology. This
property will affect what type of sheaves can be constructed. We will denote by $\mv(\mh)_A$ and $\mv(\mh)_B$,
$\mv(\mh)$ equipped with the Alexandroff topology and bucket topology
 respectively. Since
the bucket topology is weaker than the Alexandroff topology the
identity map $i : \mv(\mh)_A\rightarrow \mv(\mh)_B$ is continuous.
This gives rise to the pair of adjoint functors \ba
\iota^*: Sh(\mv(\mh)_B)&\rightarrow& Sh(\mv(\mh)_A)\\
\iota_*:Sh(\mv(\mh)_A)&\rightarrow&Sh(\mv(\mh)_B) \ea where
$\iota^*\dashv \iota_*$. 

We would now like to analyse what kind of
presheaves can be defined using the bucket topology. For example
the spectral presheaf $\us$, with associated sheaf $\bar{\us}$, was defined using the Alexandroff
topology on $\mv(\mh)$. What would happen if we define it using
the Bucket topology?  For example, one natural definition is: \be
\bar{\us}_B:=\iota_*(\bar{\us}_A) \ee 
where the subscript refers to Bucket and Alexandroff topology, respectively.

\noindent
Given an open set
$\down{\mathcal{O}}$ we then have \be
\iota_*(\bar{\us}_A)\down{\mathcal{O}}=\bar{\us}_A(\iota^{-1}\downarrow
\mathcal{O})=\bar{\us}_A(\iota^{-1}\bigcup_{g\in N}\downarrow
l_g(V))=\bar{\us}_A(\bigcup_{g\in N}\downarrow l_g(V)) \ee
where the last equation follows, since the map $\iota$ is continuous and is the identity. 

We know from previous sections that given an open set
$\mathcal{O}$ in $\mv(\mh)$ a sheaf
$\underline{\bar{A}}(\mathcal{O})$ is defined in terms of the
inverse limit, i.e.,
$$\underline{\bar{A}}(\mathcal{O})={\lim_{\longleftarrow }}_{ V\subseteq \mathcal{O}}\underline{A}_V$$ where $\underline{\bar{A}}$ represents the sheaf while $\underline{A}$ represents the corresponding presheaf.

Applying this definition to our case and denoting
$\mathcal{O}=\bigcup_{g\in N}\downarrow l_g(V)$ we have the
following: \ba\label{ali:sec}
\bar{\us}_A(\bigcup_{g\in N}\downarrow l_g(V))&=&{\lim_{\longleftarrow}}_{ l_gV\subseteq\mathcal{O} }\us_{l_g(V)}\\
&=&\{(\alpha, \beta, \cdots \rho)\in\us_{l_g(V)}\times \us_{l_{g_1}(V)}\cdots \times \us_{l_{g_n}(V)}|\nonumber\\
& &\text{ for any pair }(g, g_i)\; \alpha|_{l_gV\cap l_{g_i}V}=\beta
|_{l_gV\cap l_{g_i}V}=\cdots\}\nonumber\\
=\Gamma\us_A|_{\mathcal{O}}
\ea 
where each $ \us_{l_{g_1}(V)}$ refers to the spectral presheaf in the Alexandroff topology.

Given the set $\bigcup_{g\in N}\downarrow l_g(V)$ we need to understand if, given any two elements $g, g^{'}\in N$ then the intersection $l_gV\cap l_{g^{'}}V$ is empty or not. \\

For each pair $(g, g^{'})$ we have two distinct situation
\begin{enumerate}
\item $g, g^{'}\in G_V$.
\item $g, g^{'}\notin G_V$
\end{enumerate}
In the first case then it is trivial that $l_gV\cap l_{g^{'}}V=V$. The second case on the other hand is rather more difficult. To simplify\footnote{We thank Sander Wolters for this example.} things let us consider the situation in which the pair is $(e, g)$. We want to know what the intersection $V\cap l_gV$ is. Let us consider a three dimensional Hilbert space $\mh^3$ with orthogonal projection operators 
\[\hat{P}_1=\begin{pmatrix} 1& 0& 0\\	
0&0&0 \\
0&0&0	
  \end{pmatrix}\;\;\;\;
\hat{P}_2=\begin{pmatrix} 0& 0& 0\\
0&1&0 \\
0&0&0	
  \end{pmatrix}\;\;\;\;
\hat{P}_3=\begin{pmatrix} 0& 0& 0\\	
0&0&0 \\
0&0&1	
  \end{pmatrix}
\]
A maximal algebra would be $V=lin_{\Cl}(\hat{P}_1, \hat{P}_2, \hat{P}_3$. If we now consider the transformation induced by $g$ as a rotation around the $z$ axis, we would obtain $l_gV=lin_{\Cl}(\hat{Q}_1, \hat{Q}_2),\hat{P}_3)$. 
This implies that $V\cap l_gV=V^{'}:=lin_{\Cl}(\hat{P}_3)$. 
Thus in this very special case we would not get an empty intersection. However, equation \ref{ali:sec} requires that the intersection be non empty for all pairs $(g, g^{'})\in N$. The satisfaction or not of such a condition will depend on how `big' $N$ is. In fact keeping to our 3 dimensional example, if we consider the element $g$ as performing a rotation along the $z$ axis as before an now an element $g^{'}$ which instead performs a rotation along the $x$ axis, we would obtain that 
\be
V\cap l_gV=\hat{P}_3^{''};\;V\cap l_{g^{'}}=\hat{P}_1^{''};\;l_gV\cap l_{g^{'}}V=\Cl \hat{1}
\ee 
where $''$ represents the operation of taking the double commutant, thus $\hat{P}_3^{''}$ represents the abelian von Neuamnn algebra generated by $\hat{P}_3$. Since we are excluding the trivial algebra $\Cl\hat{1}$, this very simple example shows how easy it is for the intersection $l_gV\cap l_{g^{'}}V$ to be empty. In the case in which such intersection is empty for all pairs $g, g^{'}$ then the object $\bar{\us}_A(\bigcup_{g\in N}\downarrow l_g(V))$ becomes trivial since the condition $\alpha|_{\empty}=\beta
|_{\emptyset}$ is always satisfied. The question when such an object is not trivial is related in a way to the Kochen-Specker theorem. In fact as shown in \cite{andreas5} the topos equivalent of the Kochen-Specker theorem is that the spectral presheaf does not have any point, or equivalently that it does not allows for global sections.  However it is still the case that the spectral presheaf has local section, but the question is: how local is local? 
The answer to this question will enable us to know whether it could be the case that the bucket topology could be a viable topology to use. SInce it is equivalent to the question of what conditions should be put on the group $N$ such that $\bar{\us}_A(\bigcup_{g\in N}\downarrow l_g(V))$ is not trivial. Admittedly it does seem very likely that $\bar{\us}_A(\bigcup_{g\in N}\downarrow l_g(V))$ is almost always trivial\footnote{This possibly was first conjectured by Andreas D\"oring}, but a precise analysis has still to be done.

Although this is not a very encouraging result, it is still the
case that there will exist many interesting sheaves using the
bucket topology that could not exist in the Alexandroff space. The
reason being that the buckets are formed from the `vertical'
topologies on the fibres, and these incorporate the differential
structure of the manifold
structure of these group orbits. Thus it is possible to construct new sheaves over the fibres utilising the `vertical' topology present there.

In particular consider the map $i_w:\mathcal{O}_w\rightarrow
\mv(\mh)_B$ which is the canonical injection of the orbit
$\mathcal{O}_w$, $w\in \mv(\mh)_B/G$ in $\mv(\mh)_B$. Since the
intersection of each bucket with any given orbit is an open set,
it follows that $i_w$ is continuous. This gives rise to the
continuous bijection \be i:\coprod_{w\in
\mv(\mh)_B/G}\mathcal{O}_w\rightarrow \mv(\mh)_B \ee where
$\coprod_{w\in \mv(\mh)_B/G}\mathcal{O}_w$ is equipped with the
canonical topology for disjoint union. A moment of thought reveals
that $\coprod_{w\in \mv(\mh)_B/G}\mathcal{O}_w$ is nothing but
$\mv(\mh)$ equipped with the `vertical' topology described above.
Therefore we have a canonical continuous (but not bi-continuous)
bijection $i : \mv(\mh)_{ver}\rightarrow \mv(\mh)_B$ which gives
us the following diagram of bijections.
\[\xymatrix{
&&\mv(\mh)_{ver}\ar[dd]^{i}\\
&&\\
\mv(\mh)_A\ar[rr]_{\iota}&&\mv(\mh)_B\\
}\] Since the map $i$ is a bijection, we can define the geometric
morphism, whose direct and inverse image are respectively \ba
i_*:Sh(\mv(\mh)_{ver})&\rightarrow& Sh(\mv(\mh)_B)\\
i^*:Sh(\mv(\mh)_B)&\rightarrow&Sh(\mv(\mh)_{ver}) \ea Moreover we
also have the functor associated to individual orbit $w\in
\mv(\mh)/G$ \be i^*:Sh(\mathcal{O}_w)\rightarrow
Sh(\mv(\mh)_{ver}) \ee

Since each orbit $\mathcal{O}_w$ has a natural manifold structure
as a finite dimensional homogeneous space, $\mathcal{O}_w$ comes
equipped with a number of sheaves associated to such a structure.
These sheaves can be then pushed down to sheaves on $\mv(\mh)_B$
via the map $i_w:\mathcal{O}_w\rightarrow \mv(\mh)_B$. In
particular this setting allows to define the sheaf of continuous
differentiable functions on a topological space. These kind of
sheaves could be very useful to eventually incorporate
differential geometric constructions internally in the topos
framework. This is work in progress.

\section{In Need of a Different Base Category}
In our initial attempt to define a continuous group action which
does not lead to twisted presheaves, we tried changing the context
category from $\mv(\mh)$ to $\mv(\mh)/G$, whose elements where
equivalence classes. However such a category (poset) revealed
itself too small for our purpose. We then understood that the
correct strategy to be used was to enlarge the base category
rather than restrict it. This is precisely what we will describe
in this section.

\section{The Sheaf $\ps{G/G_F}$}
In our new approach we still use   the poset $\mv(\mh)$ as the
base category but now we `forget' the group action. To distinguish
this situation from the case in which the group does act we will
add a subscript $f$ (for fixed) and write $\mv_f(\mh)$.

We now  consider the collection, $Hom_{faithful}(\mv_f(\mh),
\mv(\mh))$, of all faithful poset representations of $\mv_f(\mh)$
in $\mv(\mh)$ that come from the action of the group of interest,
$G$. Thus we have the collection of all homomorphisms
$\phi_g:\mv_f(\mh)\rightarrow \mv(\mh)$, $g\in G$,  such that
$$\phi_{g}(V):=\hat{U}_gV\hat{U}_{g^{-1}}$$
We can `localise' $Hom_{faithful}(\mv_f(\mh), \mv(\mh))$ by
considering for each $V$, the set $Hom_{faithful}(\downarrow
V,\mv(\mh))$. It is easy to see that this actually defines a
\emph{presheaf} over $\mh$; which we will denote
$\underline{Hom}_{faithful}(\mv_f(\mh), \mv(\mh))$

Now, for each algebra $V$ there exists the fixed point group
$$G_{FV}:=\{g\in G|\forall v\in V\;\hat{U}_gv\hat{U}_{g^-1}=v\}$$
This implies that the collection of all faithful representations
for each $V$ is actually the quotient space $G/G_{FV}$. This
follows from the fact that the group homomorphisms
$\phi:G\rightarrow GL(V)$ has to be injective, but that would not
be the case if we also considered the elements of $G_{FV}$, since
each such element would give the same homomorphism.

Thus for each $V$ we have that
$$\underline{Hom}_{faithful}(\mv_f(\mh), \mv(\mh))_V:=
Hom_{faithful}(\downarrow\! V,\mv(\mh))\cong G/G_{FV}$$ As we will
shortly see, there is a presheaf, $\ps{G/G_F}$, such that, as
presheaves,
$$ \underline{Hom}_{faithful}(\mv_f(\mh), \mv(\mh))\cong \ps{G/G_F}$$
whose local components are defined above. In the rest of this
paper, unless otherwise specified, $\underline{Hom}(\mv_f(\mh),
\mv(\mh))$ will mean $\underline{Hom}_{faithful}(\mv_f(\mh),
\mv(\mh))$
\begin{Lemma}
$G_{FV}$ is a normal subgroup of $G_V$.
\end{Lemma}

\begin{Proof}
Consider an element $g\in G_{FV}$, then given any other element
$g_i\in G_V$ we consider the element $g_igg_i^{-1}$. Such an
element acts on each $v\in V$ as follows: \ba
\hat{U}_{g_igg_i^{-1}}v\hat{U}_{(g_igg_i^{-1})^{-1}}&=&\hat{U}_{g_igg_i^{-1}}v\hat{U}_{g_ig^{-1}g_i^{-1}}\\
&=&\hat{U}_{g_i}\hat{U}_g\hat{U}_{g_i^{-1}}v\hat{U}_{g_i}\hat{U}_{g^{-1}}\hat{U}_{g_i^{-1}}\nonumber\\
&=&\hat{U}_{g_i}\hat{U}_gv^{'}\hat{U}_{g^{-1}}\hat{U}_{g_i^{-1}}\nonumber\\
&=&\hat{U}_{g_i}v^{'}\hat{U}_{g_i^{-1}}\nonumber\\
&=&\hat{U}_{g_i}\hat{U}_{g_i^{-1}}v\hat{U}_{g_i}\hat{U}_{g_i^{-1}}\nonumber\\
&=&v\nonumber \ea where $v^{'}\in V$ because $g_i\in G_V$.
\end{Proof}

We then have the standard result that if $G$ is a group and $N$ a
normal subgroup of $G$ then the coset space $G/N$ has a natural
group structure. In the Lie group case, $G/N$ would only have a
Lie group structure if $N$ is a \emph{closed} subgroup of $G$.
However, it is clear from the definition of $G_{FV}$ that it is
closed, and hence for each $V$ we have a Lie group, $G/G_{FV}$. We
note \emph{en passant} that $G$ is a principal fibre bundle over
$G/G_{FV}$ with fiber $G_{FV}$.

For us, the interesting aspect of the collection $G_{FV}$,
$V\in\mv(\mh)$ is that, unlike the collection of stability groups
$G_V$, $V\in\mv(\mh)$, form the components of a presheaf over
$\mv_f(\mh)$ (or $\mv(\mh)$) defined as follows:
\begin{Definition}
The presheaf $\ug_F$ over $\mv_f(\mh)$ has as
\begin{enumerate}
\item [--] Objects: for each $V\in \mv_f(\mh)$ we define set $\ug_{FV}:=G_{FV}=\{g\in G|\forall v\in V\;\hat{U}_gv\hat{U}_{g^-1}=v\}$
\item [--] Morphisms: given a map $i:V^{'}\rightarrow V$ in $\mv_f(\mh)$ ($V^{'}\subseteq V$) then we define the morphism $\ug_F(i):\ug_{FV}\rightarrow \ug_{FV^{'}}$, as subgroup inclusion.
\end{enumerate}
\end{Definition}

The morphisms $\ug_F(i):\ug_{FV}\rightarrow \ug_{FV^{'}}$ are well defined since if   $V^{'}\subseteq V$ then clearly $G_{FV}\subseteq G_{FV^{'}}$. Associativity is obvious.

We now define the  presheaf $\ps{G/G_F}$ as follows:
\begin{Definition}
The presheaf $\ps{G/G_F}$ is defined as the presheaf with
\begin{enumerate}
\item[--] Objects: for each $V\in \mv_f(\mh)$ we assign $(\ps{G/G_F})_V:=G/G_{FV}\cong Hom(\downarrow\!V, \mv(\mh))$. An element of $G/G_{FV}$ is an orbit $w^g_V:=\{g\cdot G_{FV}\}$ which corresponds to the unique homeomorphism $\phi^g$.
\item[--] Morphisms: Given a morphisms $i_{V^{'}V}:V^{'}\rightarrow V$ ($V^{'}\subseteq V$) in $\mv_F(\mh)$ we define
\ba
\ps{G/G_F}(i_{V^{'}V}):G/G_{FV}&\rightarrow& G/G_{FV^{'}}\\
w^g_V&\mapsto&\ps{G/G_F}(i_{V^{'}V})(w^g_V)
\ea as the projection maps $\pi_{V^{'}V}$ of the fibre bundles \be
G_{FV^{'}}/G_{FV}\rightarrow G/G_{FV}\rightarrow G/G_{FV^{'}} \ee
with fibre isomorphic to $G_{FV^{'}}/G_{FV}$.

What this means is that to each $w^g_{V^{'}}=g\cdot G_{FV^{'}}\in
G/G_{FV^{'}}$ one obtains in $G/G_{FV}$ the fibre \ba
\pi^{-1}_{V^{'}V}(g\cdot G_{FV^{'}})&:=&\sigma^g_{V}=\{g_i(g\cdot G_{FV})|\forall g_i\in G_{FV^{'}}\}\\
&=&\{l_{g_i}\cdot w^g_{V}|\forall g_i\in G_{FV^{'}}\}\\
&=&\{w^{g_ig}_V|g_i\in G_{FV^{'}}\} \ea In the above expression we
have used the usual action of the group $G$ on an orbit:  \be
l_{g_i}\cdot w^g_{V}=g_i\cdot(g\cdot G_{FV})=g_i\cdot g\cdot
G_{FV}=:w^{g_ig}_V \ee The fibre $\sigma^g_{V}$ is obviously
isomorphic to $G_{FV^{'}}/G_{FV}$. Thus the projection map $\pi_{V^{'}V}$
projects \be \pi_{V^{'}V}(\sigma^g_{V})=g\cdot G_{FV^{'}}=w^g_{V^{'}} \ee
such that for individual elements we have \be
\ps{G/G_F}(i_{V^{'}V})(w^{g}_V):=\pi_{V^{'}V}(\sigma^g_{V})=w^g_{V^{'}}
\ee Note that when $g_i\in G_{FV^{'}}$ but $g_i\notin G_{FV}$ then
$w^{g}_V=g\cdot G_{FV}$ and $w^{g}_{V^{'}}=g\cdot
G_{FV^{'}}=g_igG_{FV^{'}}=w^{g_ig}_{V^{'}}$. Therefore
$\ps{G/G_F}(i_{V^{'}V})w^{g}_V=w^{g}_{V^{'}}=w^{g_ig}_{V^{'}}$

\end{enumerate}
\end{Definition}

It should be noted that the morphisms in the presheaf $\ps{G/G_F}$
can also be defined in terms of the homeomorphisms
$Hom(\mv_f(\mh), \mv(\mh))$. Namely, given an element $g_j\in
w^g_V$ we obtain the associated homomorphisms $\phi_{g_j}$, such
that \be \ps{G/G_F}(i_{V^{'}V})\phi_{g_j}:=\phi_{g_j|V^{'}} \ee

We will now define another presheaf which we will then show to be isomorphic to $\ps{G/G_F}$. To this end we first of all have to introduce the constant presheaf $\underline{G}$. This is defined as follows
\begin{Definition}
The presheaf $\underline{G}$ over $\mv_f(\mh)$ is defined on 
\begin{itemize}
\item Objects: for each context $V$, $\ps{G}_V$ is simply the entire group, i.e. $\ps{G}_V=G$
\item Morphisms: given a morphisms $i:V^{'}\subseteq V$ in $\mv(\mh)$, the corresponding morphisms $\ps{G}_V\rightarrow \ps{G}_{V^{'}}$ is simply the identity map.
\end{itemize}
\end{Definition}
We are now ready to define the new presheaf.
\begin{Definition}
The presheaf $\ps{G}/\ps{G_F}$ over $\mv_f(\mh)$ is defined on 
\begin{itemize}
\item Objects. For each $V\in \mv_f(\mh)$ we obtain $(\ps{G}/\ps{G_F})_V:=G/G_{FV}$. Since as previously explained the equivalence relation is computed context wise.
\item Morphisms. For each map $i:V^{'}\subseteq V$ we obtain the morphisms 
\ba
(\ps{G}/\ps{G_F})_V&\rightarrow& (\ps{G}/\ps{G_F})_{V^{'}}\\
G/G_{FV}&\rightarrow&G/G_{FV^{'}}
\ea
These are defined to be the projection maps $\pi_{V^{'}V}$ of the fibre bundles \be
G_{FV^{'}}/G_{FV}\rightarrow G/G_{FV}\rightarrow G/G_{FV^{'}} \ee
with fibre isomorphic to $G_{FV^{'}}/G_{FV}$.

\end{itemize}
 
\end{Definition}
From the above definition it is trivial to show the following theorem.
\begin{Theorem}
\be
\ps{G/G_F}\simeq\ps{G}/\ps{G_F}
\ee
\end{Theorem}
\begin{Proof}
We construct the map $k:\ps{G/G_F}\rightarrow \ps{G}/\ps{G_F}$ such that, for each context $V$ we have
\ba
k_V:\ps{G/G_F}_V&\rightarrow& \ps{G}/\ps{G_F}_V\\
G/G_{FV}&\mapsto&G/G_{FV}
\ea
This follows from the definitions of the individual presheaves.
\end{Proof}

\subsection{Using $\Lambda( \ps{G/G_F})$ as the Base Category }
We know that given a sheaf over a poset we obtain the corresponding
etal\'e bundle. In our case the sheaf in question is $\ps{G/G_F}$ with corresponding etal\'e bundle $p:\Lambda \ps{G/G_F}\rightarrow \mv_f(\mh)$ where $\Lambda\ps{G/G_F}$ is the etal\'e space. We will now equip the etal\'e space $\Lambda(\ps{G/G_F})=\coprod_{V\in\mv_F(\mh)}(\ps{G/G_F})_V$ with a poset structure.

The most obvious poset structure to use would be the partial order
given by restriction, i.e., $w_{V}\leq w_{V^{'}}$ iff $V\subseteq V^{'}$ and 
$w_V^g=w^g_{V^{'}}|_V$ or equivalently $g\cdot G_{FV}=g\cdot
(G_{FV}\cap G_{FV^{'}})$. We could write this last condition as an
inclusion of sets as follows:
$w_{V}\subseteq w_V^{'}$ ($g\cdot G_{FV}\subseteq g\cdot G_{FV^{'}}$). 
However this poset structure would not give a presheaf if we were
to use it as the base category, rather it would give a covariant
functor. To solve this problem we adopt the \emph{order dual} of
the partially ordered set, which is the same set but equipped with
the inverse order which is itself a partial order. We thus define
the ordering on $\Lambda( \ps{G/G_F})$ as follows:

\begin{Lemma}\label{lem:ordering}
Given two orbits $w^g_V\in G/G_{FV}$ and $w^g_{V^{'}}\in
G/G_{FV^{'}}$ we define the partial ordering $\leq$, by
defining
$$w^g_{V^{'}}\leq w^g_V$$ iff
\ba
V^{'}&\subseteq &V\\
w^g_V&\subseteq& w^g_{V^{'}} \ea Note that the last condition is
equivalent to $w^g_V=w^g_{V^{'}}|_V$ ($g\cdot G_{FV}=g\cdot
(G_{FV}\cap G_{FV^{'}})$).
\end{Lemma}
It should be noted though that if $w^g_V=w^g_{V^{'}}|_V$ then $\ps{G/G_F}(i_{V^{'}V})(w^g_V)=
\ps{G/G_F}(i_{V^{'}V})(w^g_{V^{'}}|_{V})=w^g_{V^{'}}$. In other words it is also possible to define the partial ordering 
in terms of the presheaf maps defined above, i.e., \be w^g_V\geq
w^g_{V^{'}}\text{  iff  }w^g_{V^{'}}=\ps{G/G_F}(i_{V^{'}V})w^g_V
\ee

We now show that the ordering defined on $\Lambda(\ps{G/G_F})$ is
indeed a partial order.
\begin{Proof}
\noindent
\begin{enumerate}
\item \emph{Reflexivity}. Trivially $w^g_V\leq w^g_V$ for all $w^g_V\in \Lambda \ps{G/G_F}$.
\item \emph{Transitivity}. If $w^g_{V_1}\leq w^g_{V_2}$ and $w^g_{V_2}\leq w^g_{V_3}$ then $V_1\subseteq V_2$ and $V_2\subseteq V_3$. From the partial ordering on $\mv(\mh)$ it follows that $V_1\subseteq V_3$. Moreover from the definition of  ordering on $\Lambda (\ps{G/G_F})$ we have that $w^g_{V_2}=w^g_{V_1}|_{V_2}$ and $w^g_{V_3}=w^g_{V_2}|_{V_3}$ which implies that $w^g_{V_3}=w^g_{V_1}|_{V_3}$. It follows that $w_{V_1}\leq w_{V_3}$.
\item \emph{Antisymmetry}. If $w_{V_1}\leq w_{V_2}$ and $w_{V_2}\leq w_{V_1}$, it implies that $V_1\leq V_2$ and $V_2\leq V_1$ which, by the partial ordering on $\mv(\mh)$ implies that $V_1=V_2$. Moreover the above conditions imply that $w_{V_1}= w_{V_2}|_{V_1}$ and $w_{V_2}=w_{V_1}|_{V_2}$, which by the property of subsets implies that $w_{V_1}=w_{V_2}$.
\end{enumerate}
\end{Proof}
Given the previously defined isomorphisms, 
$Hom(\down{V}, \mv(\mh))\cong (\ps{G/G_F})_V$ for each $V$, then to each equivalence class $w_V^g$ there is associated a particular homeomorphism $\phi_g:\down V\rightarrow \mv(\mh)$. Even though $w_V^g$ is an equivalence class, each element in it will give the same $\phi_g$, i.e. it will pick out the same $V_i\in \mv(\mh)$. This is because the equivalence relation is defined in terms of the fixed point group for $V$. 

Therefore it is also possible to define the ordering relation on
$\Lambda(\ps{G/G_F})$ in terms of the homeomorphisms $\phi^g_i$.
First of all we introduce the bundle space $\Lambda J\simeq \Lambda(\ps{G/G_F})$ which is essentially the same as $\Lambda(\ps{G/G_F})$, but whose elements are now the maps $\phi^g_i$, i.e., $\Lambda J=\Lambda(\underline{Hom}(\mv_f(\mh), \mv(\mh)))$. The associated bundle map is $p_J:\Lambda J\rightarrow \mv_f(\mh)$.
We then define the ordering on $\Lambda J$ as 
$\phi^g_i\leq \phi^g_j$ iff
 \be
 p_J(\phi^g_i)\subseteq p_J(\phi^g_j)
\ee and \be \phi^g_i=\phi^g_j| _{p_J(\phi^g_i)} \ee We now need to
show that this does indeed define a partial
order on $\Lambda J$.
\begin{Proof}\hspace{.2in}
\begin{enumerate}
\item \emph{Reflexivity}. Trivially $\phi^g_i\leq \phi^g_i$ since  $p_J(\phi^g_i)\subseteq p_J(\phi^g_i)$ and $\phi^g_i=\phi^g_i$.
\item \emph{Transitivity}. If $\phi^g_i\leq \phi^g_j$ and $\phi^g_j\leq \phi^g_k$ then $p_J(\phi^g_i)\subseteq p_J(\phi^g_j)$ and  $p_J(\phi^g_j)\subseteq p_J(\phi^g_k)$, therefore  $p_J(\phi^g_i)\subseteq p_J(\phi^g_k)$. Moreover we have that $\phi^g_i=\phi^g_j| _{p_J(\phi^g_i)}$ and $\phi^g_j=\phi^g_k| _{p_J(\phi^g_j)}$, therefore $\phi^g_i=\phi^g_k| _{p_J(\phi^g_i)}$.
\item \emph{Antisymmetry}.  If $\phi^g_i\leq \phi^g_j$ and $\phi^g_j\leq \phi^g_i$ it implies that  $p_J(\phi^g_i)\subseteq p_J(\phi^g_j)$ and  $p_J(\phi^g_j)\subseteq p_J(\phi^g_i)$, thus  $p_J(\phi^g_i)= p_J(\phi^g_j)$. Moreover we have that $\phi^g_i=\phi^g_j| _{p_J(\phi^g_i)}$ and $\phi^g_j=\phi^g_i| _{p_J(\phi^g_j)}$, therefore $\phi^g_i=\phi^g_j$.
\end{enumerate}
\end{Proof}
Given this ordering we can now define the corresponding ordering on $\Lambda(\ps{G/G_F})$ as
$w^g_{V_i}\leq w^g_{V_j}$ iff
$\phi^g_i\leq \phi^g_j$. We  have again used the fact that to each $w^g_{V_i}$ there is
associated a unique homeomorphism $\phi^g_V:\downarrow V\rightarrow
\mv(\mh)$.

\subsection{Topology on $\Lambda(\ps{G/G_F})$}
The next step is to give $\Lambda(\ps{G/G_F})$ a topology. A
priori, this can be done in two different ways: (i) the etal\'e
topology on $\Lambda(\ps{G/G_F})$ using the fact that
$\Lambda(\ps{G/G_F})$ is the etal\'e space of the etal\'e bundle
$p:\Lambda(\ps{G/G_F})\rightarrow \mv_f(\mh)$; and (ii) the
Alexandroff topology induced by the poset structure of
$\Lambda(\ps{G/G_F})$. As we shall see, these two topologies
are isomorphic.\\
To show the above homeomorphism we will make use of the following
result:
\begin{Lemma}
Let $\alpha: P_1\rightarrow P_2$ be a map between posets $P_1$ and
$P_2$. Then $\alpha$ is order preserving if and only if for each
lower set $L\subseteq P_2$, we have that $\alpha^{-1}(L)$ is a
lower subset of $P_1$.
\end{Lemma}
\begin{Proof}
Let us assume that $\alpha$ is order preserving and let
$L\subseteq P_2$ be lower. Now let $z\in\alpha^{-1}(L)\in P_1$,
i.e., $\alpha(z) = l$ for some $l\in L$, and suppose $y\in P_1$ is
such that $y\leq z$. Since $\alpha$ is order preserving we have
$\alpha(y)\leq \alpha(z) = l\in L$, which, since $L$ is lower,
means that $\alpha(y)\in L$,
i.e., $y\in\alpha^{-1}(L)$. Hence $\alpha^{-1}(L)$ is lower.

Conversely, suppose that for any lower set $L\in P_2$ we have that
$\alpha^{-1}(L)\in P_1$ is lower, and consider a pair $x, y\in
P_1$ such that $x\leq y$. Now $\downarrow(y)$ is lower in $P_2$
and hence $\alpha^{-1}(\downarrow \alpha(y))$ is a lower subset of
$P_1$. However $\alpha(y)\in\downarrow \alpha(y)$ and hence
$y\in\alpha^{-1}(\downarrow \alpha(y))$. Therefore, the fact that
$x\leq y$ implies that $x\in \alpha^{-1}(\downarrow \alpha(y))$,
i.e., $\alpha(x)\in\downarrow\alpha(y)$, which means that
$\alpha(x)\leq\alpha(y)$. Therefore $\alpha$ is order preserving.
\end{Proof}

The etal\'e topology on the bundle space $\Lambda( \ps{G/G_F})$ is
defined as follows:
\begin{Definition}
 Given an open set $\downarrow V$ in $\mv(\mh)$, then the corresponding open set in the etal\'e space is constructed by considering the union of the points in $\Lambda (\ps{G/G_F})$, which are defined as the germs of the elements in $(\ps{G/G_F})_{V_i}$, $V_i\in\downarrow V$. Notice that each stalk has the discrete topology.
 \end{Definition}
Since in our case the base space has the Alexandroff topology the
situation simplifies. In fact, given any point $V\in\mv(\mh)$,
there is a unique smallest open set, namely $\downarrow V$, to
which V belongs. If we then consider two open neighbourhoods
$\mathcal{O}_1$ and $\mathcal{O}_2$ of $V\in\mv(\mh)$ with $w_1\in
\overline{\ps{G/G_F}}(\mathcal{O}_1)$ and\footnote{Recall that $
\overline{\ps{G/G_F}}$ denotes the sheaf that is associated with
the presheaf  $\ps{G/G_F}$. Here $w_i$ represents a general element $w^{g_i}_{V_i}\in \Lambda(\ps{G/G_F})$.} $w_2\in
\overline{\ps{G/G_F}}(\mathcal{O}_2)$, $w_1$ and $w_2$ have the
same germ at $V$ if there is some open set $\mathcal{O}\subseteq
\mathcal{O}_1\cap\mathcal{O}_2$, such that
$w_1|_{\mathcal{O}}=w_2|_{\mathcal{O}}$, where restriction is given in terms of the sheaf maps (see below). However, since $\mv(\mh)$
is equipped with the the Alexandroff topology, the smallest such
open set will be $\downarrow V$. It follows that $w_1$ and $w_2$
have the same germ at $V$ iff
$$w_1|_{\downarrow V}:=\{\pi_{V_iV}w_1|V_i\in\downarrow V\}=w_2|_{\downarrow V}:=\{\pi_{V_iV}w_2|V_i\in\downarrow V\}$$
Therefore if $V\in \mathcal{O}$ and
$w\in\overline{\ps{G/G_F}}(\mathcal{O})$, then $germ_V w =
w|_{\downarrow V}$. Thus \ba
\Lambda(\ps{G/G_F})_V&=&\overline{\ps{G/G_F}}(\down V )\\
&\simeq& (\ps{G/G_F})_V \ea
where  $\ps{G/G_F}$ denotes the presheaf and $\overline{\ps{G/G_F}}$ in the first equation represents the corresponding sheaf.

Moreover we have the general result that \be
\Lambda(\ps{G/G_F})_V = {\lim_{\longrightarrow}}_{ V\in
\mathcal{O}}\overline{\ps{G/G_F}}(\mathcal{O}) \ee Given the above
discussion, the construction of the etal\'e topology on our bundle
$\Lambda(\ps{G/G_F})$ simplifies. In fact consider the open set
$\down V$, then for each $w_V^g\in  (\ps{G/G_F})_V$ we get a
set of points in $\Lambda( \ps{G/G_F})$ defined as follows \be
\{germ_{V_i}w_V^g:=(w_V^g)_{|V_i}|V_i\in\down
V\}=\{\pi_{V_iV}w_V^g|V_i\in\downarrow V\}=\{w_{V_i}^g|V_i\in\down V\} \ee
 Each point will come from a different fibre $\Lambda(\ps{G/G_{F}})_{V_i}$. The collection of all these points is open in $\Lambda(\ps{G/G_F})$.
From this it follows that the topology on each stalk is discrete since, given the definition of open sets above, the only intersection between them and a stalk is a point.

 Our aim is to show that the above defined etal\'e topology on
  $\Lambda({\ps{G/G_F}})$ is isomorphic to the Alexandroff topology on $\Lambda(\ps{G/G_F})$.
Although the definition of the Alexandroff topology is already
known from the case of the poset $\mv(\mh)$, for the sake of completeness 
we will define the Alexandroff topology for the poset
$\Lambda(\ps{G/G_F})$.
\begin{Definition}
The Alexandroff Topology on $\Lambda(\ps{G/G_F})$ is the topology
whose basis are the open sets $\down w_v^g$ with partial
ordering defined in the Lemma \ref{lem:ordering}.
\end{Definition}
\begin{Theorem}
The Alexandroff topology on $\Lambda(\ps{G/G_F})$ is isomorphic to
the etal\'e topology.
\end{Theorem}
\begin{Proof}
Let us consider an open set $U$ in the etal\'e topology of
$\Lambda (\ps{G/G_F})$. Since $p:\Lambda(\ps{G/G_F})\rightarrow \mv(\mh)$ is
a local homeomorphism\footnote{In the sense that for each element
$w_V^g\in\Lambda(\ps{G/G_F})_V$, given the open neighbourhood $U$,
$p(U)$ is open in $\mv(\mh)$ and $p$ restricted to $U\ni w_V^g $  is a
homomorphisms, i.e., $p_U:U\rightarrow p(U)$ is a homomorphisms.}
then $p(U)$ is open in $\mv(\mh)$, i.e., is a lower set in the Alexandroff topology. However, by the definition of the poset structure on $\Lambda(\ps{G/G_F})$, $p$ is order preserving, thus $p^{-1}\circ p(U)$ is a lower set in $\Lambda(\ps{G/G_F})$. Moreover since $p$ is a local homeomorphism then $p^{-1}\circ p(U)=U$ is a lower set in $\Lambda(\ps{G/G_F})$. \\
\\
Conversely, let $U$ be an open set in the Alexandroff topology on
$\Lambda(\ps{G/G_F})$. Since $p$ is order preserving then $p(U)$ is
a lower set in $\mv(\mh)$. Now since $p:\Lambda(\ps{G/G_{F}})\rightarrow
\mv(\mh)$ is an etal\'e bundle we know that $p$ is a local
homeomorphism in the etal\'e topology. Thus, restricting only to
open sets, we have that $p^{-1}(p(U))$ is an open set in the
etal\'e topology. However $p^{-1}\circ p(U)=U$, i.e., $U$ is open in the etal\'e topology.
\end{Proof}

\section{Group Action on $\Lambda (\ps{G/G_F})$}
Since we are planning to use the poset $\Lambda(\ps{G/G_F})$ as our
new base category we would like to analyse the action of the group
$G$ on it. Thus we will perform a similar analysis which was done
for the category $\mv(\mh)$ and check for which topologies the
action is continuous. As can be expected the answer will coincide
with the case of $\mv(\mh)$.
\subsection{Alexandroff Topology}
We would like to check whether the action of the group $G$ is
continuous with respect to the Alexandroff topology on
$\Lambda(\ps{G/G_F})$. We first check for individual group
elements $g$. The action is then defined as follows: \ba
g\leadsto l_g:\Lambda(\ps{G/G_F})&\rightarrow &\Lambda(\ps{G/G_F})\\
w_V^{g_1}&\mapsto&l_g(w_V^{g_1}):=w_V^{gg_1}=g(g_iG_{FV}) \ea
Alternatively we can define the group action in terms of the
homeomorphisms as follows: \ba
g\leadsto l_g:\Lambda J&\rightarrow &\Lambda J\\
\phi&\mapsto&l_g(\phi) \ea
such that $l_g(\phi)(V):=l_g(\phi(V))$.

We need to check whether the above maps preserve the partial
ordering on $\Lambda(\ps{G/G_F})$, i.e. we need to check that if
$w_V^{g_1}\leq w_{V^{'}}^{g_1}$ then $l_g(w_V^{g_1})\leq
l_g(w_{V^{'}}^{g_1})$ or alternatively if $\phi_1\leq\phi_2$ then
$l_g(\phi_1)\leq l_g(\phi_2)$.
\begin{Proof}
We assume that $w_V^{g_1}\leq w_{V^{'}}^{g_1}$ which implies that
$V^{'}\leq V$ and $w_{V^{'}}^{g_1}\subseteq w_{V}^{g_1}$
($g_1G_{FV}\subseteq g_1G_{FV^{'}}$). We then have \be
l_g(w_V^{g_1})=l_g\{g_i\cdot g_1|g_i\in
G_{FV}\}=g(g_1G_{FV})=gg_1G_{FV}=w^{gg_1}_V \ee and \be
l_g(w_{V^{'}}^{g_1})=l_g\{g_i\cdot g_1|g_i\in
G_{FV^{'}}\}=g(g_1G_{FV^{'}})=gg_1G_{FV^{'}}=w^{gg_1}_{V^{'}} \ee
It follows trivially that $l_g(w_{V^{'}}^{g_1})\geq
l_g(w_{V}^{g_1})$ since $w^{gg_1}_V=gg_1G_{FV}=gg_1(G_{FV}\cap
G_{FV^{'}})=w^{gg_1}_{V^{'}}|_{G_{FV}}$. Therefore the action of the
maps $l_g$ for all $g\in G$ is continuous.
\end{Proof}

We would also like to show that these maps are open. Thus we need
to show that $l_g(\down w^{g_1}_{V})$ is open. This follows
trivially from the definition of the group actions \be
l_g(\down w^{g_1}_{V})=\down w^{gg_1}_{V} \ee However,
similarly as was the case for the category $\mv(\mh)$, the global
group action is not continuous, i.e., the map \ba
\Phi:G\times \Lambda(\ps{G/G_F})&\rightarrow &\Lambda(\ps{G/G_F})\\
\langle g, w_V^{g_1}\rangle&\mapsto&l_g(w_V^{g_1}):=w_V^{gg_1} \ea
is not continuous.

Since open sets of the form $\down w^{g_i}_V$, $w^{g_i}_V\in
\Lambda(\ps{G/G_F})$ (for some $V\in\mv_f(\mh)$), form a basis for
the Alexandroff topology of $\Lambda(\ps{G/G_F})$, it suffices to
look at \ba
\Phi^{-1}(\down w^{g_i}_V)& =& \{(g, w_{V_i}^{g_i}) | l_g(w_{V_i}^{g_i})\in \down w_V \} \\
&=& \{(g, w_{V_i}^{g_i}) | l_g(w_{V_i}^{g_i})\leq w_V \} \ea A
necessary condition for this to be continuous is that, for each
$w_V^{g_i}\in \Lambda(\ps{G/G_F})$, the induced map
\ba\label{ali:continuous}
f_{w^{g_i}_V}: G&\rightarrow&\Lambda(\ps{G/G_F})\\
g&\mapsto&l_g(w^{g_i}_V) \ea is continuous. Now consider the open set
$\down w^{g_i}_V$ in $ \Lambda(\ps{G/G_F})$. Then, in particular, \ba
f^{-1}_{w^{g_i}_V} (\down w^{g_i}_V)& =&\{g\in G | l_g(w^{g_i}_V)\in\down w_V\}\\
&=& \{g\in G |l_g(w^{g_i}_V)\leq w_V\}\\
&= &\{g\in G |l_g(w^{g_i}_V) = w^{g_i}_V\} =: G_{w_V} \ea where $G_{w^{g_i}_V}$
represents the stabiliser of the coset $w^{g_i}_V$, i.e. all the group
elements which leave the entire coset unchanged. This should be
equivalent to the fixed point group of $V$. In fact we have that
if \be l_g(w^{g_i}_V) = w^{g_i}_V \ee then \be \{g_j\cdot(g\cdot
g_i)|g_i\in G_{FV}\}=\{g_j\cdot g_i|g_i\in G_{FV}\} \ee or
equivalently \be g(g_iG_{FV})=g_iG_{FV} \ee
This can only be true iff $g\in G_{FV}$.

Alternatively we can do the proof using the bundle $\Lambda\ps{Hom}(\mv_f(\mh), \mv(\mh))$. In particular for
each $\phi:\downarrow V\rightarrow \mv(\mh)$, the analogue of
\ref{ali:continuous} is \ba
f_{\phi}:G&\rightarrow& \Lambda\ps{Hom}(\mv_f(\mh), \mv(\mh))\\
g&\mapsto&l_g(\phi) \ea Thus, in order to show that such a map is
continuous we need to show that the following is open \ba
f_{\phi}^{-1}(\down\phi)&=&\{g\in G|l_g(\phi)\in\down\phi\}\\
&=&\{g\in G|l_g(\phi)\leq\phi\}\nonumber\\
&=&\{g\in G|l_g(\phi)=\phi\}\nonumber\\
\ea
where the last equation follows from the definition of partial ordering on $ \Lambda\ps{Hom}(\mv_f(\mh), \mv(\mh))=\lambda J$. In fact if $l_g(\phi)\leq\phi$ then $p_J(l_g(\phi))\subseteq p_J(\phi)$ and $l_g(\phi)=\phi|_{p_J(l_g(\phi))}$. However, from the definition of the group action on $\Lambda J$ it follows that $p_J(l_g(\phi))=p_J(\phi)$, therefore $l_g(\phi)=\phi$.

Since we are only considering faithful representations, 
The only equivalent representations are those defined by elements $g\in G_{FV}$, i.e. $\phi^g=\phi^{g_i}$ iff $g, g_i\in G_{FV}$.

\noindent
Therefore in the case at hand $l_g(\phi)=\phi$ iff $g\in G_{FV}$.
Thus $\{g\in G|l_g(\phi)=\phi\}=G_{FV}$.

The question is now whether the fixed point group is open or
closed. We have seen previously that the stability group for a
given $V$ is closed. What about the fixed point group? If the
topology on $\Lambda(\ps{G/G_F})$ was Hausdorff, then from
corollary 16.1 in the appendix it would follow immediately that
$G_w$ is closed. However the topology on $\Lambda(\ps{G/G_F})$ is
not Hausdorff. So we will show that $G_{FV}$ is closed using the
weak and strong operator topology.
\begin{Lemma}
For each $V\in \mv(\mh)$ the fixed point group $G_{FV}$ is a
closed subgroup of the topological group $G$.
\end{Lemma}
\begin{Proof}

Since the fixed point group $G_{FV}$ of $V$ is defined as \be
G_{FV}:=\{g\in G|\forall v\in V, \hat{U}_gv\hat{U}_{g^{-1}}=v\}
\ee
it follows that it is the intersection of the stability groups $G_{\hat{A}}$ of each $\hat{A}\in V$. Therefore
we need to show that all such stability groups are closed. 

For each $G_{\hat{A}}$, consider the net of elements
$\{g_{\nu}\}$ in $G_{\hat{A}}$ for some index $\nu\in I$, i.e. \be
\hat{U}_{g_{\nu}}\hat{A}\hat{U}_{g_{\nu}^{-1}}=\hat{A} \ee
We then assume that the limit of such a sequence is $g$, i.e. $\lim_{\nu\in I}g_{\nu}=g$.

The series
$\hat{U}_{g_{\nu}}\hat{A}\hat{U}_{g_{\nu}^{-1}}$ for all $\nu\in
I$ is actually the constant series whose only value is $\hat{A}$.
Given the general result that a constant series $(a,a,a,a,
\cdots, a)$ converges to $a$ we should expect
that $\hat{U}_{g}\hat{A}\hat{U}_{g^{-1}}=\hat{A}$.

In the case at hand, since $g\in G_{\hat{A}}$ then $g\in G_V$ and we know from theorem
\ref{lem:closedV} that we have the following weak convergence for
all $\nu\in I$ \be
\hat{U}_{g_{\nu}}\hat{A}\hat{U}_{g_{\nu}^{-1}}=\hat{A}\mapsto_w
\hat{U}_{g}\hat{A}\hat{U}_{g^{-1}} \ee

which means that \be \langle
\hat{U}_{g_{\nu}}\hat{A}\hat{U}_{g_{\nu}^{-1}}(x),
y\rangle=\langle \hat{A}(x),
y\rangle\rightarrow\langle\hat{U}_{g}\hat{A}\hat{U}_{g^{-1}}(x),y\rangle
\ee for all $x, y\in \mathcal{H}$. Or equivalently we can write
the above as \be
|l(\hat{U}_{g_{\nu}}\hat{A}\hat{U}_{g_{\nu}^{-1}}(x))-l(\hat{U}_{g}\hat{A}\hat{U}_{g^{-1}}(x))|=|l(\hat{A}(x))-l(\hat{U}_{g}\hat{A}\hat{U}_{g^{-1}}(x))|\rightarrow
0 \ee
for all $l\in \mathcal{H}^*$ and $x\in \mathcal{H}$.

Since for all $g_{\nu}$, $\hat{U}_{g_{\nu}}\hat{A}\hat{U}_{g_{\nu}^{-1}}=\hat{A}$, weak convergence of this constant series implies that $\hat{U}_{g}\hat{A}\hat{U}_{g^{-1}}=\hat{A}$. Thus $g\in G_{\hat{A}}$ and $G_{\hat{A}}$ is closed.

The intersection of closed groups is closed, therefore $G_{FV}$ is
closed.
\end{Proof}

\subsection{Vertical Topology}
We will now define the `vertical' topology on $\Lambda(\ps{G/G_F})$ in a similar way as was done for $\mv(\mh)$, this will be the coset topology, as defined for each orbit $G/G_{FV}$. Such a coset topology is nothing but the identification topology, i.e., the finest topology on $G/G_{FV}$, such that the projection map $p:G\rightarrow G/G_{FV}$ is continuous. The basis consists of the sets $\{U\subseteq G/G_{FV}|p^{-1}(U)\text{ open in }G\}$.

Obviously, for each $V\in \mv(\mh)$ the action of the group is
continuous, i.e., the map
$$G\times\Lambda( \ps{G/G_F})_{V}\rightarrow \Lambda(\ps{G/G_F})_V$$ is continuous. 

It is straightforward to see that the basis sets for the `vertical'
topology on $\Lambda(\ps{G/G_F})$ have the same form as those for
the `vertical' topology on $\mv(\mh)$, i.e., \be \mathcal{O}(w^{g_i}_V,
N)=U:=\{l_g w^{g_i}_V|g\in N\subseteq G\} \ee
For some open set $N\subseteq G$. Since the action of $G$ on
$\Lambda(\ps{G/G_F})_{V}$ is transitive one can consider $N$ as a
neighbourhood of the identity.

Consider a set $U\subseteq G/G_{FV}$ open in the `vertical'
topology. We know by definition that $p^{-1}(U)=N$ is open in $G$.
We now would like to know what is the explicit form of $U$ in
terms of the elements $w^{g_i}_v\in \Lambda(\ps{G/G_F})_{V}$. $U=
p(p^{-1}(U))=p(N)$ ($pp^{-1}=id$ for any surjection) and by
definition the action of the projection map is such that
$p(N):=\bigcup_{g\in N}l_gw^{g_i}_V$. We then obtain: \be
p(N):=\bigcup_{g\in
N}l_{g}w^{g_i}_V=:\mathcal{O}(w^{g_i}_V, N)=U \ee The sets
$\mathcal{O}(w^{g_i},N)$ are then a basis of the neighbourhood filter of
$w^{g_i}_V$. Given the above, even in this case we have the following
result: \ba
\mathcal{O}(w^{g_i}_V,N_1)\cap \mathcal{O}(w^{g_i}_V,N_2)& =& \{l_g(w^{g_i}_V) |g\in N_1\subseteq G\}\cap \{l_g(w^{g_i}_V) | g\in N_2\subseteq G\}\\
&=& \{l_g(w^{g_i}_V) | g\in N_1\cap N_2\}\\
&=&\mathcal{O}(w_V^{g_i},N_1\cap N_2) \ea
\begin{Lemma}
The action of the group $G$, i.e., $G\times \Lambda
(\ps{G/G_F})\rightarrow \Lambda (\ps{G/G_F}) $ is continuous in
the `vertical' topology.
\end{Lemma}
This proof is similar to the case of the `vertical' topology on
$\mv(\mh)$, but for sake of completeness, we will report it below.
\begin{Proof}
The poset $\Lambda(\ps{G/G_F})$ can be written as the following
disjoint union: \be \Lambda(\ps{G/G_F})=\coprod_{V\in
\mv(\mh)}G/G_{FV} \ee Thus the $G$ action is \be \Theta:G\times
\coprod_{V\in \mv(\mh)}G/G_{FV}\rightarrow \coprod_{V\in
\mv(\mh)}G/G_{FV} \ee We know that such an action is continuous
iff the restrictions $\Theta|_{G/G_{FV}}:G\times
G/G_{FV}\rightarrow G/G_{FV}$ are continuous, which they are.
\end{Proof}

\subsection{Bucket Topology}
In this section will define the \emph{bucket topology} for
$\Lambda( \ps{G/G_F}$) as the combination of the Alexandroff
topology and the `vertical' topology. The basis sets are of the form \be \down{\mathcal{O}} (w^{g_i}_V,
N):=\bigcup_{g\in N}\down l_g w^{g_i}_V \ee and represent the basis
of the neighbourhood filter of $w^{g_i}_V$ in the bucket topology.
Similarly, as was the case for $\mv(\mh)$, we have that \be
\down{\mathcal{O}} (w^{g_i}_V, N_1)\cap \down{\mathcal{O}} (w^{g_i}_V,
N_2)=\down{\mathcal{O}}(w^{g_i}_V, N_1\cap N_2) \ee We will now prove
the analogue of Lemma \ref{lem:orbit} for the poset
$\Lambda\ps{(G/G_F)}$.
\begin{Lemma}
if $w_{V_1}^{g_i}\leq w_V^{g_i}$ then $\down{\mathcal{O}}(w^{g_i}_V, N)\cap G
w_{V_1}^{g_i}$ is open in the `vertical' topology.
\end{Lemma}
The proof of this lemma is similar to the case of the poset
$\mv(\mh)$, but for clarity reasons we will nonetheless report it.

\begin{Proof}
Given an element $w_{V_0}^g\in \downarrow \mathcal{O}(w^{g_i}_V, N)\cap
G w^{g_i}_{V_1}$, then there exists an element $w^g_{V_i}\in
\mathcal{O}(w^{g_i}_V, N)$ such that $w_{V_0}^g\leq w^g_{V_i}$. Since $
\mathcal{O}(w^{g_i}_V, N)$ is open in the `vertical' topology, there
exists some $N_0$ such that $w^g_{V_i}\in\mathcal{O}(w^g_{V_i},
N_0)\subseteq \mathcal{O}(w^{g_i}_V, N)$. We have seen above that the
action of each $g_j\in G$ on $\Lambda(\ps{G/G_F})$ is order preserving, thus
$w_{V_0}^g\leq w^g_{V_i}$ implies that $l_{g_j}w_{V_0}^g\leq
l_{g_j}w^g_{V_i}$ for all $g_j\in G$. Therefore we have \be w_{V_0}^g\in
\mathcal{O}(w^g_{V_0}, N_0)\subseteq \downarrow \mathcal{O}(w^g_{V_i},
N_0)\cap G w^{g_i}_{V_1}\subseteq \downarrow \mathcal{O}(w^{g_i}_V, N)\cap G
w^{g_i}_{V_1} \ee
\end{Proof}
From the definition of the bucket topology\footnote{It is worth noting that the bucket topology on $\Lambda(\ps{G/G_F})$ is not Hausdorff.} it follows that it is nothing but the union of Alexandroff open sets ($\down l_g(w^g_V)$). Therefore every open set in the bucket topology is open in the Alexandroff topology, but the converse is not true. This implies that the bucket topology is weaker than the Alexandroff topology.

\section{Sheaves on $\Lambda(\ps{G/G_F})$ with Respect to the Bucket Topology}
Similarly as was the case for the base category $\mv(\mh)$, it is
now possible to define a map between the poset
$\Lambda(\ps{G/G_F})$, equipped with the Alexandroff topology, to
the same poset equipped with the bucket topology. Such an identity
map $l:\Lambda(\ps{G/G_F})_A\rightarrow \Lambda(\ps{G/G_F})_B$ is
continuous since the bucket topology is weaker. Moreover it gives
rise to the geometric morphism (which we again denote as $l$)
$l:Sh(\Lambda(\ps{G/G_F})_A)\rightarrow Sh(\Lambda(\ps{G/G_F})_B)$
with direct and reverse image, respectively \ba
l_*:Sh(\Lambda(\ps{G/G_F})_A)&\rightarrow& Sh(\Lambda(\ps{G/G_F})_B)\\
l^*:Sh(\Lambda(\ps{G/G_F})_B)&\rightarrow
&Sh(\Lambda(\ps{G/G_F})_A) \ea Similarly, as was the case for
$\mv(\mh)$, we would like to analyse the push forward of the
spectral sheaf (see definition below), which is defined for each $\bigcup_{g_i\in N}\down w^g_V$ as
\be l_*(\bar{\us})\big(\bigcup_{g_i\in N}\down l_{g_i}
w^g_V\big):=\bar{\us}(l^{-1}(\bigcup_{g_i\in N}\down l_{g_i}
w^g_V))=\bar{\us}(\bigcup_{g_i\in N}\down l_{g_i} w_V) \ee So the
first issue is to understand what the sheaf $\bar{\us}$, with associated presheaf $\us$ really is.
In the definition we will use, the
isomorphisms $Hom(\down V,
\mv(\mh))\simeq(\ps{G/G_F})_V$.

\begin{Definition}
The presheaf $\us$ is defined:
\begin{itemize}
\item On objects: for each $w^g_V\in \Lambda (\ps{G/G_F})$ we have
\be \us_{w^g_V}:=\us_{\phi^g_V(p_J(\phi^g_V))} \ee where $\phi^g_V$
is the unique homeomorphism acting on $V$ associated to the coset
$w^g_V$ and $V=p_J(\phi^g_V)$.
\item On morphisms: given $w^g_V\leq w^g_{V^{'}}$ which is equivalent to $\phi^g_V\leq \phi_{V^{'}}^g$ the corresponding morphisms is
\ba
\us(i_{w^g_V,w^g_{V^{'}}}):\us_{w^g_V}&\rightarrow& \us_{w^g_{V^{'}}}\\
\us_{\phi^g_V(V)}&\rightarrow&\us_{\phi^g_{V^{'}}(V^{'})}\nonumber \ea
Therefore \be \us(i_{w^g_V,w^g_{V^{'}}}):=
\us_{\phi^g_V(V),\phi^g_{V^{'}}(V^{'})} \ee
\end{itemize}
where $V=p_J(\phi^g_V)$ and $V^{'}=p_J(\phi^g_{V^{'}})$
\end{Definition}
As we will see later on the spectral presheaf $\us$ (or corresponding spectral sheaf $\bar{\us}$) defined on $\Lambda(\ps{G/G_F})$ is nothing but the spectral presheaf $\us$ on $\mv(\mh)$ mapped via the functor, yet to be defined,  $I:Sh(\mv(\mh))\rightarrow Sh(\Lambda(\ps{G/G_F})$.
What then is $l_*(\bar{\us})$? It is defined for each open set
$\down{\mathcal{O}}(w^{g_i}_V, N):=\bigcup_{g\in N}\down l_g w^{g_i}_V$ as
\be l_*(\bar{\us})\Big(\bigcup_{g\in N}\down l_g
w^{g_i}_V\Big):=\bar{\us}(\bigcup_{g\in N}\down l_g w^{g_i}_V) \ee We now
evaluate such a set in terms of inverse limit.
We thus obtain \ba
& &\bar{\us}(\bigcup_{g\in N}\down l_g w^{g_i}_V)={\lim_{\longleftarrow}}_{ l_{g} w^{g_i}_V\in \mathcal{O}(w^{g_i}_V, N)}\us_{l_{g}w^{g_i}_V}\\
&=&\{(\alpha, \beta,\cdots,\rho)\in  \us_{l_{g_1}w^{g_i}_V}\times \us_{l_{g_2}w^{g_i}_V}\cdots\times\us_{l_{g_n}w^{g_i}_V}|\forall\;  (g, g^{'})\in N; \alpha|_{(l_{g}w^{g_i}_V\cap l_{g^{'}}w^{g_i}_V)}
=\beta|_{(l_{g}w^{g_i}_V\cap l_{g^{'}}w^{g_i}_V)}
=\nonumber\\
&&\cdots=\rho|_{(l_{g}w^{g_i}_V\cap l_{g^{'}}w^{g_i}_V)}
\}\nonumber \ea
Now, any two cosets (or equivalence classes) are either equal or disjoint. Since the group action is to move coset (or one equivalence class) to another, the intersection $l_{g}w^{g_i}_V\cap l_{g^{'}}w^{g_i}_V$ seems to be always empty. This implies that the condition $\alpha|_{(l_{g}w^{g_i}_V\cap l_{g^{'}}w^{g_i}_V) }=\beta|_{(l_{g}w^{g_i}_V\cap l_{g^{'}}w^{g_i}_V)}=\cdots$ is always satisfied thus $l_*(\bar{\us})$ is trivial.

We are then in the same, not so reassuring situation, as we were
for the base category $\mv(\mh)$. However, even in this case we
can define the map $i_V:\Lambda(\ps{G/G_F})_V\rightarrow
\Lambda(\ps{G/G_F})_B$ which represents the canonical injection of
a single orbit into the orbit space. Since the intersection of
each bucket with any given orbit is an open set, $i_V$ is
continuous. This gives rise to the bijection \be
\coprod_{V\in\mv(\mh)}\Lambda(\ps{G/G_F})_V\rightarrow
\Lambda(\ps{G/G_F})_B \ee where
$\coprod_{V\in\mv(\mh)}\Lambda(\ps{G/G_F})_V$ is equipped with the
canonical topology of disjoint union. It is straight forward to understand that $\coprod_{V\in\mv(\mh)}\Lambda(\ps{G/G_F})_V$ is nothing but $\Lambda(\ps{G/G_F})$ equipped with the `vertical' topology previously defined. Therefore we obtain a canonical
continuous (but not bi-continuous) bijection $i:
\Lambda(\ps{G/G_F})_{ver}\rightarrow \Lambda(\ps{G/G_F})_B$ with
induced diagram
\[\xymatrix{
&&\Lambda(\ps{G/G_F})_{ver}\ar[dd]^{i}\\
&&\\
\Lambda(\ps{G/G_F})_A\ar[rr]_{l}&&\Lambda(\ps{G/G_F})_B\\
}\] Similar as was the case for the base category $\mv(\mh)$ we
can define the following direct and inverse image geometric
morphisms: \ba
i_*:Sh(\Lambda(\ps{G/G_F})_{ver})&\rightarrow& Sh(\Lambda(\ps{G/G_F})_{B})\\
i^*:Sh(\Lambda(\ps{G/G_F})_{B})&\rightarrow&Sh(\Lambda(\ps{G/G_F})_{ver})
\ea The latter can be restricted to individual orbits
$\Lambda(\ps{G/G_F})_V$ obtaining \be
i^*:Sh(\Lambda(\ps{G/G_F})_V)\rightarrow
Sh(\Lambda(\ps{G/G_F})_{ver}) \ee Even in this case the manifold
structure on each of these orbits would allow us to define types
of sheaves which could not have been defined in the Alexandroff
topology. Of particular importance will be sheaves relating to differentiable
structures.

\section{Group Action on $\ps{G/G_F}$}
We would now like to analyse how the topos analogue $\ug$ of the
group $G$ acts on the presheaf $\ps{G/G_F}$. 
The action of the group
is defined by the map \be \ug\times \ps{G/G_F}\rightarrow
\ps{G/G_F} \ee such that for each context $V\in \mv_f(\mh)$ we
obtain \ba
\ug_V\times (\ps{G/G_{F}})_V&\rightarrow & (\ps{G/G_{F}})_V\\
\langle g, w^{g_1}_V\rangle&\mapsto& l_g(w^{g_1}_V) \ea where
$w^{g_1}_V=g_1 G_{FV}$. Therefore we get \ba
g(w^{g_1}_V)&=&g(g_1G_{FV})\\
&=&(gg_1)G_{FV}\\
&=&w^{gg_1}_V \ea Thus the action of the group is to move elements
along the stalk, but it never switches elements
between different stalks.
This is precisely what we were looking for in order to avoid the twisted presheaves. It is easy to see that such an action is transitive on the orbits of the sheaf $\ps{G/G_F}$.

If instead we considered the elements $\ps{Hom}(\mv_f(\mh), \mv(\mh))_V$,
the $G$-action is then defined as follows: \be
(l_g\phi^{g_i})V:=l_g(\phi^{g_i}(V))=\hat{U}_g\hat{U}_{g_i}(V)\hat{U}_{g_i^{-1}}\hat{U}_{g^{-1}}
\ee Now the homeomorphism $\phi^{g_i}$ (we will omit the $V$ subscript unless it is not clear from the context which base element we are considering ) is the unique
homeomorphism associated to the coset $w^{g_i}_V$. On the other
hand $\hat{U}_g\hat{U}_{g_i}(V)\hat{U}_{g_i}\hat{U}_{g^{-1}}$ is
identified with the homeomorphism $\phi^{gg_i}$, since by
definition \ba
\phi^{gg_i}V&=&\hat{U}_{gg_i}(V)\hat{U}_{(gg_i)^{-1}}\\
&=&\hat{U}_g\hat{U}_{g_i}(V)\hat{U}_{g_i^{-1}}\hat{U}_{g^{-1}} \ea
Therefore $\phi^{gg_i}$ is the unique homeomorphism associated to
the orbit $w^{gg_i}_V$. Thus it follows that the group action, as
defined with respect to the homeomorphisms or with respect to the
coset elements, coincides.
\subsection{Topological Considerations on the Group Action}
It is now interesting to understand whether the group presheaf $\ug$ acts continuously on the quotient presheaf $\ps{G/G_F}$. To this end we need to define what are the topologies on the individual presheaves. Since the presheaf $\ug$ is the constant presheaf which assigns to each $V\in\mv_f(\mh)$ the group $G$, the topology on each of the stalks is the topology of $G$. To combine such a `vertical' topology with the `horizontal' topology one uses the presheaf maps which, as seen in the previous sections, are simply the identity maps. It follows that each open set in $\ug$ is a kind of tube whose base would be an open set in the fibre $G$, which is then mapped to the same open set in another fibre through the presheaf maps.

Thus a typical open set would be of the form \be \coprod_{V_i\in
\downarrow V} \underline{H}_{V_i}=\coprod_{V_i\in \downarrow V}H_i
\ee
where $H_i\subseteq G$ is an open subset of $G$.

\noindent
In this way we have managed to combine the `horizontal' Alexandroff topology on the base category with the `vertical' topology of the fibres. Such a topology will be called the \emph{tube topology}.

We now consider the set
$\tilde{G}:=\coprod_{V\in\mv(\mh)}\underline{G}_V$ with
corresponding projection map $p_G:\tilde{G}\rightarrow
\mv_f(\mh)$. One now needs to check whether the map
$p_G:\tilde{G}\rightarrow \mv_f(\mh)$ is continuous in the tube
topology.
\begin{Lemma}
The map $p_G:\tilde{G}\rightarrow \mv_f(\mh)$ is continuous with
respect to the Alexandroff topology on $\mv_f(\mh)$ and the tube
topology on $\tilde{G}$.
\end{Lemma}
\begin{Proof}
$p^{-1}_G(\downarrow V)=\coprod _{V^{'}\in \downarrow V} G_{V^{'}}$.
Such a set is associated with the open subset of $\tilde{G}$ which has
value $G$ for each $V^{'}\in \downarrow V$ and $\emptyset$
everywhere else.
\end{Proof}
It is also possible to equip $\ug$ with the canonical topology on the disjoint union. Such a topology is the finest (strongest) topology on $\ug$, such that the group topology is induced on each stalk. Given such a topology the map $p_G:\tilde{G}\rightarrow \mv_f(\mh)$ is continuous. In fact we have $p_G^{-1}(\downarrow V)=\coprod_{V^{'}\in \downarrow V}G_{V^{'}}$ and each $G_{V^{'}}$ is an open subset of $\coprod _{V^{'}\in \downarrow V} G_V$.

\noindent
This canonical topology is stronger than the topology we defined earlier, since the open sets of the latter are unions of open sets of the former.

It should be noted that it is not possible to define the bucket topology on $\mv_f(\mh)$, since we have assumed that the group does not act on it.

What about the topology in the presheaf $\ps{G/G_F}$?  We know that for each $V$, $\Lambda(\ps{G/G_F})_V\simeq G/G_{FV}$, thus the topology on the presheaf space should be related to the topology on the bundle space $\Lambda(\ps{G/G_F})$. 

Let us first construct the \emph{tube topology} on the sheaf $\ps{G/G_F}$. This is by definition constructed by first considering the open subsets of each fibre, then extending them `horizontally' using the presheaf maps. However, by analysing the definition of the presheaf maps it turns out that the \emph{tube topology} is nothing but the bucket topology on $\Lambda(\ps{G/G_F})$.\\
It follows that with respect to such a topology, the map
$$\Phi:\ug\times \ps{G/G_F}\rightarrow \ps{G/G_F}$$ is indeed
continuous.

\section{From Sheaves on $\mv(\mh)$ to Sheaves on $\Lambda(\ps{G/G_F})$}
In what follows we will move freely between the language of presheaves and that of sheaves which we will  both denote as $\underline{A}$. Which of the two is being used should be clear from the context. The reason we are able to do this is because our base categories are posets (see discussion at the end of section 2).

We are now interested in `transforming' all the physically
relevant sheaves on $\mv(\mh)$ to sheaves over
$\Lambda(\ps{G/G_F})$. Therefore we are interested in finding a
functor $I: Sh(\mv(\mh))\rightarrow Sh(\Lambda(\ps{G/G_F}))$. As a
first attempt we define: \ba
I: Sh(\mv(\mh))&\rightarrow& Sh(\Lambda(\ps{G/G_F}))\\
\underline{A}&\mapsto&I(\underline{A}) \ea such that for each
context $w_V^g\simeq \phi^g$ we define \be
\big(I(\underline{A})\big)_{w_V^g}:=\underline{A}_{\phi^g(V)}=\Big((\phi^g)^*(\underline{A})\Big)(V)
\ee
where $\phi^g:\down V\rightarrow \mv(\mh)$ is the unique homeomorphism associated with the equivalence class $w^g_V=g\cdot G_{FV}$.

We then need to define the morphisms. Thus, given $i_{w_{V^{'}}^g,w_V^g}:w_{V^{'}}^g\rightarrow w_{V}^g$ ($w_{V^{'}}^g\leq
w_{V}^g$) with corresponding homeomorphisms $\phi_2^g\leq \phi_2^g$ ($\phi_1^g\in Hom(\down V, \mv(\mh))$ and $\phi_2^g\in Hom(\downarrow V^{'}, \mv(\mh))$) we have the associated morphisms
$I\underline{A}(i_{w_{V^{'}}^g,w_V^g}):\big(I(\underline{A})\Big)_{w_V^g}\rightarrow
\big(I(\underline{A})\Big)_{w_{V^{'}}^g}$ defined as \be
(I\underline{A}(i_{w_{V^{'}}^g,w_V^g}))(a)=(I\underline{A}(i_{\phi_2^g,\phi_1^g}))(a):=\underline{A}_{\phi^g_1(V),\phi^g_2(V^{'})}(a)
\ee
for all $a\in \underline{A}_{\phi^g(V)}$. In the above equation $V=p_J(\phi_1^g)$ and $V^{'}=p_J(\phi_2^g)$\footnote{Recall that $p_J:\Lambda J=\Lambda(\underline{Hom}(\mv_f(\mh),\mv(\mh))\rightarrow\mv_f(\mh)$.}. 
Moreover, since $\phi_2^g\leq \phi^g_1$ is equivalent to the condition $w_{V^{'}}^g\leq w_{V}^g$, then $\phi^g_2(V^{'})\subseteq \phi^g_1(V)$ and $\phi^g_2=\phi^g_1|_{V^{'}}$.
\begin{Theorem}
The map $I:Sh(\mv(\mh))\rightarrow Sh(\Lambda( \ps{G/G_F}))$ is a
functor defined as follows:
\begin{enumerate}
\item [(i)] Objects: $\big(I(\underline{A})\big)_{w_V^g}:=\underline{A}_{\phi^g_1(V)}=\Big((\phi^g)^*(\underline{A})\Big)(V)$. If $w_{V^{'}}^g\leq w_{V}^g$ with associated homeomorphisms $\phi_2^g\leq \phi^g_1$ ($\phi_1^g\in Hom(\down V, \mv(\mh))$ and $\phi_2^g\in Hom(\downarrow V^{'}, \mv(\mh))$), then
$$(I\underline{A}(i_{w_{V^{'}}^g,w_V^g}))=I\underline{A}(i_{\phi_2^g,\phi_1^g}):=\underline{A}_{\phi^g_1(V),\phi^g_2(V^{'})}:\underline{A}_{\phi^g_1(V)}\rightarrow \underline{A}_{\phi^g_2(V^{'})}$$
where $V=p_J(\phi_1^g)$ and $V^{'}=p_J(\phi_2^g)$.
\item [(ii)] Morphisms: if we have a morphisms $f:\underline{A}\rightarrow\underline{B}$ in $Sh(\mv(\mh))$ we then define the corresponding morphisms in $Sh(\Lambda (\ps{G/G_F}))$ as
\ba
I(f)_{w_V^g}:I(\underline{A})_{w_V^g}&\rightarrow& I(\underline{B})_{w_V^g}\\
f_{\phi_1^g}:\underline{A}_{\phi_1^g(p_J(\phi_1^g))}&\rightarrow
&\underline{B}_{\phi_1^g(p_J(\phi_1^g))} \ea
\end{enumerate}
\end{Theorem}
\begin{Proof}
Consider an arrow $f:\underline{A}\rightarrow\underline{B}$ in
$Sh(\mv(\mh))$ so that, for each $V\in \mv(\mh)$, the local
component is $f_V:\underline{A}_V\rightarrow\underline{B}_V$ with
commutative diagram
\[\xymatrix{
\underline{A}_{V_1}\ar[rr]^{f_{V_1}}\ar[dd]_{\underline{A}_{V_1V_2}}&&\underline{B}_{V_1}\ar[dd]^{\underline{B}_{V_1V_2}}\\
&&\\
\underline{A}_{V_2}\ar[rr]_{f_{V_2}}&&\underline{B}_{V_2}\\
}\] for all pairs $V_1$, $V_2$ with $V_2\leq V_1$. Now suppose
that $w_{V_2}^g\leq w_{V_1}^g$ with associated homeomorphisms $\phi^g_2\leq\phi^g_1$, such
that (i) $p_J(\phi^g_2)\subseteq p_J(\phi^g_1)$; and (ii) $\phi^g_2 =
\phi^g_1|_{p_J(\phi^g_2)}$. We want to show that the action of the $I$
functor gives the commutative diagram
\[\xymatrix{
I(\underline{A})_{w_{V_1}^g}\ar[rr]^{I(f)_{w_{V_1}^g}}\ar[dd]_{I(\underline{A})(i_{w_{V_1}^g,w_{V_2}^g})}&&I(\underline{B})_{w_{V_1}^g}\ar[dd]^{I(\underline{B})(i_{w_{V_1}^g,w_{V_2}^g})}\\
&&\\
I(\underline{A})_{w_{V_2}^g}\ar[rr]_{I(f)_{w_{V_2}^g}}&&I(\underline{B})_{w_{V_2}^g}\\
}\] for all $V_2\subseteq V_1$. By applying the definitions we get

\[\xymatrix{
\underline{A}_{\phi^g_1(p_J(\phi^g_1))}\ar[rr]^{f_{\phi^g_1(p_J(\phi^g_1))}}\ar[dd]_{\underline{A}_{\phi^g_1(p_J(\phi^g_1)),\phi^g_2(p_J(\phi^g_2))}}&&\underline{B}_{\phi^g_1(p_J(\phi^g_1))}\ar[dd]^{\underline{B}_{\phi^g_1(p_J(\phi^g_1)),\phi^g_2(p_J(\phi^g_2))}}\\
&&\\
\underline{A}_{\phi^g_2(p_J(\phi^g_2))}\ar[rr]_{f_{\phi^g_2(p_J(\phi^g_2))}}&&\underline{B}_{\phi^g_2(p_J(\phi^g_2))}\\
}\]
which is commutative. Therefore $I(f)$ is a well defined arrow in $Sh(\Lambda\ps{G/G_F})$ from $I(\underline{A})$ to $I(\underline{B})$.

Given two arrows $f,g$ in $Sh(\mv(\mh))$ then it follows that: \be
I(f\circ g)=I(f)\circ I(g) \ee This proves that $I$ is a functor
from $Sh(\mv(\mh))$ to $Sh(\Lambda(\ps{G/G_F}))$.
\end{Proof}
From the above definition of the functor $I$ we immediately have
the following corollary:
\begin{Corollary}
The functor $I$ preserves monic arrows.
\end{Corollary}
\begin{Proof}
Given a monic arrow $f:\underline{A}\rightarrow \underline{B}$ in
$Sh(\mv(\mh))$ then by definition \ba
I(f)_{w_{V_2}^g}:I(\underline{A})_{w_{V_2}^g}&\rightarrow& I(\underline{B})_{w_{V_2}^g}\\
f_{\phi_2^g(p_J(\phi^g_2))}:\underline{A}_{\phi_2^g(p_J(\phi^g_2))}&\rightarrow&\underline{B}_{\phi_2^g(p_J(\phi^g_2))}
\ea The fact that such a map is monic is straightforward.
\end{Proof}
Similarly we can show that
\begin{Corollary}
The functor $I$ preserves epic arrows.
\end{Corollary}
\begin{Proof}
Given an epic arrow $f:\underline{A}\rightarrow \underline{B}$ in
$Sh(\mv(\mh))$ then by definition \ba
I(f)_{w_{V_2}^g}:I(\underline{A})_{w_{V_2}^g}&\rightarrow& I(\underline{B})_{w_{V_2}^g}\\
f_{\phi_2^g(p_J(\phi^g_2))}:\underline{A}_{\phi_2^g(p_J(\phi^g_2))}&\rightarrow&\underline{B}_{\phi_2^g(p_J(\phi^g_2))}
\ea The fact that such a map is epic is straightforward.
\end{Proof}
We would now like to know how such a functor behaves with respect
to the terminal object. To this end we define the following
corollary:
\begin{Corollary}
The functor $I$ preserves the terminal object.
\end{Corollary}
\begin{Proof}
The terminal object in $Sh(\mv(\mh))$ is the objects
$\underline{1}_{Sh(\mv(\mh))}$ such that to each element $V\in
\mv(\mh)$ it associates the singleton set $\{*\}$. We now apply
the $I$ functor to such an object obtaining \be
I(\underline{1}_{Sh(\mv(\mh))})_{w_V^g}:=(\underline{1}_{Sh(\mv(\mh))})_{\phi^g(p_J(\phi^g))}=\{*\}
\ee
where $\phi^g$ is the unique homeomorphism associated to the coset $w^g_V$.\\
Thus it follows that
$I(\underline{1}_{Sh(\mv(\mh))})=\underline{1}_{Sh(\Lambda(\ps{G/G_F}))}$
\end{Proof}
We now check whether $I$ preserves the initial object. We recall that
the initial object in $Sh(\mv(\mh))$ is simply the sheaf
$\underline{O}_{Sh(\mv(\mh))}$ which assigns to each element $V$
the empty set $\{\emptyset\}$. We then have \be
I(\underline{O}_{Sh(\mv(\mh))})_{w_V^g}:=(\underline{O}_{Sh(\mv(\mh))})_{\phi^g(p_J(\phi^g))}=\{\emptyset\}
\ee
where $\phi^g\in Hom(\down V, \mv(\mh))$ is the unique homeomorphism associated with the coset $w^g_V$.\\
It follows that: \be
I(\underline{O}_{Sh(\mv(\mh))})=\underline{O}_{Sh(\Lambda(\ps{G/G_F}))}
\ee

From the above proof it transpires that the reason the functor $I$ preserves monic, epic, terminal object, and initial object is manly due to the fact that the action of $I$ is defined component-wise as $(I(\underline{A}))_{\phi}:=\underline{A}_{\phi(V)}$ for $\phi\in Hom(\downarrow V,\mv(\mh))$. In particular, it can be shown that $I$ preserves all limits and colimits.
\begin{Theorem}\label{the:lim}
The functor $I$ preserves limits.
\end{Theorem}

In order to prove the above theorem we first of all have to recall some general results and definitions. To this end consider two categories $\mc$ and $\md$, such that there exists a functor between them $F:\mc\rightarrow \md$. For a small index category $J$, we consider diagrams of type $J$ in both $\mc$ and $\md$, i.e. elements in $\mc^J$ and $\md^J$, respectively. The functor $F$ then induces a functor  between these diagrams as follows:
\ba
F^J:\mc^J&\rightarrow& \md^J\\
A&\mapsto&F^J(A)
\ea
such that $(F^J(A))(j):=F(A(j))$. Therefore, if limits of type $J$ exist in $\mc$ and $\md$ we obtain the diagram
\[\xymatrix{
\mc^J\ar[rr]^{\lim_{\leftarrow J}}\ar[dd]_{F^J}&&\mc\ar[dd]^{F}\\
&&\\
\md^J\ar[rr]_{\lim_{\leftarrow J}}&&\md\\
}\]
where the map
\ba
\lim_{\leftarrow J}:\mc^J&\rightarrow&\mc\\
A&\mapsto&\lim_{\leftarrow J}(A)
\ea
assigns, to each diagram $A$ of type $J$ in $\mc$, its limit $\lim_{\leftarrow J}(A)\in \mc$.
By the universal properties of limits we obtain the \emph{natural transformation} 
\be
\alpha_{J}:F\circ \lim_{\leftarrow J}\rightarrow \lim_{\leftarrow J}\circ F^J
\ee
We then say that $F$ preserves limits if $\alpha_J$ is a \emph{natural isomorphisms}. \\
For the case at hand, in order to show that the functor $I$ preserves limits we need to show that there exists a map
\be
\alpha_J:I\circ \lim_{\leftarrow J}\rightarrow \lim_{\leftarrow J}\circ I^J
\ee
which is a natural isomorphisms. Here $I^J$ represents the map
\ba
I^J:(Sh(\mv(\mh))\Big)^J&\rightarrow& \Big(Sh(\Lambda(\ps{G/G_F}))\Big)^J\\
A&\mapsto&I^J(A)
\ea
where $(I^J(A)(j))_{\phi}:=I(A(j))_{\phi}$.

\noindent
The proof of $\alpha_J$ being a natural isomorphisms will utilise a result derived in \cite{topos7} where it is shown that for any diagram $A:J\rightarrow \mc^{\md}$ of type $J$ in $\mc^{\md}$ the following isomorphisms holds
\be\label{equ:dual}
\Big( \lim_{\leftarrow J} A\Big)D\simeq \lim_{\leftarrow J}A_D\;\forall\; D\in \md
\ee
where $A_D:J\rightarrow \mc$ is a diagram in $\md$.
With these results in mind we are now ready to prove theorem \ref{the:lim}
\begin{Proof}
Let us consider a diagram $A:J\rightarrow Sets^{\mv(\mh)}$ of type $J$ in $Sets^{\mv(\mh)}$:
\ba
A:J&\rightarrow& Sets^{\mv(\mh)}\\
j&\mapsto&A(j)
\ea
where $A(j)(V):=A_V(j)$ for $A_V:j\rightarrow Sets$ a diagram in $Sets$. Assume that $L$ is a limit of type $J$ for $A$, i.e. $L:\mv(\mh)\rightarrow Sets$ such that $\lim_{\leftarrow J}A=J$. We then construct the diagram
\[\xymatrix{
\Big(Sets^{\mv(\mh)}\Big)^J\ar[rr]^{\lim_{\leftarrow J}}\ar[dd]_{I^J}&&Sets^{\mv(\mh)}\ar[dd]^{I}\\
&&\\
\Big(Sets^{\Lambda(\ps{G/G_F})}\Big)^J\ar[rr]_{\lim_{\leftarrow J}}&&Sets^{\Lambda(\ps{G/G_F})}\\
}\]
and the associated natural transformation
\be
\alpha_J:I\circ \lim_{\leftarrow J}\rightarrow \lim_{\leftarrow J}\circ I^J
\ee
For each diagram $A:J\rightarrow Sets^{\mv(\mh)}$ and $\phi\in\Lambda(\ps{G/G_F})$ we obtain
\be
\Big(I\circ \lim_{\leftarrow J}(A)\Big)_{\phi}=\Big(I\big(\lim_{\leftarrow J}A\big)\Big)_{\phi}:=\big(\lim_{\leftarrow J}A\big)_{\phi(V)}\simeq \lim_{\leftarrow J} A_{\phi(V)}
\ee
where $A_{\phi(V)}:J\rightarrow Sets$, such that $A_{\phi(V)}(j)=A(j)(\phi{V})$\footnote{Recall that $A:J\rightarrow Sets^{\mv(\mh)}$ is such that $A_{V}(j)=A(j)(V)$, therefore $\big(I(A(j))\big)_{\phi}:=A(j)_{\phi(V)}=A_{\phi(V)}(j)$}

On the other hand
\be
\Big(\big(\lim_{\leftarrow J}\circ I^J\big)A\Big)_{\phi}=\Big(\lim_{J}(I^J(A))\Big)_{\phi}\simeq \lim_{\leftarrow J}(I^J(A))_{\phi}=\lim_{\leftarrow J}A_{\phi(V)}
\ee
where 
\ba
I^J(A):J&\rightarrow& Sets^{\Lambda(\ps{G/G_F})}\\
j&\mapsto&I^J(A)(j)
\ea
such that for all $\phi\in \Lambda(\ps{G/G_F})$ we have $\big(I^J(A(j))\big)_{\phi}=\big(I(A(j))\big)_{\phi}=A(j)_{\phi(V)}$.

\noindent
It follows that 
\be
I\circ \lim_{\leftarrow J}\simeq \lim_{\leftarrow J}\circ I^J
\ee
\end{Proof}
Similarly one can show that
\begin{Theorem}
The functor $I$ preserves all colimits
\end{Theorem}
Since colimits are simply duals to the limits, the proof of this theorem is similar to the proof given above. However, for completeness sake we will, nonetheless, report it here.
\begin{Proof}
We first of all construct the analogue of the diagram above:
\[\xymatrix{
\Big(Sets^{\mv(\mh)}\Big)^J\ar[rr]^{\lim_{\rightarrow J}}\ar[dd]_{I^J}&&Sets^{\mv(\mh)}\ar[dd]^{I}\\
&&\\
\Big(Sets^{\Lambda(\ps{G/G_F})}\Big)^J\ar[rr]_{\lim_{\rightarrow J}}&&Sets^{\Lambda(\ps{G/G_F})}\\
}\]
where $\lim_{\rightarrow J}:\Big(Sets^{\mv(\mh)}\Big)^I\rightarrow Sets^{\mv(\mh)}$ represents the map which assigns colimits to all diagrams in $ \Big(Sets^{\mv(\mh)}\Big)^I$.

\noindent
We now need to show that the associated natural transformation
\be
\beta_J:I\circ \lim_{\rightarrow J} \rightarrow \lim_{\rightarrow J}\circ I^J
\ee
is a natural isomorphisms.

\noindent
For any diagram $A\in \Big(Sets^{\mv(\mh)}\Big)^I$ and $\phi\in \Lambda(\ps{G/G_F})$ we compute
\be
\Big(I\circ \lim_{\rightarrow J}(A)\Big)_{\phi}=\Big(I( \lim_{\rightarrow J}A)\Big)_{\phi}=\Big(\lim_{\rightarrow J}A\Big)_{\phi(V)}\simeq \lim_{\rightarrow J}A_{\phi(V)}
\ee
where $\Big(\lim_{\rightarrow J}A\Big)_{\phi(V)}\simeq \lim_{\rightarrow J}A_{\phi(V)}$ is the dual of \ref{equ:dual}.
On the other hand 
\be
\Big(( \lim_{\rightarrow J}\circ I^J)(A)\Big)_{\phi}=\Big( \lim_{\rightarrow J}(I^J(A))\Big)_{\phi}\simeq \lim_{\rightarrow J}(I^J(A))_{\phi}=\lim_{\rightarrow J}A_{\phi(V)}
\ee
It follows that indeed $\beta_J$ is a natural isomorphisms.
\end{Proof}


Now we would like to check whether $I$ is a left adjoint. To this end we need to
construct its right adjoint and show that indeed they form an adjoint pair. Unfortunately, the existence of this putative right adjoint can not be proven. In the next section we will show why this is the case. Despite this unfortunate result, the functor $I$ still has very important properties, which allow us to map the relevant objects from the topos $Sh(\mv(\mh))$ to $Sh(\Lambda(\ps{G/G_F}))$.

\subsection{The Right Adjoint $J$?}
As a first guess we define the right adjoint
$J:Sh(\Lambda(\ps{G/G_F}))\rightarrow Sh(\mv(\mh))$ to be such
that, given a sheaf $\underline{A}\in Sh(\Lambda(\ps{G/G_F}))$ we
obtain \ba J(\underline{A})_{V}:=\coprod_{w^{g_i}_V\in
\ug/\ug_{FV}}\underline{A}_{w^{g_i}_V} =\coprod_{\phi_i\in
Hom(\down V,\mv(\mh))}\underline{A}_{\phi_i} \ea 
where from now on, for notational simplicity, we will simply denote by $\phi_i$ the unique homomorphism associated to the coset $w_V^{g_i}$.

\noindent
We then need to
define the morphisms. Thus given $i_{V^{'}V}:V^{'}\rightarrow V$
we have that the associated morphisms \ba
J(\underline{A})(i_{V^{'}V}):J(\underline{A})_{V}&\rightarrow& J(\underline{A})_{V^{'}}\nonumber\\
\coprod_{w^{g_i}_V\in (\ps{G/G_F})_V}\underline{A}_{w^{g_i}_V}&\rightarrow& \coprod_{w^{g_j}_{V^{'}}\in(\ps{G/G_F})_{V^{'}}}\underline{A}_{w^{g_j}_{V^{'}}}\nonumber\\
\coprod_{\phi_i\in
Hom(\down V,\mv(\mh))}\underline{A}_{\phi_i}&\rightarrow&\coprod_{\phi_j\in
Hom(\down V^{'},\mv(\mh))}\underline{A}_{\phi_j} \ea are
defined\footnote{In the definition of morphisms we utilise the
concept of arrows between co-products. Let us assume we have the
arrows $f:A\rightarrow B$ and $g:C\rightarrow D$ in some category.
We then define the co-products $A\sqcup C$ and $B\sqcup C$. The
co-product map $h:A\sqcup C\rightarrow B\sqcup D$, is then defined
by the following commutative diagram:
\[\xymatrix{
A\ar[rr]^{i_A}\ar[dd]_{f}&&A\sqcup C\ar[dd]^{[i_{B}\circ f, i_D\circ g]=h}&&C\ar[dd]^g\ar[ll]_{i_C}\\
&&&&\\
B\ar[rr]_{i_B}&&B\sqcup D&&D\ar[ll]^{i_D}\\
}\] where the arrows $i_A$ etc are the injection maps. } as \be
J(\underline{A})(i_{V^{'}V})(a_i):=\underline{A}(i_{w^{g_i}_V,w^{g_j}_{V^{'}}})(a_i)=\underline{A}_{\phi_i\phi_j}(a_i)
\ee for all $a_i\in \underline{A}_{w^{g_i}_V}$. Here 
$V=p_J(\phi_i)$,
$V^{'}=p_J(\phi_j)$ and
$\phi_j(V^{'}):=\phi_{i|V^{'}}(V^{'})$. Thus
$$J(\underline{A})(i_{V^{'}V})=[i_{\underline{A}_{w^{g_1}_V}}\circ \underline{A}(i_{w^{g_1}_V,w^{g_1}_{V^{'}}})\cdots i_{\underline{A}_{w^{g_n}_V}}\circ \underline{A}(i_{w^{g_n}_V,w^{g_n}_{V^{'}}})]$$
We need to show that indeed $J$ is a functor.
\begin{Theorem}
The map $J:Sh(\Lambda(\ps{G/G_F}))\rightarrow Sh(\mv(\mh))$ is a
functor defined as follows:
\begin{enumerate}
\item [(i)] Objects: $J(\underline{A})_V:=\coprod_{w^{g_i}_V\in (\ps{G/G_F})_{V}}\underline{A}_{w^{g_i}_V} =\coprod_{\phi_i\in Hom(\down V, \mv(\mh))}\underline{A}_{\phi_i}$ where $V=p_J(\phi_i)$. If $V^{'}\leq V$ then:
\ba J(\underline{A})(i_{V^{'}V}):J(\underline{A})_V&\rightarrow& J(\underline{A})_{V^{'}}\ea
is defined as 
\ba
[i_{\underline{A}_{w^{g_1}_V}}\circ
\underline{A}(i_{w^{g_1}_V,w^{g_1}_{V^{'}}})\cdots
i_{\underline{A}_{w^{g_n}_V}}\circ
\underline{A}(i_{w^{g_n}_V,w^{g_n}_{V^{'}}})]:\coprod_{w^{g_i}_V\in
(\ps{G/G_F})_{V}}\underline{A}_{w^{g_i}_V}&\rightarrow&
\coprod_{w^{g_i}_{V^{'}}\in
(\ps{G/G_F})_{V^{'}}}\underline{A}_{w^{g_i}_{V^{'}}} \\
\coprod_{\phi_i\in Hom(\down V, \mv(\mh))}\underline{A}_{\phi_i(V)}&\rightarrow& \coprod_{\phi_j\in Hom(\down V^{'}, \mv(\mh))}\underline{A}_{\phi_j}\nonumber
\ea
where $V=p_J(\phi_i)$ and $V^{'}=p_J(\phi_j)$.
\item Morphisms: Given a morphisms $f:\underline{A}\rightarrow\underline{B}$ in $Sh(\Lambda(\ps{G/G_F}))$ with local components $f_{w^{g_i}_V}:\underline{A}_{w^{g_i}_V}\rightarrow\underline{B}_{w^{g_i}_V}$, for all $w^{g_i}_V\in \Lambda(\ps{G/G_F})$, then the corresponding morphisms in $Sh(\mv(\mh))$ would be
\ba
J(f)_V:J(\underline{A})_V&\rightarrow& J(\underline{B})_V\\
\;[i_{\underline{A}_{w^{g_i}_V}}\circ
f_{w^{g_i}_V},\cdots,i_{\underline{A}_{w^{g_n}_V}}\circ
f_{w^{g_n}_V}]:\coprod_{w^{g_i}_V\in
(\ps{G/G_F})_{V}}\underline{A}_{w^{g_i}_V}&\rightarrow&\coprod_{w^{g_i}_V\in
(\ps{G/G_F})_{V}}\underline{B}_{w^{g_i}_V} \nonumber\\
\coprod_{\phi_i\in Hom(\down V, \mv(\mh))}\underline{A}_{\phi_i}&\rightarrow&\coprod_{\phi_i\in Hom(\down V, \mv(\mh))}\underline{B}_{\phi_i}\nonumber
\ea
\end{enumerate}
\end{Theorem}
\begin{Proof}
Let us consider a map $f:\underline{X}\rightarrow \underline{Y}$
in $Sh(\Lambda(\ps{G/G_F})$, then, for each elements $w^{g}_{V}\in
\Lambda(\ps{G/G_F})$ we obtain the function
$f:\underline{X}_{w^{g}_{V}}\rightarrow \underline{Y}_{w^{g}_{V}}$
with commutative diagram
\[\xymatrix{
\underline{X}_{w^g_V}\ar[rr]^{f_{w^g_V}}\ar[dd]_{\underline{X}_{w^g_V,w^g_{V^{'}}}}&&\underline{Y}_{w^g_V}\ar[dd]^{\underline{Y}_{w^g_V,w^g_{V^{'}}}}\\
&&\\
\underline{X}_{w^g_{V^{'}}}\ar[rr]_{f_{w^g_{V^{'}}}}&&\underline{Y}_{w^g_{V^{'}}}\\
}\]
for all pairs $w^g_{V}, w^g_{V^{'}}$ such that $w^g_{V}\leq w^g_{V^{'}}$.\\
From the definition of the ordering relation we know that
$w^g_{V}\leq w^g_{V^{'}}$ implies $V\leq V^{'}$. We now want to
show that by applying the $J$ functor we obtain the following
commutative diagram in $Sh(\mv(\mh))$.
\[\xymatrix{
J(\underline{X})_{V}\ar[rr]^{J(f)_V}\ar[dd]_{J(\underline{X})_{V,V^{'}}}&&J(\underline{Y})_{V}\ar[dd]^{J(\underline{Y})_{V,V^{'}}}\\
&&\\
J(\underline{X})_{V^{'}}\ar[rr]_{J(f)_{V^{'}}}&&J(\underline{Y})_{V^{'}}\\
}\] By applying the definition of the $J$ functor we get
\[\xymatrix{
\coprod_{w^{g_i}_V\in(\ps{G/G_F})_{V}}\underline{X}_{w^{g_i}_V}\ar[rr]^{J(f)_V}\ar[dd]_{J(\underline{X})_{V,V^{'}}}&&\coprod_{w^{g_1}_V\in (\ps{G/G_F})_{V}}\underline{Y}_{w^{g_i}_V}\ar[dd]^{J(\underline{Y})_{V,V^{'}}}\\
&&\\
\coprod_{w^{g_i}_{V^{'}}\in (\ps{G/G_F})_{V^{'}}}\underline{X}_{w^{g_i}_{V^{'}}}\ar[rr]_{J(f)_{V^{'}}}&&\coprod_{w^{g_i}_{V^{'}}\in (\ps{G/G_F})_{V^{'}}}\underline{Y}_{w^{g_i}_{V^{'}}}\\
}\]
where the definition of the maps $J(\underline{X})_{V,V^{'}}$, $J(\underline{Y})_{V,V^{'}}$, $J(f)_V$ and $J(f)_{V^{'}}$ were given above. 
Clearly this diagram commutes.

From this it follows that given two morphisms $f, g\in
Sh(\Lambda(\ps{G/G_F})$, then \be J(f\circ g)=J(f)\circ J(g) \ee
\end{Proof}
Now that we have defined the functor $J$, in order to show that it
is indeed the right adjoint of $I$ we need to show that the
following isomorphisms exists: \be
Hom_{Sh(\Lambda(\ps{G/G_F})}(I(\underline{Y}),
\underline{X})\simeq Hom_{Sh(\mv(\mh))}(\underline{Y},
J(\underline{X})) \ee Thus we need to define an isomorphic map
\ba\label{ali:isoadj}
i:Hom_{Sh(\Lambda(\ps{G/G_F})}(I(\underline{Y}), \underline{X})&\rightarrow &Hom_{Sh(\mv(\mh))}(\underline{Y}, J(\underline{X}))\\
f&\mapsto& i(f) \ea First of all let us analyse the map
$f:I(\underline{Y})\rightarrow \underline{X}$ which has as
individual components \be
f_{w^g_V}:I(\underline{Y})_{w^g_V}\rightarrow
\underline{X}_{w^g_V} \ee 
for each $w_V^{g_i}\in \ps{G/G_F}$.

Utilising the homeomorphism $G/G_{FV}\simeq Hom(\down V,
\mv(\mh))$ for each $V\in\mv_f(\mh)$ the above map $f$ can be written as \be
f_{w^{g_i}_V}:=f_{\phi_i}:I(\underline{Y})_{\phi_i}\rightarrow
\underline{X}_{\phi_i} \ee Since $\phi_i\in  Hom(\down V, \mv(\mh))$ and
since
$I(\underline{Y})_{\phi_i}:=\underline{Y}_{\phi_i(p_J(\phi_i))}$
we can write $f$ as \be
f_{\phi_i}:\underline{Y}_{\phi_i(p_J(\phi_i))}\rightarrow
\underline{X}_{\phi_i} \ee 
where $\phi_i(p_J(\phi_i))=V^{'}\in\mv(\mh)$. 

We now need to define the
action of the $i$ map. This is defined for all $V\in \mv(\mh)$ as
\ba
(i(f))_V:\underline{Y}_V&\rightarrow& J(\underline{X})_V\\
\underline{Y}_V&\rightarrow&\coprod_{w^{g_i}_V\in
(\ps{G/G_{F}})_V}\underline{X}_{w^{g_i}_V} \ea however
$\coprod_{w^{g_i}_V\in \ps{G/G_{F}}_V}\underline{X}_{w^{g_i}_V}\simeq
\coprod_{\phi_i\in Hom(\down V,
\mv(\mh))}\underline{X}_{\phi_i}$. Therefore we get \ba
(i(f))_V:\underline{Y}_V&\rightarrow&\coprod_{\phi_i\in Hom(\down V,
\mv(\mh))}\underline{X}_{\phi_i} \ea Then $i(f)$ is defined as the following map in $Sh(\mv(\mh))$: 
\ba
(i(f))_{V}&:=&i_{\underline{X}_{\phi_i}}
\circ f_{\phi_i} 
\ea
where
$i_{\underline{X}_{\phi_i}}$ is the injection map, and $\phi_i\in Hom(\downarrow V, \mv(\mh))$. In
other words $i(f)_{V}$ is the composite map \be
\underline{Y}_{V}\xrightarrow{f_{\phi_i}
}\underline{X}_{\phi_i}\xrightarrow{i_{\underline{X}_{\phi_i}}
}\coprod_{\phi_i\in Hom(\down V,
\mv(\mh))}\underline{X}_{\phi_i} \ee Given
this definition we need to check wether $i$ is an
isomorphism.
\begin{Conjecture}
The map $i$ defined in \ref{ali:isoadj} is an isomorphism
\end{Conjecture}

\begin{enumerate}
\item [i)] The map $i$ is 1:2:1. In fact if $i(f)=i(f^{'})$ then for all $V\in \mv(\mh)$ we have that
\be
i(f)_V=i(f^{'})_V= i_{\underline{X}_{\phi_i}}\circ
f_{\phi_i}=i_{\underline{X}_{\phi_i}}\circ
f^{'}_{\phi_i} \ee
where $\phi_i\in Hom(\downarrow V,\mv(\mh))$. However $i_{\underline{X}_{\phi_i}}$ is monic thus left cancellable. From this it follows that
$f=f^{'}$.
\item [ii)] The map $i$ is onto. This is true by construction.
\item [iii)] We need to check wether a possible inverse exists. A first guess would be to define 
\ba
j:Hom_{Sh(\mv(\mh))}(\underline{Y}, J(\underline{X}))&\rightarrow &Hom_{Sh(\Lambda(\ps{G/G_F})}(I(\underline{Y}), \underline{X})\\
g&\mapsto& j(g) \ea Where 
\be
j(g)_{\phi}:=pr_{\phi}\circ g_{p_J(\phi)}\circ pr_{p_J(\phi)}\circ i_{\underline{Y}_{\phi(p_J(\phi))}}
\ee
The graphical representation of the above arrow is the following:
\be
\underline{Y}_{\phi(p_J(\phi))}\xrightarrow{i_{\underline{Y}_{\phi(p_J(\phi))}}}\coprod_{\phi\in Hom(\downarrow p_J(\phi), \mv(\mh))}\underline{Y}_{\phi(p_J(\phi))}\xrightarrow{pr_{p_J(\phi)}}\underline{Y}_{p_J(\phi)}\xrightarrow{g_{p_J(\phi)}}\coprod_{\phi^{'}\in Hom(\downarrow p_J(\phi),\mv(\mh))}\underline{X}_{\phi^{'}}\xrightarrow{pr_{\phi}}\underline{X}_{\phi}
\ee
However this does not give us the desired inverse. The reason being that the effect of the composition of $i$ and $j$ is to change the context one starts from. In fact, given a map $g\in Hom(I(\underline{Y},\underline{X})$, $(i\circ j(g))_V=g_{\phi(V)}$ for some $\phi\in Hom(\downarrow V, \mv(\mh))$. However if $i$ and $j$ where inverse of each other we should obtain $ (i\circ j(g))_V=g_V$ which we don't.
In fact we obtain
\be
(i\circ j(g))_V=i_{\underline{X}_{\phi}}\circ pr_{\phi}\circ g_{p_J(\phi)}\circ pr_{p_J(\phi)}\circ i_{\underline{Y}_{\phi(p_J(\phi))}}
\ee
Here $\phi\in Hom(\downarrow V, \mv(\mh))$ and $p_J(\phi)=V$, therefore $\phi(p_J(\phi))=V$ iff $\phi$ represents the transformation the transformation associated to the identity element or to any group element belonging to the stability group of $V$.
\end{enumerate}

\noindent
It follows that $J\nvdash I$.

\section{The Left Adjoint $p_J!$ of $p_J^*$ }
It is a standard result that, given a map $f : X\rightarrow Y$
between topological spaces $X$ and $Y$, we obtain a
geometric morphisms\footnote{\begin{Definition} \cite{topos7}, \cite{sv}
A \emph{geometric morphism} $\phi:\tau_1\rightarrow \tau_2$
between topoi $\tau_1$ and $ \tau_2$ is defined to be a pair of
functors $\phi_*:\tau_1\rightarrow \tau_2$ and
$\phi^*:\tau_2\rightarrow \tau_1$, called respectively the
\emph{direct image} and the \emph{inverse image} part of the
geometric morphism, such that
\begin{enumerate}
\item $\phi^*\dashv  \phi_*$ i.e., $\phi^*$ is the left adjoint of $\phi_*$
\item $\phi^*$ is left exact, i.e., it preserves all finite limits.
\end{enumerate}
\end{Definition}}
 \ba
f^*: Sh(Y )&\rightarrow& Sh(X)\\
f_* : Sh(X)&\rightarrow& Sh(Y ) \ea and we know that $f^*\dashv
f_*$, i.e., $f^*$ is the left-adjoint of $f_*$. If $f$ is an
etal\'e map, however, there also exists the left adjoint $f!$ to
$f^*$, namely \be f! : Sh(X)\rightarrow Sh(Y ) \ee
with $f!\dashv f^*\dashv f_*$ (see \cite{topos7}).\\
In the appendix we will show that \be f!(p_A : A\rightarrow X) =
f\circ p_A : A\rightarrow Y \ee so that we combine the etal\'e
bundle $p_A:A\rightarrow X$ with the etal\'e map $f : X\rightarrow
Y$ to give the
etal\'e bundle $f\circ p_A : A\rightarrow Y$. Here we have used the fact that sheaves can be defined in terms of etal\'e bundles. In fact in \cite{topos7} is was shown that there exists an equivalence of categories $Sh(X)\simeq Etale(X)$ for any topological space $X$.

Given a map $\alpha:A\rightarrow B$ of etal\'e bundles over $X$,
we obtain the map $f!(\alpha) : f!(A)\rightarrow f!(B)$ which is
defined as follows. We start with the collection of fibre maps $\alpha_x :
A_x\rightarrow B_x$,  $x\in X$, where $A_x := p^{-1}A (\{x\})$.
Then, for each $y\in Y$ we want to define the maps $f!(\alpha)_y :
f!(A)_y\rightarrow f!(B)_y$,  i.e., $f!(\alpha)_y : p^{-1}\big( A
(f^{-1}\{y\})\big)\rightarrow p^{-1}\big( B (f^{-1}\{y\})\big)$.
This are defined as \be\label{equ:shreack} f!(\alpha)_y(a) :=
\alpha_{p_A(a)}(a) \ee
for all $a\in f!(A)_y = p^{-1}\big( A (f^{-1}\{a\})\big)$.

For the case of interest we obtain the left adjoint functor
$p_J!:Sh(\Lambda(\ps{G/G_F}))\rightarrow Sh(\mv_f(\mh))$ of
$p_J^*:Sh(\mv_f(\mh))\rightarrow Sh(\Lambda(\ps{G/G_F}))$. The
existence of such a functor enables us to define the composite
functor \be F:=p_J!\circ I:Sh(\mv(\mh))\rightarrow Sh(\mv_f(\mh))
\ee Such a functor sends all the original sheaves we had defined
over $\mv(\mh)$ to new sheaves over $\mv_f(\mh)$. Thus, denoting
the sheaves over $\mv_f(\mh)$ as $\underline{\breve{A}}$ we have
\be \breve{\us}:=F(\us)=p_J!\circ I(\us) \ee

What happens to the terminal object? Given
$\underline{1}_{\mv(\mh)}$ we obtain \be
F(\underline{1}_{\mv(\mh)})=p_J!\circ
I(\underline{1}_{\mv(\mh)})=p_J!(\underline{1}_{\Lambda(\ps{G/G_F})})
\ee Now the etal\'e bundle associated to the sheaf
$\underline{1}_{\Lambda(\ps{G/G_F})}$ is
$p_1:\Lambda(\{*\})\rightarrow \Lambda(\ps{G/G_F}))$ where
$\Lambda(\{*\})$ represents the collection of singletons, one for
each $w_V^g\in \Lambda(\ps{G/G_F})$. Obviously the etal\'e bundle
$p_1:\Lambda(\{*\})\rightarrow \Lambda(\ps{G/G_F}))$ is nothing
but $\Lambda(\ps{G/G_F})$. Thus by applying the definition of
$p_J!$ we then get \be
p_J!(\underline{1}_{\Lambda(\ps{G/G_F})})=\ps{G/G_F} \ee
It follows that the functor $F$ does not preserve the terminal object therefore it can not be a right adjoint. In fact we would like $F$ to be left adjoint. However so far that does not seem the case. 
%
We have seen above that the functor $I$ preserves \emph{colimits} (initial object) and \emph{limits}. Since $F=p_J!\circ I$ and $p_J!$ is left adjoint thus preserves \emph{colimits}, it follows that $F$ will preserve \emph{colimits}.

Of particular importance to us is the following: each object
$\underline{A}\in Sh(\mv(\mh))$ has associated to it the unique
arrow $!\underline{A}:\underline{A}\rightarrow
\underline{1}_{\mv(\mh)}$. This arrow is epic thus
$F(!\underline{A}):F(\underline{A})\rightarrow
F(\underline{1}_{\mv(\mh)})$ is also epic. In particular we obtain
\ba
F(!\underline{A}):F(\underline{A})&\rightarrow &F(\underline{1}_{\mv(\mh)})\\
\underline{\breve{A}}&\rightarrow&\ps{G/G_F} \ea such that for
each $V\in \mv(\mh)$ we get \ba
\underline{\breve{A}}_V&\rightarrow&(\ps{G/G_F})_V\\
\coprod_{w^g_V\in(\ps{
G/G_F})_V}\underline{A}_{w^g_V}&\rightarrow&G/G_{FV} \ea 

However,
since we are considering sub-objects of the state object presheaf
$\breve{\us}$ we would like the $F$ functor to also preserve monic
arrows. And indeed it does.
\begin{Lemma}
The functor $F:Sh(\mv(\mh))\rightarrow Sh(\mv_f(\mh))$ preserves
monics.
\end{Lemma}
\begin{Proof}
Let $i:\underline{A}\rightarrow \underline{B}$ be a monic arrow in
$Sh(\mv(\mh))$, then we have that \be F(i) =p_J!(I(i)) \ee
However, the $I$ functor preserves monics, as a consequence $I(i)$
is monic in $Sh(\Lambda(\ps{G/G_F}))$.
\\
Moreover, from the defining equation \ref{equ:shreack}, it follows
that if $f : X\rightarrow Y$ is etal\'e and $p_A:A\rightarrow X$
is etal\'e then, since $i : A\rightarrow B$ is monic then so is
$f!(i) : f!(A)\rightarrow f!(B)$. Therefore applying this
reasoning to our case it follows that $F(i) =p_J!(I(i))$ is monic.
\end{Proof}
\section{From Sheaves over $\mv(\mh)$ to Sheaves over $\mv(\mh_f)$}
Now that we have defined the left adjoint functor $F$ we will map
all the sheaves in our original formalism ($Sh(\mv(\mh))$) to
sheaves over $\mv_f(\mh)$. We will then analyse how the truth
values behave under such mappings.
\subsection{Spectral Sheaf}
Given the spectral sheaf $\us\in Sh(\mv(\mh))$ we define the
following: \be \breve{\us}:=F(\us)=p_{I}!\circ I (\us) \ee This will
be our new spectral sheaf. The definition given below will be in
terms of the corresponding presheaf (which we will still denote
$\breve{\us)}$), where we have used the correspondence between
sheaves and presheaves induced by the fact that the base category
is a poset (see equation \ref{equ:correspondence})
\begin{Definition}
The spectral presheaf $\breve{\us}$ is defined on
\begin{itemize}
\item [--] Objects: For each $V\in\mv_f(\mh)$ we have
\be
\breve{\us}_V:=\coprod_{w^{g_i}_V\in\Lambda(\ps{G/G_F})_V}\us_{w^{g_i}_V}\simeq\coprod_{\phi_i\in
Hom(\down V,\mv(\mh))}\us_{\phi(V)} \ee which represents the disjoint
union of the Gel'fand spectrum of all algebras related to $V$ via
a group transformation
\item[--] Morphisms: Given a morphism $i:V^{'}\rightarrow V$, ($V^{'}\subseteq V)$ in $\mv_f(\mh)$ the corresponding spectral presheaf morphism is
\ba
\breve{\us}(i_{V^{'}V}):\breve{\us}_{V}&\rightarrow& \breve{\us}_{V^{'}}\\
\coprod_{\phi_i\in
Hom(\down V,\mv(\mh))}\us_{\phi_i(V)}&\rightarrow&\coprod_{\phi_j\in
Hom(\down V^{'},\mv(\mh))}\us_{\phi(V^{'})} \ea
such that given $\lambda\in\us_{\phi_i(V)}$ we obtain $\breve{\us}(i_{V^{'}V})(\lambda):=\us_{\phi_i(V),\phi_j(V^{'})}\lambda=\lambda_{|\phi_j(V^{'})}$\\
Thus in effect $\breve{\us}(i_{V^{'}V})$ is actually a co-product of morphisms $\us_{\phi_i(V),\phi_j(V^{'})}$, one for each 
$\phi_i \in Hom(\down V,\mv(\mh))$.
\end{itemize}
\end{Definition}
From the above definition it is clear that the new spectral sheaf
contains the information of all possible representations of a
given abelian von-Neumann algebra at the same time. It is such an
idea that will reveal itself fruitful when considering how
quantisation is defined in a topos.

\subsubsection{Topology on The State Space}
We would now like to analyse what kind of topology the sheaf
$\breve{\us}:=F(\us)$ has. We know that for each $V\in\mv_f(\mh)$
we obtain the collection $\coprod_{w^{g_i}_V\in
G/G_{FV}}\us_{w^{g_i}_V}$, where each
$\us_{w^{g_i}_V}:=\us_{\phi^{g_i}(V)}$ is equipped with the spectral
topology. Thus, similarly as was the case of the sheaf $\us\in
Sh(\mv(\mh))$, we could equip $\breve{\us}$ with the disjoint
union topology or with the spectral topology. In order to
understand the spectral topology we should recall that the functor
$F:Sh(\mv(\mh))\rightarrow Sh(\mv_f(\mh))$ preserves monics, thus
if $\underline{S}\subseteq \us$, then
$\breve{\underline{S}}:=F(\underline{S})\subseteq
\breve{\us}:=F(\us)$. We can then define the spectral topology on
$\breve{\us}$ as follows
\begin{Definition}
The spectral topology on $\breve{\us}$ has as basis the collection
of clopen sub-objects $\breve{\underline{S}}\subseteq \breve{\us}$
which are defined for each $V\in \mv_f(\mh)$ as \be
\breve{\underline{S}}_V:=\coprod_{w_{V}^{g_i}\in
G/G_{FV}}\underline{S}_{w_{V}^{g_i}}=\coprod_{\phi_i\in Hom(\down V,
\mv(\mh))}\underline{S}_{\phi_i(V)} \ee
\end{Definition}
From the definition it follows that on each element $\us_{w^{g_i}_V}$ of the stalks we retrieve the standard spectral topology.

It is easy to see that the map $p:\coprod_{w^{g_i}_V\in \Lambda(\ps{G/G_F})}\us_{w^{g_i}_V}\rightarrow \mv_f(\mh)$ is continuous since
$p^{-1}(\downarrow V):=\coprod_{w^{g_i}_{V^{'}}\in \downarrow w^{g_i}_V|\forall w^{g_i}_V\in G/G_{FV}}\us_{w^{g_i}_{V^{'}}}$ is the clopen sub-object which has value $\coprod_{w^{g_i}_{V^{'}}\in G/G_{FV^{'}}}\us_{w^{g_i}_{V^{'}}}$ at each context $V^{'}\in \downarrow V$ and $\emptyset $ everywhere else.

Similarly, as was the case for the topology on $\us\in
Sh(\mv(\mh))$, the spectral topology defined above is weaker than
the product topology and it has the advantage that if takes into account both the `vertical' topology on the fibres and the `horizontal' topology on the base space $\mv_f(\mh)$. 

A moment of thought will reveal that also with respect to the disjoint union topology the map $p$ is continuous, however because of the above argument, from now on we will use the spectral
topology on the spectral presheaf.
\subsection{Quantity Value Object}
We are now interested in mapping the quantity value objects
$\underline{\Rl}^{\leftrightarrow}\in Sh(\mv(\mh))$ to an object
in $Sh(\mv_f(\mh))$ via the $F$ functor. We thus define:
\begin{Definition}
The quantity value objects
$\breve{\underline{R}}^{\leftrightarrow}:=F(\underline{\Rl}^{\leftrightarrow})=p_{I}!\circ
I(\underline{\Rl}^{\leftrightarrow})$ is an $\Rl$-valued presheaf
of order-preserving and order-reversing functions on $\mv_f(\mh)$
defined as follows:
\begin{itemize}
\item[--] On objects $V\in\mv_f(\mh)$ we have
\be (F(\underline{\Rl}^{\leftrightarrow}))_V:=\coprod_{\phi_i\in
Hom(\down V,\mv(\mh))}\underline{\Rl}^{\leftrightarrow}_{\phi_i(V)} \ee
where each \be
\underline{\Rl}^{\leftrightarrow}_{\phi_i(V)}:=\{(\mu, \nu)|\mu\in
OP(\downarrow \phi_i(V), \Rl)\;,\; \mu\in OR(\downarrow \phi_i(V),
\Rl),\; \mu\leq\nu\} \ee The downward set $\downarrow \phi_i(V)$
comprises all the sub-algebras $V^{'}\subseteq \phi_i(V)$. The
condition $\mu\leq\nu$ implies that for all $V^{'}\in\down
\phi_i(V)$, $\mu(V^{'})\leq\nu(V^{'})$.
\item[--] On morphisms $i_{V^{'}V}:V^{'}\rightarrow V$ ($V^{'}\subseteq V)$ we get:
\ba
\breve{\underline{R}}^{\leftrightarrow}(i_{V^{'}V}):\breve{\underline{R}}^{\leftrightarrow}_V&\rightarrow& \breve{\underline{R}}^{\leftrightarrow}_{V^{'}}\\
\coprod_{\phi_i\in
Hom(\down V,\mv(\mh))}\underline{\Rl}^{\leftrightarrow}_{\phi_i(V)}&\rightarrow
&\coprod_{\phi_j\in
Hom(\down V^{'},\mv(\mh))}\underline{\Rl}^{\leftrightarrow}_{\phi_j(V^{'})}
\ea where for each element $(\mu,\nu)\in
\underline{\Rl}^{\leftrightarrow}_{\phi_i(V)}$ we obtain \ba
\breve{\underline{R}}^{\leftrightarrow}(i_{V^{'}V})(\mu,\nu)&:=&\underline{R}^{\leftrightarrow}(i_{\phi_i(V),\phi_j(V^{'})})(\mu,\nu)\\
&=&(\mu_{|\phi_i(V^{'})},\nu_{|\phi_j(V^{'})}) \ea where
$\mu_{|\phi_i(V^{'})}$ denotes the restriction of $\mu$ to
$\downarrow \phi_j(V^{'})\subseteq\downarrow\phi_i(V)$, and
analogously for $\nu_{|\phi_j(V^{'})}$.
\end{itemize}

\end{Definition}
\subsubsection{Topology on the Quantity Value Object}
We are now interested in defining a topology for our newly defined quantity value object $\breve{\underline{\Rl}}$. Similarly, as was done for the spectral sheaf, we define the set
\be
\mathcal{R}=\coprod_{V\in\mv_f(\mh)}\breve{\underline{\Rl}}^{\leftrightarrow}_V=\bigcup_{V\in\mv_f(\mh)}\{V\}\times\breve{\underline{\Rl}}^{\leftrightarrow}_V
\ee
where each $\breve{\underline{\Rl}}^{\leftrightarrow}_V:=\coprod_{\phi_i\in Hom(\down V, \mv(\mh))}\underline{\Rl}^{\leftrightarrow}_{\phi_i(V)}$.\\
The above represents a bundle over $\mv_f(\mh)$ with bundle map $p_{\mathcal{R}}:\mathcal{R}\rightarrow \mv_f(\mh)$ such that $p_{\mathcal{R}}(\mu, \nu)=V=p_J(\phi_i)$, where $V$ is the context such that $(\mu, \nu)\in\underline{\Rl}^{\leftrightarrow}_{\phi_i(V)}$. 
In this setting
$p^{-1}_{\mathcal{R}}(V)=\breve{\underline{\Rl}}^{\leftrightarrow}_{V}$
are the fibres of the map $p_{\mathcal{R}}$.

We would like to define a topology on $\mathcal{R}$ with the minimal require that the map $p_{\mathcal{R}}$ is continuous. We know that the category $\mv_f(\mh)$ has the Alexandroff topology whose basis open sets are of the form $\downarrow V$ for some $V\in\mv_f(\mh)$. Thus we are looking for a topology such that the pullback 
$p_{\underline{\Rl}}^{-1}(\downarrow V):=\coprod_{V^{'}\in\downarrow V}\underline{\breve{\Rl}}_{V^{'}}$ is open in $\mathcal{R}$. 

Following the discussion at the end of section 2.1 we know that each $\underline{\Rl}^{\leftrightarrow}$ is equipped with the discrete topology in which all sub-objects are open (in particular each $\underline{\Rl}^{\leftrightarrow}_V$ has the discrete topology).
Since the $F$ functor preserves monics, if $\underline{Q}\subseteq\underline{\Rl}^{\leftrightarrow}$ is open then $F(\underline{Q})\subseteq F(\underline{\Rl}^{\leftrightarrow})$ is open, where $F(\underline{Q}):=\coprod_{\phi_i\in Hom(\down V, \mv(\mh))}\underline{Q}_{\phi_i(V)}$. 

Therefore we define a sub-sheaf, $\underline{\breve{Q}}$, of 
$\breve{\underline{\Rl}}^{\leftrightarrow}$ to be \emph{open}
if for each $V\in\mv_f(\mh)$ the set $\underline{\breve{Q}}_V\subseteq \breve{\underline{\Rl}}_V$ is open, i.e., each $\underline{Q}_{\phi_i(V)}\subseteq \underline{\Rl}^{\leftrightarrow}_{\phi_i(V)}$ is open in the discrete topology on $\underline{\Rl}^{\leftrightarrow}_{\phi_i(V)}$. It follows that the sheaf $\breve{\underline{\Rl}}^{\leftrightarrow}$ gets induced the discrete topology in which all sub-objects are open. In this setting the `horizontal' topology on the base category $\mv_f(\mh)$ would be accounted for by the sheave maps.

For each $\downarrow V$ we then obtain the open set $p_{\underline{\Rl}}^{-1}(\downarrow V)$ which has value
$\underline{\breve{\Rl}}_{V^{'}}$ at contexts $V^{'}\in\downarrow
V$ and $\emptyset$ everywhere else.

\subsection{Truth Values}
We now want to see what happens to the truth
values when they are mapped via the functor $F$. In particular,
given the sub-object classifier $\uom^{\mv(\mh)}\in Sh(\mv(\mh))$ we want to know
what $F(\uom^{\mv(\mh)})$ is. Since \ba
F(\uom^{\mv(\mh)})=p_J!\circ I (\uom^{\mv(\mh)}) \ea we first of
all need to analyse what $I (\uom^{\mv(\mh)})$ is. Applying the
definition for each $w^{g_i}_V\in \Lambda(\ps{G/G_F})$ we obtain 
\be (I(\uom^{\mv(\mh)}))_{w^{g_i}_V}:=\uom^{\mv(\mh)}_{\phi_i(V)} \ee Where
$\phi_i\in Hom(\down V, \mv(\mh))$ is the unique homeomorphism
associated to the equivalence class $w^{g_i}_V\in G/G_{FV}$. If we
then consider another element $w^{g_j}_V\in G/G_{FV}$, we then
have \be (I
(\uom^{\mv(\mh)}))_{w^{g_j}_V}:=\uom^{\mv(\mh)}_{\phi_j(V)} \ee
where now $\phi_i(V)\neq \phi_j(V))$. What this implies is
that once we apply the functor $p_J!$ to push everything down to
$\mv_f(\mh)$, the distinct elements $\uom^{\mv(\mh)}_{\phi_i(V)}$
and $\uom^{\mv(\mh)}_{\phi_j(V)}$ will be pushed down to the
same $V$, since both $\phi_i, \phi_j\in Hom(\down V, \mv(\mh))$. It
follows that, for every $V\in \mv_f(\mh)$, $F(\uom^{\mv(\mh)})$ is
defined as \be F(\uom^{\mv(\mh)})_V:=\coprod_{w^{g_i}_V\in
G/G_{FV}}\uom^{\mv(\mh)}_{w^{g_i}_V}\simeq\bigcup_{w^{g_i}_V\in
G/G_{FV}}\{w^{g_i}_V\}\times\uom^{\mv(\mh)}_{w^{g_i}_V}\simeq\bigcup_{\phi_i\in
Hom(\down V,\mv(\mh))}\{\phi_i\}\times \uom^{\mv(\mh)}_{\phi_i(V)}\ee

Thus it seems that for each $V\in \mv_f(\mh)$,
$F(\uom^{\mv(\mh)})_V$ assigns the disjoint union of the collection
of sieves for each algebra $V_{i}\in \mv(\mh)$ such that $V_{i}=\phi_i(V)$,
where $\phi_i$ are the unique homeomorphisms associated to each
$w^{g_i}_V\in G/G_{FV}$. This leads to the
following conjecture:
\begin{Conjecture}
$F(\uom^{\mv(\mh)})\simeq\ps{G/G_F}\times \uom^{\mv(\mh)}$
\end{Conjecture}
It should be noted that $\mv_f(\mh)\simeq \mv(\mh)$ since
$\mv_f(\mh)$ and  $\mv(\mh)$ are in fact the same categories only
that in the former there is no group action on it. Thus it also
follows trivially that $\uom^{\mv_f(\mh)}\simeq \uom^{\mv(\mh)}$. Having said that we can now prove the above conjecture
\begin{Proof}
For each $V\in \mv_f(\mh)$ we define the map \ba
i_V: F(\uom^{\mv(\mh)})_V&\rightarrow&G/G_{FV}\times \uom^{\mv(\mh)}_V\\
S&\mapsto&(w^{g_i}_V, l_{g_i^{-1}}S) \ea
where $S\in \uom^{\mv(\mh)}_{w^{g_i}_V}=\uom^{\mv(\mh)}_{\phi_i(V)}$ for $\phi_i\in Hom(\downarrow V, \mv(\mh))$ and $\phi_i(V):=l_{g_i}V$ while $l_{g_i^{-1}}S\in \uom^{\mv(\mh)}_V$.

Such a map is one to one since if $(w^{g_i}_V, l_{g_i^{-1}}S_1)=(w^{g_i}_V, l_{g_i^{-1}}S_2)$ then $l_{g_i^{-1}}S_1=l_{g_i^{-1}}S_2$ and $S_1=S_2$.
The fact that it is onto follows form the definition.

We now construct, for each $V\in \mv(\mh)$ the map \ba
j:G/G_{FV}\times \uom^{\mv(\mh)}_V&\rightarrow&F(\uom^{\mv(\mh)})_V\\
(w^{g_i}_V, S)&\mapsto&l_{g_i}(S) \ea 
where $S\in \uom^{\mv(\mh)}_V$ and $l_{g_i}S\in \uom^{\mv(\mh)}_{l_{g_i}V}$ for $l_{g_i}V=\phi_i(V)$ thus $l_{g_i}S\in \uom^{\mv(\mh)}_{w^{g_i}_V}$

A moment of thought reveals that
$j=i^{-1}$
\end{Proof}
From the above result we obtain the following conjecture:
\begin{Conjecture}\label{con:trueiso}
$\uom^{\mv_f(\mh)}\simeq F(\uom^{\mv(\mh)})/\ug$
\end{Conjecture}
Before proving the above conjecture we, first of all, need to define what a quotient presheaf is. This is simply a presheaf in which the quotient is computed context wise, thus, in the case at hand the quotient is computed for each $V\in\mv_f(\mh)$. In order to understand the definition of the quotient presheaf we will analyse what the equivalence classes look like.

We already know that for presheaves over $\mv_f(\mh)$ the group
action is at the level of the base category $\Lambda(\ps{G/G_F})$.
In particular for each $g\in G$ we have \be
(l^*_g(\uom^{\mv(\mh)}))_{\phi (V)}:=\uom^{\mv(\mh)}_{l_g(\phi (V))}
\ee where $\phi\in Hom(\down V, \mv(\mh))$. Therefore by
defining for each $V\in \mv(\mh)$ the equivalence relation on
$\coprod_{\phi_i\in
Hom(\down V,\mv(\mh))}\uom^{\mv(\mh)}_{\phi_i(V)}=:(F(\uom^{\mv(\mh)}))_V$ by
the action of $G$, the elements in $(F(\uom^{\mv(\mh)}))_V/G_V=\Big(\coprod_{\phi_i\in
Hom(\down V,\mv(\mh))}\uom^{\mv(\mh)}_{\phi_i(V)}\Big)/G$ will be
equivalence classes of sieves, i.e., \be [S_i]:=\{l_g(S_i)|g\in
G\} \ee
for each $S_i\in \uom^{\mv(\mh)}_{\phi_i(V)}\in \coprod_{{\phi_i\in Hom(\down V,\mv(\mh))}}\uom^{\mv(\mh)}_{\phi_i(V)}$. In the above we used the action of the group $G$ on sieves which is defined as $l_gS:=\{l_gV^{'}|V^{'}\in S\}$.
We are now ready to define the presheaf $F(\uom^{\mv(\mh)})/\ug$.

\begin{Definition}
The Presheaf $F(\uom^{\mv(\mh)})/\ug$ is defined:
\begin{itemize}
\item On objects: for each context $V\in \mv_f(\mh)$ we have the object 
\be
(F(\uom^{\mv(\mh)}))_V/G_V:=\Big(\coprod_{\phi_i\in Hom(\down V,\mv(\mh))}\uom^{\mv(\mh)}_{\phi_i(V)}\Big)/(G)\ee
whose elements are equivalence classes of sieves $[S_i]$, i.e., $S_1, S_2\in [S_i]$ iff $S_1:=\{l_gS_2|g\in G\}$ and $S_2\in\uom^{\mv(\mh)}_{\phi_i(V)}$ and $S_1=\uom^{\mv(\mh)}_{l_g\phi_i(V)}$, i.e. each equivalence class will contain only one sieve for each algebra. This definition of equivalence condition follows from the fact that the group action of $G$ moves each set $\uom^{\mv(\mh)}_{\phi_i(V)}$ to another set $\uom^{\mv(\mh)}_{l_g\phi_i(V)}$ in the same stork $F(\uom^{\mv(\mh)}))_V$, i.e. the group action is at the level of the base category $\Lambda(\ps{G/G_F})$.
\item On morphisms: for each $V^{'}\subseteq V$ we then have the corresponding morphisms
\ba
\alpha_{VV^{'}}:\Big(\coprod_{\phi_i\in Hom(\down V,\mv(\mh))}\uom^{\mv(\mh)}_{\phi_i(V)}\Big)/(G)&\rightarrow& \Big(\coprod_{\phi_j\in Hom(\down V^{'},\mv(\mh))}\uom^{\mv(\mh)}_{\phi_j(V^{'})}\Big)/(G)\\
\;[S]&\mapsto&\alpha_{VV^{'}}([S]):=[S\cap V^{'}] \ea 

where
$[S\cap V^{'}]:=\{l_g(S\cap V^{'})|g\in G\}$,  and we choose as the representative for the equivalence class $S\in\uom^{\mv(\mh)}_V$ for $V=\phi_i(V)$ where $\phi_i\in Hom(\downarrow V, \mv(\mh))$ is associated to some $g\in G_V$
\end{itemize}
\end{Definition}
We can now prove the above conjecture (\ref{con:trueiso}), i.e.,
we will show that the functor \be \beta:\uom^{\mv_f(\mh)}\rightarrow
F(\uom^{\mv(\mh)})/(\ug) \ee
is an isomorphism.

In particular for each context $V\in \mv(\mh)$ we define \ba
\beta_V:\uom^{\mv_f(\mh)}_V&\rightarrow& F(\uom^{\mv(\mh)})_V/(\ug)_{V}\\
S&\mapsto&[S] \ea
where $[S]$ denotes the equivalence class to which the sieve $S$ belongs to, i.e., $[S]:=\{l_g S|g\in G\}$.

First we need to show that $\beta$ is indeed a functor,
i.e., we need to show that the following diagram commutes
\[\xymatrix{
\uom^{\mv_f(\mh)}_V\ar[rr]^{\beta_V}\ar[dd]_{\uom^{\mv_f(\mh)}(i_{V^{'}V})}&&F(\uom^{\mv(\mh)})_V/G\ar[dd]^{\alpha_{VV^{'}}}\\
&&\\
\uom^{\mv_f(\mh)}_{V^{'}}\ar[rr]^{\beta_{V^{'}}}&&F(\uom^{\mv(\mh)})_V/G\\
}\] Thus for each $S$ we obtain for one direction
\be\big(\beta_{V^{'}}\circ\uom^{\mv_f(\mh)}(i_{V^{'}V})\big)(S)=\beta_{V^{'}}(S\cap
V^{'})=[S\cap V^{'}] \ee 
where the first equality follows from the definition of the sub-object classifier $\uom^{\mv_f(\mh)}$ \cite{andreas5}.

Going the opposite direction we get
\be \big(\alpha_{VV^{'}}\circ
\beta_V\big)S=\alpha_{VV^{'}}[S]=[S\cap V^{'}] \ee It follows that
indeed the above diagram commutes. Now that we have showed that
$\beta$ is a functor we need to show that it is an isomorphisms. We consider each individual component $\beta_V$, $V\in  \mv_f(\mh)$.
\begin{enumerate}
\item
\emph{The map $\beta_V$ is one-to-one}.

Given $S_1, S_2\in\uom^{\mv_f(\mh)}_{V}$, 
if $\beta_V(S_1)=\beta_V(S_2)$ then $[S_1]=[S_2]$, thus both $S_1$
and $S_2$ belong to the same equivalence class. Each equivalence
class is of the form $[S]=\{l_gS|g\in G\}$, therefore $S_1=l_gS_2$ for some $g\in G$.
However, the definition of the equivalence classes of sieves implied that for each equivalence class there is one and only one sieve for each algebra. Thus if $[S_1]=[S_2]$ and both $S_1, S_2\in\uom^{\mv_f(\mh)}_V$, then $S_1=S_2$.

%
%
%
\item
\emph{The map $\beta_V$ is onto}. This follows at once from the definition.
\item
\emph{The map $\beta_V$ has an inverse}.

We now need to define an inverse. We choose \be
\gamma:F(\uom^{\mv(\mh)})/G\rightarrow \uom^{\mv_f(\mh)} \ee such
that for each context we get \ba
\gamma_V:F(\uom^{\mv(\mh)})_V/G&\rightarrow& \uom^{\mv_f(\mh)}_V\\
\;[S]&\mapsto&[S]\cap V \ea where $[S]\cap V:=\{l_g(S)\cap V|g\in
G\}$ represents the only sieve in the equivalence class which belongs to $\uom^{\mv_f(\mh)}_V$ . We first of all have to show that this is indeed a functor.
Thus we need to show that, for each $V^{'}\subseteq V$ the following diagram commutes
\[\xymatrix{
F(\uom^{\mv(\mh)})_V/G\ar[dd]_{\alpha_{VV^{'}}}\ar[rr]^{\gamma_V}&&\uom^{\mv_f(\mh)}_V\ar[dd]^{\uom^{\mv_f(\mh)}(i_{V^{'}V})}\\
&&\\
F(\uom^{\mv(\mh)})_V/G\ar[rr]^{\gamma_{V^{'}}}&&\uom^{\mv_f(\mh)}{V^{'}}
}\] Chasing the diagram around for each $S$ we obtain \be
\uom^{\mv_f(\mh)}(i_{V^{'}V})\circ
\gamma_V([S])=\uom^{\mv_f(\mh)}(i_{V^{'}V})([S]\cap V)=([S]\cap
V)\cap V^{'}=[S]\cap V^{'} \ee

On the other hand we have \be \gamma_{V^{'}}\circ
\alpha_{VV^{'}}[S]=\gamma_{V^{'}}[S\cap V^{'}]=[S\cap V^{'}]\cap
V^{'}=[S]\cap V^{'} \ee
where the last equality follows since $[S\cap V^{'}]\cap V^{'}:=\{l_g(S\cap V^{'})|g\in G\}\cap V^{'}$ and the only sieve in $[S]$ belonging to $\uom^{\mv_f(\mh)}_{V^{'}}$ is $S\cap V^{'}$. Therefore the map $\gamma$ is a functor. 

It now remains to show that, for each $V\in \mv_f(\mh)$ and each
$S\in\uom^{\mv(\mh)}_V$, $\gamma_V$  is the inverse of $\beta_V$.
Thus 
\be \gamma_V\circ\beta_V(S)=\gamma_V([S])=[S]\cap V=S \ee
where the last equality follows from the fact that in each
equivalence class of sieves there is one and only one referred to
each context $l_gV$. On the other hand we have \be \beta_V\circ
\gamma_V([S])=\beta_V\circ ([S]\cap V)=\beta_V(S)=[S] \ee
\end{enumerate}
The functor $\beta$ is indeed an isomorphism.

\section{No More Twisted Presheaves}
In this section we will briefly analyse the problem of twisted
presheaves present in the old formalism which utilised the topos
$\Sets^{\mv(\mh)^{\op}}$. We will then show how, by changing the topos to
$Sh(\mv(\mh_f))$, such problem is overcome.

In previous sections we defined the action of the group $G$
on the base category $\mv(\mh)$ as $l_g(V):=\hat{U}_gV\hat{U}_g^{-1}:=\{\hat{U}_g\hat{A}\hat{U}_g^{-1}|\hat{A}\in
V\}$, $g\in G$. When considering the topos $\Sets^{\mv(\mh)^{\op}}$, each $g$ we obtain the
functor $l_{\hat{U}_g}:\mv(\mh)\rightarrow \mv(\mh)$ with 
induces a geometric morphisms \be
l_{\hat{U}_g}:\Sets^{\mv(\mh)^{\op}}\rightarrow \Sets^{\mv(\mh)^{\op}} \ee whose
inverse image part is \ba
l^*_{\hat{U}_g}:\Sets^{\mv(\mh)^{\op}}&\rightarrow& \Sets^{\mv(\mh)^{\op}}\\
\underline{F}&\mapsto&l^*_{\hat{U}_g}(\underline{F}):=\underline{F}\circ
l_{\hat{U}_g} \ea In \cite{andreas} it was shown how the above
geometric morphism acted on the spectral presheaf
$\us^{\mv(\mh)}$, the quantity value object
$\underline{\Rl}^{\leftrightarrow}$, truth values and
daseinisation. Let us analyse each of such actions in detail.
\subsection{Group Action on the Old Presheaves}
In this section we will describe how the old presheaves where defined in the topos $\Sets^{\mv(\mh)}$. We will then show how the group action gave rise to the twisted presheaves.
\subsubsection{Spectral Presheaf}
Given the speactral presheaf $\us\in \Sets^{\mv(\mh)^{\op}}$ (\cite{andreas}), the action of each element of the group is
given by the following theorem:
\begin{Theorem}
For each $\hat{U}\in\mathcal{U}(\mh)$, there is a natural
isomorphism $\iota^{\hat{U}}:\us\rightarrow\us^{\hat{U}}$ which is
defined through the following diagram:
\[\xymatrix{
\us_V\ar[rr]^{\iota^{\hat{U}}_V}\ar[dd]_{\us_V(i_{V^{'}V})}&&\us^{\hat{U}}_V\ar[dd]^{\us^{\hat{U}}_V(i_{V^{'}V})}\\
&&\\
\us_{V^{'}}\ar[rr]_{\iota^{\hat{U}}_{V^{'}}}&&\us^{\hat{U}}_{V^{'}}\\
}\] where, at each stage $V$ \be
(\iota^{\hat{U}}_V(\lambda))(\hat{A}):=\langle \lambda,
\hat{U}\hat{A}\hat{U}^{-1}\rangle \ee for all $\lambda\in \us_V$
and $\hat{A}\in V_{sa}$.
\end{Theorem}
The presheaf $\us^{\hat{U}}$ is the twisted presheaf associated to
the unitary operator $\hat{U}$. Such a presheaf is defined as
follows:
\begin{Definition}
The twisted presheaf $\us^{\hat{U}}$ has as:
\begin{itemize}
\item[--] Objects: for each $V\in \mv(\mh)$ it assigns the Gel'fand spectrum of the algebra $\hat{U}V\hat{U}^{-1}$, i.e., $\us^{\hat{U}}_V:=\{\lambda:\hat{U}V\hat{U}^{-1}\rightarrow\Cl|\lambda(\hat{1})=1\}$.
\item[--] Morphisms: for each $i_{V^{'}V}:V^{'}\rightarrow V$ ($V^{'}\subseteq V$) it assigns the presheaf maps
\ba
\us^{\hat{U}}(i_{V^{'}V}):\us^{\hat{U}}_V&\rightarrow&\us^{\hat{U}}_{V^{'}}\\
\lambda&\mapsto&\lambda_{|\hat{U}V^{'}\hat{U}^{-1}} \ea
\end{itemize}
\end{Definition}
\subsubsection{Quantity Value Object}
Similarly, for the quantity value object we obtain the following
theorem:
\begin{Theorem}
For each $\hat{U}\in \mathcal{U}(\mh)$, there exists a natural
isomorphism
$k^{\hat{U}}:\underline{\Rl}^{\leftrightarrow}\rightarrow(\underline{\Rl}^{\leftrightarrow})^{\hat{U}}$,
such that for each $V\in \mv(\mh)$ we obtain the individual
components
$k^{\hat{U}}:\underline{\Rl}^{\leftrightarrow}_V\rightarrow(\underline{\Rl}^{\leftrightarrow})^{\hat{U}}_V$
defined as \be\label{grouponR} k^{\hat{U}}_V(\mu,
\nu)(l^{\hat{U}}(V^{'})):=(\mu(V^{'}),\nu(V^{'})) \ee for all
$V^{'}\subseteq V$
\end{Theorem}
Here, $\mu\in \mathcal{R}^{\leftrightarrow}_V$ is an order
preserving function
$\mu :\downarrow V\rightarrow \Rl$ such that, if $V_2\subseteq V_1\subseteq V$, then $\mu(V_2)\geq\mu(V_1)\geq\mu(V)$, while $\nu$ is an order reversing function $\nu :\downarrow V\rightarrow \Rl$ such that, if $V_2\subseteq V_1\subseteq V$, then $\nu(V_2)\leq\nu(V_1)\leq\nu(V)$.

\noindent
In the equation \ref{grouponR} we have used the bijection between
the sets $\downarrow l^{\hat{U}}(V )$ and $\downarrow V$ .
\subsubsection{Daseinisation}
We recall the concept of daseinisation: 
given a projection operator $\hat{P}$ its daseinisation with
respect to each context $V$ is \be
\delta^o(\hat{P})_V:=\bigwedge\{\hat{Q}\in\mathcal{P}(V)|\hat{Q}\geq\hat{P}\}
\ee 
where $P(V)$ represents the collection of projection operators in $V$.

If we then act upon it by any $\hat{U}$ we obtain \ba
\hat{U}\delta^o(\hat{P})_V\hat{U}^{-1}&:=&\hat{U}\bigwedge\{\hat{Q}\in\mathcal{P}(V)|\hat{Q}\geq\hat{P}\}\hat{U}^{-1}\\
&=&\bigwedge\{\hat{U}\hat{Q}\hat{U}^{-1}\in\mathcal{P}(l_{\hat{U}}(V))|\hat{Q}\geq\hat{P}\}\\
&=&\bigwedge\{\hat{U}\hat{Q}\hat{U}^{-1}\in\mathcal{P}(l_{\hat{U}}(V))|\hat{U}\hat{Q}\hat{U}^{-1}\geq\hat{U}\hat{P}\hat{U}^{-1}\}\\
&=&\delta^o(\hat{U}\hat{P}\hat{U}^{-1})_{l_{\hat{U}}(V)} \ea
where the second and third equation hold since the map $\hat{Q}\rightarrow \hat{U}\hat{Q}\hat{U}^{-1}$ is weakly continuous.

What this implies is that the clopen sub-objects which represent
propositions, i.e., $\underline{\delta(\hat{P})}$, get mapped to
one another by the action of the group.
\subsubsection{Truth Values}
Now that we have defined the group action on daseinisation we can
define the group action on the truth values. We recall that for
pure states the truth object at each stage $V$ is defined as
\ba\label{ali:truthob}
\underline{\mathbb{T}}^{|\psi\rangle}_V&:=&\{\hat{\alpha}\in\mathcal{P}(V)|Prob(\hat{\alpha};|\psi\rangle)=1\}\\
&=&\{\hat{\alpha}\in\mathcal{P}(V)|\langle\psi|\hat{\alpha}|\psi\rangle=1\}
\ea For each context $V\in \mv(\mh)$ the truth value is \ba
v(\underline{\delta(\hat{P})}\in \underline{\mathbb{T}}^{|\psi\rangle})_V&:=&\{V^{'}\subseteq V|\delta^o(\hat{ P})_{V^{'}}\in\mathbb{T}^{|\psi\rangle}_{V^{'}}\}\\
&=&\{V^{'}\subseteq V|\langle\psi|\delta^o(\hat{
P})_{V^{'}}|\psi\rangle=1\} \ea we now act upon it with a group
element $\hat{U}$ obtaining \ba
l_{\hat{U}}\Big(v(\delta^o(\hat{P})\in
\underline{\mathbb{T}}^{|\psi\rangle})_V\Big)
&:=&l_{\hat{U}} \{V^{'}\subseteq V|\langle\psi|\delta^o(\hat{ P})_{V^{'}}|\psi\rangle=1\}\\
&=&\{l_{\hat{U}} V^{'}\subseteq l_{\hat{U}} V|\langle\psi|\delta^o(\hat{ P})_{V^{'}}|\psi\rangle=1\}\\
&=&\{l_{\hat{U}} V^{'}\subseteq l_{\hat{U}} V|\langle\psi|\hat{U}^{-1}\hat{U}\delta^o(\hat{ P})_{V^{'}}\hat{U}^{-1}\hat{U}|\psi\rangle=1\}\\
&=&\{l_{\hat{U}} V^{'}\subseteq l_{\hat{U}} V|\langle\psi|\hat{U}^{-1}\delta^o(\hat{U}\hat{P}\hat{U}^{-1})_{l_{\hat{U}}(V)}\hat{U}|\psi\rangle=1\}\\
&=&v(\delta^o(\hat{U}\hat{P}\hat{U}^{-1})\in
\underline{\mathbb{T}}^{\hat{U}|\psi\rangle})_{l_{\hat{U}}(V)} \ea
We thus obtain the following equality: \be
l_{\hat{U}}\Big(v(\delta^o(\hat{P})\in
\underline{\mathbb{T}}^{|\psi\rangle})_V\Big)=v(\delta^o(\hat{U}\hat{P}\hat{U}^{-1})\in
\underline{\mathbb{T}}^{\hat{U}|\psi\rangle})_{l_{\hat{U}}(V)} \ee
Thus truth values are invariant under the group transformations. This is the topos analogue of Dirac covariance, i.e., given a state $|\psi\rangle$ and a physical quantity $\hat{A}$, we would obtain the same predictions if we replaced the state by $\hat{U}|\psi\rangle$ and the quantity by $\hat{U}\hat{A}\hat{U}^{-1} $

A similar result holds if we consider mixed states
$\rho=\sum_{i=1}^Nr_i|\psi_i\rangle\langle\psi_i|$. However, in
this case, as explained in \cite{probabilities}, the topos to
utilise is $Sh(\mv(\mh)\times(0,1)_L)$ rather than $Sh(\mv(\mh))$.
In order to relate these two topoi one utilises the projection map $pr_1: \mv(\mh)\times(0,1)_L\times\rightarrow \mv(\mh)$, which induces the inverse image geometric morphism $p^*_1:Sh(\mv(\mh))\rightarrow Sh(\mv(\mh)\times(0,1)_L)$. In this way any object defined in the old formalism can be mapped to an object in the new topos. 

In this setting, for each context $(V, r)\in \mv(\mh)\times
(0,1)_L$ the truth object is \ba \underline{\mathbb{T}}_{V,
r}^{\rho}:=\{\underline{S}\in Sub_{cl}(\us_{\downarrow V})|\forall
V^{'}\subseteq V, tr(\rho\hat{P}_{\underline{S}_{V^{'}}})\geq r\}
\ea While the truth values of a proposition
$p^*_2(\underline{\delta(\hat{P})})$ is \be
v(p^*_2(\underline{\delta(\hat{P})})\in
\underline{\mathbb{T}}^{\rho})_{(V, r)}:=\{\langle V^{'},
r^{'}\rangle\leq\langle V,
r\rangle|\mu^{\rho}(\underline{\hat{P}})_{V^{'}}\geq r^{'}\} \ee If
we then perform a group transformation on it we obtain \ba
l_{\hat{U}}\Big(v(p^*_2(\underline{\delta(\hat{P})})\in \underline{\mathbb{T}}^{\rho})_{(V, r)}\Big)&:=&l_{\hat{U}}\{\langle V^{'}, r^{'}\rangle\leq\langle V, r\rangle|\mu^{\rho}(\underline{\hat{P}})_{V^{'}}\geq r^{'}\}\\
&=&\{\langle l_{\hat{U}}V^{'}, r^{'}\rangle\leq\langle l_{\hat{U}}V, r\rangle|\mu^{\rho}(\underline{\delta(\hat{P})})_{V^{'}}\geq r^{'}\}\\
&=&\{\langle l_{\hat{U}}V^{'}, r^{'}\rangle\leq\langle l_{\hat{U}}V, r\rangle|\mu^{\rho}(\delta^o(\hat{P}))_{V^{'}}\geq r^{'}\}\\
&=&\{\langle l_{\hat{U}}V^{'}, r^{'}\rangle\leq\langle l_{\hat{U}}V, r\rangle|\mu^{\hat{U}\rho\hat{U}^{-1}}(\delta^o(\hat{U}\hat{P}\hat{U}^{-1}))_{l_{\hat{U}}(V)}\geq r^{'}\}\\
&=&v(p^*_2(\underline{\delta(\hat{U}\hat{P}\hat{U}^{-1})})\in
\underline{\mathbb{T}}^{\hat{U}\rho\hat{U}^{-1}})_{(l_{\hat{U}}V,
r)} \ea Obtaining the important result \be
l_{\hat{U}}\Big(v(p^*_2(\underline{\delta(\hat{P})})\in
\underline{\mathbb{T}}^{\rho})_{(V,
r)}\Big)=v(p^*_2(\underline{\delta(\hat{U}\hat{P}\hat{U}^{-1})})\in
\underline{\mathbb{T}}^{\hat{U}\rho\hat{U}^{-1}})_{(l_{\hat{U}}(V),
r)} \ee
\subsection{Group Action on the New Sheaves}
We would now like to analyse what the group action on the new
sheaves is. In particular we will show how the action of the group
$\ug$ on the sheaves define on $\mv_f(\mh)$ via the $F$ functor
will not induce twisted sheaves.
\subsubsection{Spectral Sheaf}
The action of the group $\ug$ on the new spectral sheaf
$\breve{\us}:=F(\us)$ is given by the following map: \be \ug\times
\breve{\us}\rightarrow\breve{\us} \ee defined for each context
$V\in\mv_f(\mh)$ as \ba\label{ali:gonvalue}
\ug_V\times \breve{\us}_V&\rightarrow&\breve{\us}_V\\
(g, \lambda)&\mapsto&l_g\lambda \ea where
$\breve{\us}_V:=\coprod_{\phi_i\in Hom(\down V,
\mv(\mh))}\us_{\phi_i(V)}$ such that if $\lambda\in \us_{\phi_i(V)}$
we define $l_g\lambda\in l_g\us_{\phi_i(V)}:=\us_{l_g(\phi_i(V))}$ by
\be (l_g(\lambda))\hat{A}:=\langle
\lambda,\hat{U}(g)^{-1}\hat{A}\hat{U}(g)\rangle \ee
for all $g\in G$, $\hat{A}\in V_{sa}$(self adjoint operators in $V$) and $V\in \mv(\mh)$.

However from the definition of $\breve{\us}$, both $ \us_{\phi_i(V)}$ and $\us_{l_g(\phi_i(V))}$ belong to the same stalk, i.e., belong to $\breve{\us}_V$.

\noindent
We thus obtain a well defined group action which does not induce twisted presheaves. 

We would now like to check whether such a group action is
continuous with respect to the spectral topology, i.e., if the map
\be \rho:\ug\times \breve{\us}\rightarrow\breve{\us} \ee is
continuous. In particular we want to check if for each
$V\in\mv_f(\mh)$ the local component \be \rho_V:\ug_V\times
\breve{\us}_V\rightarrow\breve{\us}_V \ee is continuous, i.e., if
$\rho^{-1}_V\breve{\underline{S}}_V=\rho^{-1}_V\Big(\coprod_{\phi_i\in
Hom(\down V, \mv(\mh))}\underline{S}_{\phi_i(V)}\Big)$ is open for
$\breve{\underline{S}}_V$ open. \ba
\rho^{-1}_V\Big(\coprod_{\phi_i\in Hom(\down V, \mv(\mh))}\underline{S}_{\phi_i(V)}\Big)&=&\{(g_j, \underline{S}_{\phi_i(V)})|l_{g_j}(\underline{S}_{\phi_i(V)})\in \underline{\breve{S}}_V\}\\
&=& (G, \underline{\breve{S}}_V) \ea where
$l_{g_j}(\underline{S}_{\phi_i(V)}):=\underline{S}_{l_{g_j}\phi_i(V)}=\underline{S}_{l_{g_j}(\phi_i(V))}$.
It follows that the action is continuous.

Moreover it seems that the sub-objects $\breve{\underline{S}}$ actually remain invariant under the group action. In fact, for each $V\in \mv_f(\mh)$, $\breve{\underline{S}}_V=\coprod_{\phi_i\in Hom(\down V, \mv(\mh))}\underline{S}_{\phi_i(V)}$ where the set $Hom(\down V, \mv(\mh))$ contains all $G$ related homeomorphisms, i.e. all $l_{g_j}(\phi_i)\;\forall \; g_j\in G$, ($l_{g_j}(\phi)(V):=l_{g_j}(\phi(V)))$.

\noindent
It follows that the sub-objects $\breve{\underline{S}}\subseteq\breve{\us}$ are invariant under the group action.

This is an important result when considering propositions which
are identified with clopen sub-objects coming from
daseinisation. In this context the group action is defined, for each
$V\in\mv_f(\mh)$, as: \ba
\underline{G}_V\times \underline{\delta\breve{P}}_V&\rightarrow& \underline{\delta\breve{P}}_V\\
\underline{G}_V\times \coprod_{\phi_i\in Hom(\down V, \mv(\mh))}\delta^o(\hat{P})_{\phi_i(V)}&\rightarrow&\coprod_{\phi_i\in Hom(\down V, \mv(\mh))}\delta^o(\hat{P})_{\phi_i(V)}\\
(g,
\delta^o(\hat{P})_{\phi_i(V)})&\mapsto&\delta^o(\hat{U}_g\hat{P}\hat{U}_g^{-1})_{l_g(\phi_i(V))}
\ea Thus for each $g\in G$ we get a collection of transformations each similar to those obtained in the original formalism. However, since the effect
of such a transformation is to move the objects around within a
stalk, when considering the action of the entire $G$, the
stalk, as an entire set,  remains invariant, i.e., the collection
of local component of the propositions stays the same.

Moreover the fact that individual sub-objects $\breve{\underline{S}}\subseteq\breve{\us}$ are invariant under the group action, implies that the action $\ps{G}\times\breve{\us}\rightarrow \breve{\us}$ is not transitive. In fact the transitivity of the action of a group sheaf is defined as follows
\begin{Definition}
Given a group $\ps{G}$, we say that the action of $\ps{G}$ on any other sheaf $\underline{A}$ is transitive iff there are no invariant sub-objects of $A$.
\end{Definition}

Thus although the group actions moves the elements around in each stalk, it nev
\subsubsection{Sub-object Classifier}
We now are interested in defining the group action on the
sub-object classifier $\uom^{\mv_f(\mh)}$. However, by definition,
there is no action on such object. The only action which could be
defined would be the action on $\breve{\uom}:=F(\uom^{\mv(\mh)})$.
In this case, for each $V\in \mv_f(\mh)$, we have \ba
\alpha_V:\ug_V\times \breve{\uom}_V&\rightarrow&\breve{\uom}_V\\
\ug_V\times \coprod_{w^{g_i}_V\in G/G_{FV}}\uom_{w^{g_i}_V}&\rightarrow&\coprod_{w^{g_i}_V\in G/G_{FV}}\uom_{w^{g_i}_V}\nonumber\\
\ug_V\times \coprod_{\phi_i\in Hom(\down V,\mv(\mh))}\uom_{\phi_i(V)}&\rightarrow&\coprod_{\phi_i\in Hom(\down V,\mv(\mh))}\uom_{\phi_i(V)}\nonumber\\
(g, S)&\mapsto&l_g(S) \ea
where $l_g(S):=\{l_gV|V\in S\}$. 

If $S\in \uom_{\phi_i(V)}\in \coprod_{\phi_i\in Hom(\down V,\mv(\mh))}\uom_{\phi_i(V)}$, then $l_g(S)$ is a sieve on $l_g\phi_i(V)$, i.e., $l_g(S)\in
\uom_{l_g\phi_i(V)}\in \coprod_{\phi_i\in Hom(\down V,\mv(\mh))}\uom_{\phi_i(V)}$. 

\noindent 
It follows that the action of the group $\ug$ is to move sieves around in each stalk but never to move sieves to different stalks.

The next question is to define a topology on $\breve{\uom}$ and
check whether the action is continuous or not.

\noindent 
A possible topology would be the topology whose basis are the collection of open sub-sheaves of $\breve{\uom}$. If we assume that each $\uom_{\phi(V)}$ has the discrete topology, coming from the fact that it can be seen as an etal\'e bundle, then the topology on $\breve{\uom}$ will be the topology in which each sub-sheaf is open, i.e., the discrete topology. 

Given such a topology we would like to check if the group action
is continuous. To this end we need to show that
$\alpha_V^{-1}(\underline{\breve{S}}_V)$ is open for $\breve{S}_V$
open sub-object. We recall that $\underline{\breve{S}}_V=\coprod_{\phi_i\in
Hom(\down V, \mv(\mh)}\underline{S}_{\phi_i(V)}$. We then obtain \ba
\alpha_V^{-1}(\underline{\breve{S}}_V)&=&\{(g, S)|l_g(S)\in \underline{\breve{S}}_V\}\\
&=&(\underline{G}_V, \underline{\breve{S}}_V) \ea which is open.
\subsubsection{Quantity Value Object}
We would now like to analyse how the group acts on the new
quantity value object $\breve{\underline{\Rl}}^{\leftrightarrow}$.
This is defined via the map \ba \ug\times
\breve{\underline{\Rl}}^{\leftrightarrow}\rightarrow
\breve{\underline{\Rl}}^{\leftrightarrow} \ea which, for each
$V\in\mv_f(\mh)$, has local components \ba \ug_V\times
\breve{\underline{\Rl}}^{\leftrightarrow}_V&\rightarrow&
\breve{\underline{\Rl}}^{\leftrightarrow}_V\\\nonumber \ug_V\times
\coprod_{\phi_i\in
Hom(\down V,\mv(\mh))}\underline{\Rl}^{\leftrightarrow}_{\phi_i(V)}&\rightarrow&\coprod_{\phi_i\in
Hom(\down V,\mv(\mh))}\underline{\Rl}^{\leftrightarrow}_{\phi_i(V)}\\\nonumber
\Big(g, (\mu,\nu)\Big)&\mapsto&(l_g\mu, l_g\nu) \ea where
$(\mu,\nu)\in \underline{\Rl}^{\leftrightarrow}_{\phi_i(V)}$,
while $(l_g\mu,
l_g\nu)\in\underline{\Rl}^{\leftrightarrow}_{l_g(\phi_i(V))}$.
Therefore $l_g\mu:\downarrow l_g(\phi_i(V))\rightarrow\Rl$ and $l_g\nu:\downarrow l_g(\phi_i(\nu))\rightarrow \Rl$.

As it can be easily deduced, even in this case the action of the $\ug$ group is to map elements around in the same stalk but never to map elements between different stalks. Thus yet again we do not obtain twisted sheaves.

We would now like to check whether the group action is continuous
with respect to the discrete topology on $\breve{\underline{\Rl}}$ defined in section 12.2.1.
Thus  we have to check whether for $V\in \mv_f(\mh)$ the following
map is continuous \ba
\Phi_V:\underline{G}_V\times \breve{\underline{\Rl}}_V&\rightarrow &\breve{\underline{\Rl}}_V\\
(g, (\mu,\nu))&\rightarrow&(l_g\mu,l_g\nu)
\ea
A typical open set in $\breve{\underline{\Rl}}_V$ is of the form $\underline{\breve{Q}}_{V}:=\coprod_{\phi_i\in Hom(\down V, \mv(\mh)}\underline{Q}_{\phi_i(V)}$ where each $\underline{Q}_{\phi_i(V)}\subseteq \underline{\Rl^{\leftrightarrow}}_{\phi_i(V)}$ is open.
Therefore \ba
\Phi^{-1}_V(\underline{\breve{Q}}_{V})&=&\{g_i, (\mu, \nu)|(l_{g_i}\mu, l_{g_i}\nu)\in \underline{\breve{Q}}_{V}\}\\
&=&(G, \underline{\breve{Q}}_{V}) \ea 
Therefore the group action with respect to the discrete topology is continuous. 
\subsubsection{Truth Object}
The new truth value object for pure states obtained through the
action of the $F$ functor is \be
\underline{\breve{\mathbb{T}}}^{|\psi\rangle}:=F(\underline{\mathbb{T}}^{|\psi\rangle})
\ee  which is defined as follows:
\begin{Definition}
The truth object $F(\underline{\mathbb{T}}^{|\psi\rangle})$ is the
presheaf defined on
\begin{itemize}
\item [--] Objects: for each $V\in\mv_f(\mh)$  we get
\be F(\underline{\mathbb{T}}^{|\psi\rangle}):=\coprod_{\phi_i\in
Hom(\down V, \mv(\mh))}\underline{\mathbb{T}}^{|\psi\rangle}_{\phi_i(V)}
\ee where
$\underline{\mathbb{T}}^{|\psi\rangle}_{\phi_i(V)}:=\{\hat{\alpha}\in
P(\phi_i(V))|\langle\psi|\hat{\alpha}|\psi\rangle=1\}$ and $P(\phi_i(V))$ denotes the collection of all projection operators in $\phi_i(V)$.
\item[--] Morphisms: given $V^{'}\subseteq V$ the corresponding map is
\be
\underline{\breve{\mathbb{T}}}^{|\psi\rangle}(i_{V^{'}V}):\coprod_{\phi_i\in
Hom(\down V,
\mv(\mh))}\underline{\mathbb{T}}^{|\psi\rangle}_{\phi_i(V)}\rightarrow
\coprod_{\phi_j\in Hom(\down V^{'},
\mv(\mh))}\underline{\mathbb{T}}^{|\psi\rangle}_{\phi_j(V^{'})}
\ee such that, given $\underline{S}\in
\underline{\mathbb{T}}^{|\psi\rangle}_{\phi_i(V)}$, then \be
\underline{\breve{\mathbb{T}}}^{|\psi\rangle}(i_{V^{'}V})\underline{S}:=\underline{\mathbb{T}}^{|\psi\rangle}(i_{\phi_i(V),\phi_j(V^{'})})\underline{S}=\underline{S}_{|\phi_j(V^{'})}
\ee where $\phi_j\leq \phi_i$ thus $\phi_j(V^{'})\subseteq
\phi_i(V)$ and $\phi_j(V^{'})={\phi_i}_{|V^{'}}(V^{'})$.
\end{itemize}
\end{Definition}
In order to define the truth object for density matrices we need to change the topos as was done in \cite{probabilities}, but this time replacing the category $\mv(\mh)$ with $\mv_f(\mh)$. In particular we need to go to the topos $Sh(\mv_f(\mh)\times (0,1)_L)$. This is done by first defining the map $pr_1:\mv_f(\mh)\times (0,1)_L\rightarrow \mv_f(\mh)$, which gives rise to the geometric morphisms whose inverse image part is $pr_1^*:Sh(\mv_f(\mh))\rightarrow Sh(\mv_f(\mh)\times (0,1)_L)$.

We then compose the two functors \be
Sh(\mv(\mh))\xrightarrow{F}Sh(\mv_f(\mh))\xrightarrow{pr_1^*}Sh(\mv_f(\mh)\times
(0,1)_L) \ee It is such a functor which is used to map our
original truth object (for each density matrix $\rho$ )
$\underline{\mathbb{T}}^{\rho}\in Sh(\mv(\mh))$ to our new truth
object $\breve{\underline{\mathbb{T}}}^{\rho}\in
Sh(\mv_f(\mh)\times (0,1)_L)$: \be
\breve{\underline{\mathbb{T}}}^{\rho}:=pr_1^*\circ
F(\underline{\mathbb{T}}^{\rho}) \ee The definition is as follows:
\begin{Definition}
The truth object presheaf $\breve{\underline{\mathbb{T}}}^{\rho}$
is defined on
\begin{itemize}
\item[--] Objects: for each pair $(V, r)\in \mv_f(\mh)\times (0,1)_L$ we obtain
\be
\breve{\underline{\mathbb{T}}}^{\rho}_{(V,r)}:=\coprod_{\phi_i\in
Hom(\down V,\mv(\mh))}\underline{\mathbb{T}}^{\rho_{\phi_i}}_{(\phi_i(V),
r)} \ee where $\rho_{\phi_i}:\phi_i(V)\rightarrow \Cl$ represents
a state on the algebra $\phi_i(V)$ and each \be
\underline{\mathbb{T}}^{\rho_{\phi_i}}_{(\phi_i(V),
r)}=\{\underline{S}\in Sub(\us_{\downarrow \phi_i(V)})|\forall
V_k\subseteq \phi_i(V),
tr(\rho_{\phi_i}\hat{P}_{\underline{S}_{V_k}})\geq r\} \ee
\item[--] Morphisms: given a map $i:(V^{'}, r^{'})\leq (V, r)$ (iff $V^{'}\subseteq V$ and $r^{'}\leq r$), then the corresponding map is
\be \breve{\underline{\mathbb{T}}}^{\rho}(i):\coprod_{\phi_i\in
Hom(\down V,\mv(\mh))}\underline{\mathbb{T}}^{\rho_{\phi_i}}_{(\phi_i(V),
r)}\rightarrow \coprod_{\phi_j\in
Hom(\down V^{'},\mv(\mh))}\underline{\mathbb{T}}^{\rho_{\phi_j}}_{(\phi_j(V^{'}),
r^{'})} \ee such that given a sub-object $\underline{S}\in
\underline{\mathbb{T}}^{\rho_{\phi_i}}_{(\phi_i(V), r)} $ we get
\be
\breve{\underline{\mathbb{T}}}^{\rho}(i)\underline{S}:=\underline{\mathbb{T}}^{\rho_{\phi_i}}(i_{\phi_i(V),\phi_j(V^{'})})\underline{S}=\underline{S}_{|\phi_j(V^{'})}
\ee where $\phi_j\leq \phi_i$ thus $\phi_j(V^{'})\subseteq
\phi_i(V)$ and $\phi_j(V^{'})=(\phi_i)_{|V^{'}}(V^{'})$. Obviously now
the condition on the restricted sub-object is
$tr(\rho\hat{P}_{S_{V^{''}}})\geq r^{'}$ where $V^{''}\subseteq
\phi_j(V^{'})$. However such a condition is trivially satisfied since
$r^{'}\leq r$.
\end{itemize}
\end{Definition}
We would now like to define the group action on such an object.
Thus we define the following \be\label{equ:actiont}
\ug\times\breve{\underline{\mathbb{T}}}^{\rho}\rightarrow
\breve{\underline{\mathbb{T}}}^{\rho} \ee such that for each
context $V\in\mv_f(\mh)$ we obtain \ba
\ug_V\times\breve{\underline{\mathbb{T}}}^{\rho}_V&\rightarrow& \breve{\underline{\mathbb{T}}}^{\rho}_V\\
(g, \underline{S})&\mapsto&l_g(\underline{S}) \ea
where $\underline{S}\in\underline{\mathbb{T}}^{\rho_{\phi_i}}_{\phi_i(V)}\in \breve{\underline{\mathbb{T}}}^{\rho}_V$, while $l_g(\underline{S})\in\underline{\mathbb{T}}^{\rho_{l_g(\phi_i)}}_{l_g(\phi_i(V))}\in \breve{\underline{\mathbb{T}}}^{\rho}_V$. Here $\underline{\mathbb{T}}^{\rho_{l_g(\phi_i)}}_{l_g(\phi_i(V))}=l_g\underline{\mathbb{T}}^{\rho_{\phi_i}}_{\phi_i(V)}$.\\
Therefore also for the truth object the action of the group is to map elements around in a given stalk, but never to map elements in between stalks.


\section{Group Action on Physical Quantities}
We would now like to analyse how a possible group action can be
defined on physical quantities. To this end we should recall how
physical quantities are represented in the topos $\mv(\mh)$ and
then understand how they get mapped to elements in
$Sh(\mv_f(\mh))$, via the $F$ functor.
\subsection{Old Topos Representation of Physical Quantities}
In classical theory, given a state space $S$ a physical quantity
is represented by a function $A:\Sigma\rightarrow \mathcal{R}$.
The analogue of such a representation in the context of the topos
$\Sets^{\mv(\mh)^{op}}$ is via a functor $
\breve{\delta}(\hat{A}):\us\rightarrow \ps{\Rl^{\leftrightarrow}}
$ which, at each context $V$, is defined as \ba
\breve{\delta}(\hat{A})_V:\us_V&\rightarrow&
\ps{\Rl^{\leftrightarrow}}_V\\\nonumber
(\breve{\delta}^i(\hat{A})_V(\cdot),
\breve{\delta}^o(\hat{A})_V(\cdot)):\us_V&\rightarrow&
\ps{\Rl^{\leftrightarrow}}_V\\\nonumber
\lambda&\mapsto&(\mu_{\lambda},\nu_{\lambda}) \ea
where $\breve{\delta}^i(\hat{A})_V(\lambda):=\overline{\breve{\delta}^i(\hat{A})_V}(\lambda)=\lambda(\delta^i(\hat{A})_V)$ and $\breve{\delta}^o(\hat{A})_V(\lambda):=\overline{\breve{\delta}^o(\hat{A})_V}(\lambda)=\lambda(\delta^o(\hat{A})_V)$. Here $\overline{\breve{\delta}^i(\hat{A})_V}$ and $\overline{\breve{\delta}^o(\hat{A})_V}$ represent the Gel'fand transforms associated with the operators $\delta^i(\hat{A})_V$ and $\delta^o(\hat{A})_V$, respectively (not to be confused with how we denoted sheaves in pervious sections).

In order to understand the precise way in which the functor
$\breve{\delta}(\hat{A})$ is defined, we need to introduce the
notion of \emph{spectral order}. The reason why this order was
chosen, rather than the standard operator ordering\footnote{Recall
that the standard operator ordering, is given as follows:
$\hat{A}\leq\hat{B}$ iff
$\langle\psi|\hat{A}|\psi\rangle\leq\langle\psi|\hat{B}|\psi\rangle$
for all $|\psi\rangle$.} is because the former preserves the
relation between the spectra of the operator, i.e., if
$\hat{A}\leq_s\hat{B}$, then $sp(\hat{A})\subseteq sp(\hat{B})$.
This feature will be very important when defining the values for
physical quantities.

We will now define what the spectral order is. Consider two
self-adjoint operators $\hat{A}$ and $\hat{B}$ with spectral
families $(\hat{E}^{\hat{A}}_r)_{r\in\Rl}$ and
$(\hat{E}^{\hat{B}}_r)_{r\in\Rl}$, respectively. Then the spectral
order is defined as follows: \be \hat{A}\leq_s\hat{B}\;\;\text{
iff }\;\;\forall
r\in\Rl\;\;\;\hat{E}^{\hat{A}}_r\geq\hat{E}^{\hat{B}}_r \ee From
the definition it follows that the spectral order implies the
usual order between operators, i.e. if  $\hat{A}\leq_s\hat{B}$
then $\hat{A}\leq\hat{B}$, but the converse is not true.

We are now ready to define the  functor $\breve{\delta}(\hat{A})$.
To this end, let us consider the self-adjoint operator $\hat{A}$
and a context $V$, such that $\hat{A}\notin V_{sa}$ ($V_{sa}$
denotes the collection of self-adjoint operators in $V$). We then
need to approximate $\hat{A}$ so as to be in $V$. However, since
we eventually want to define an interval of possible values of
$\hat{A}$ at $V$ we will approximate $\hat{A}$, both from above
and from below. In particular, we consider the pair of operators
\ba\label{ali:order} \delta^o(\hat{A})_V:=\bigwedge\{\hat{B}\in
V_{sa}|\hat{A}\leq_s\hat{B}\}\;;\;\;\;
\delta^i(\hat{A})_V:=\bigvee\{\hat{B}\in
V_{sa}|\hat{A}\geq_s\hat{B}\} \ea In the above equation
$\delta^o(\hat{A})_V$ represents the smallest self-adjoint
operator in $V$, which is spectrally larger or equal to $\hat{A}$,
while $\delta^i(\hat{A})_V$ represents the biggest self-adjoint
operator in $V_{sa}$, that is spectrally smaller or equal to
$\hat{A}$. The process represented by $\delta^i$ is called
\emph{inner daseinisation}, while $\delta^o$ represents the
already encountered \emph{outer daseinisation}.

From the definition of $\delta^i(\hat{A})_V$ it follows that if
$V^{'}\subseteq V$ then
$\delta^i(\hat{A})_{V^{'}}\leq_s\delta^i(\hat{A})_V$. Moreover,
from \ref{ali:order} it follows that: \be
sp(\delta^i(\hat{A})_V)\subseteq
sp(\hat{A}),\;\;\;\;\;sp(\delta^o(\hat{A})_V)\subseteq sp(\hat{A})
\ee which, as mentioned above,  is precisely the reason why the
spectral order was chosen.

It is interesting to note that the definition of spectral order,
when applied to inner and outer daseinisation implies the
following: \be
\delta^i(\hat{A})_{V}\leq_s\delta^o(\hat{A})_V\;\;\text{ iff
}\forall
r\in\Rl\;\;\;\hat{E}^{\delta^i(\hat{A})_{V}}_r\geq\hat{E}^{\delta^o(\hat{A})_{V}}_r
\ee 
Therefore for all $r\in\Rl$ we define \ba
\hat{E}^{\delta^i(\hat{A})_{V}}_r:=\delta^o(\hat{E}^{\hat{A}}_r)_V\;;\;\;
\hat{E}^{\delta^o(\hat{A})_{V}}_r:=\delta^i(\hat{E}^{\hat{A}}_r)_V
\ea The spectral family described by the second equation is
right-continuous, while the first is not. To overcome this problem
we define the following: \be
\hat{E}^{\delta^i(\hat{A})_{V}}_r:=\bigwedge_{s>
r}\delta^o(\hat{E}^{\hat{A}}_s)_V \ee Putting together all the
results above we can write inner and outer daseinisation of
self-adjoint operators as follows: \ba
\delta^o(\hat{A})_V:=\int_{\Rl}\lambda
d\Big(\delta^i(\hat{E}_{\lambda}^{\hat{A}})\Big)\text{ and }
\delta^i(\hat{A})_V:=\int_{\Rl}\lambda
d\Big(\bigwedge_{\mu>\lambda}\delta^o(\hat{E}_{\mu}^{\hat{A}})\Big)
\ea where the  integrals are Riemann-Stieltjes
integrals\footnote{This is why right continuity was needed.}.
We can now define the order-reversing and order-preserving
functions as follows: \ba \mu_{\lambda}:\downarrow V\rightarrow
\Rl\;;\;\;
V^{'}\mapsto\lambda_{|V^{'}}(\delta^i(\hat{A})_{V^{'}})=\lambda(\delta^i(\hat{A})_{V^{'}})
\ea The order-reversing functions are defined as follows: \ba
\nu_{\lambda}:\downarrow V\rightarrow \Rl\;;\;\;
V^{'}\mapsto\lambda_{|V^{'}}(\delta^o(\hat{A})_{V^{'}})=\lambda(\delta^o(\hat{A})_{V^{'}})
\ea
\subsection{New Representation of Physical Quantities}
We are now interested in understanding the action of the $F$
functor on physical quantities. We thus define the following \be
F(\breve{\delta}(\hat{A})):\breve{\us}\rightarrow
\ps{\breve{\Rl}^{\leftrightarrow}} \ee which, at each context $V$,
is defined as \be F(\breve{\delta}(\hat{A}))_V:\coprod_{\phi_i\in
Hom(\down V,\mv(\mh))}\us_{\phi_i(V)}\rightarrow\coprod_{\phi_i\in
Hom(\down V,\mv(\mh))} \ps{\Rl^{\leftrightarrow}}_{\phi_i(V)} \ee such
that for a given $\lambda\in \us_{\phi_i(V)}$ we obtain \ba
F(\breve{\delta}(\hat{A}))_V(\lambda)&:=&\breve{\delta}(\hat{A})_{\phi_i(V)}(\lambda)\\\nonumber
&=&(\breve{\delta}^i(\hat{A})_{\phi_i(V)}(\cdot),
\breve{\delta}^o(\hat{A})_{\phi_i(V)}(\cdot))(\lambda)=(\mu_{\lambda},\nu_{\lambda})
\ea
Thus in effect the map $F(\breve{\delta}(\hat{A}))_V$ is a co-product of maps of the form $F(\breve{\delta}(\hat{A}))_{\phi_i(V)}$ for all $\phi_i\in Hom(\down V, \mv(\mh))$.

From this definition it is straightforward to understand how the
group acts on such physical quantities. In particular, for each context $V\in\mv_f(\mh)$ we obtain a collection of
maps \be F(\breve{\delta}(\hat{A}))_V:\coprod_{\phi_i\in
Hom(\down V,\mv(\mh))}\us_{\phi_i(V)}\rightarrow\coprod_{\phi_i\in
Hom(\down V,\mv(\mh))} \ps{\Rl^{\leftrightarrow}}_{\phi_i(V)} \ee and
the group action is to map individual maps in such a collection into one
another. Thus, for example, if we consider the component \be
\breve{\delta}(\hat{A})_{\phi_i(V)}:\us_{\phi_i(V)}\rightarrow
\ps{\Rl^{\leftrightarrow}}_{\phi_i(V)} \ee by acting on it by an
element of the group we would obtain \ba
l_g\Big(\breve{\delta}(\hat{A})_{\phi_i(V)}\Big):l_g\us_{\phi_i(V)}&\rightarrow& l_g \ps{\Rl^{\leftrightarrow}}_{\phi_i(V)}\\
\us_{l_g(\phi_i(V))}&\rightarrow&
\ps{\Rl^{\leftrightarrow}}_{l_g(\phi_i(V))} \ea
Let us now analyse what exactly is $l_g\breve{\delta}(\hat{A}))_{\phi_i(V)}$.\\
We know that it is comprised of two functions, namely \be
l_g\Big(\breve{\delta}(\hat{A})_{\phi_i(V)}\Big)=\Big(l_g\Big(\breve{\delta}^i(\hat{A}))_{\phi_i(V)}\Big)(\cdot),\Big(l_g\breve{\delta}^o(\hat{A}))_{\phi_i(V)}\Big)(\cdot)\Big)
\ee We will consider each of them separately. Given
$\lambda\in\us_{l_g(\phi_i(V))}$ we obtain \ba
l_g\Big(\breve{\delta}^i(\hat{A})_{\phi_i(V)}\Big)(\lambda)&=&\lambda\Big(l_g\big(\delta^i(\hat{A})_{\phi_i(V)}\big)\Big)\\\nonumber
&=&\lambda\Big(\hat{U}_g\big(\delta^i(\hat{A})_{\phi_i(V)}\big)\hat{U}_g^{-1}\Big)\\\nonumber
&=&\lambda\Big(\delta^i(\hat{U}_g\hat{A}\hat{U}_g^{-1})_{l_g(\phi_i(V))}\Big)\\\nonumber
&=&\breve{\delta}^i(\hat{U}_g\hat{A}\hat{U}_g^{-1})_{l_g(\phi_i(V))}(\lambda)
\ea Similarly for the order reversing function we obtain \be
l_g\Big(\breve{\delta}^o(\hat{A})_{\phi_i(V)}\Big)(\lambda)=\breve{\delta}^o(\hat{U}_g\hat{A}\hat{U}_g^{-1})_{l_g(\phi_i(V))}(\lambda)
\ee Thus putting the two results together we have \be
l_g\Big(\breve{\delta}(\hat{A})_{\phi_i(V)}\Big)=\Big(\breve{\delta}(\hat{U}_g\hat{A}\hat{U}_g^{-1})_{l_g(\phi_i(V))}\Big)
\ee This is the topos analogue of the standard transformation of
self adjoint operators in the canonical formalism of quantum
theory. In particular, given a self adjoint operator
$\breve{\delta}(\hat{A})$ its local component in the context $V$
is $\breve{\delta}(\hat{A})_{V}$.
 This `represents' the pair of self adjoint operators $(\delta^i(\hat{A})_{V}, \delta^o(\hat{A})_{V})$ which live in $V$. By acting with a unitary transformation we obtain the transformed quantity $l_g\Big(\breve{\delta}(\hat{A})\Big)$ with local components $\Big(\breve{\delta}(\hat{U}_g\hat{A}\hat{U}_g^{-1})_{l_gV}\Big)$, $V\in \mv_f(\mh)$. Such a quantity represents the pair $(\delta^i(\hat{U}_g\hat{A}\hat{U}_g^{-1})_{l_g(V)}, \delta^o(\hat{U}_g\hat{A}\hat{U}_g^{-1})_{l_g(V)})$ of self adjoint operators living in the transformed context $l_g(V)$.
%
\section{Conclusions}
In this paper we have shown how it is possible to introduce a
group and its respective group action in the topos formulation of
quantum theory. The aim of the topos formalism is to render
quantum theory more ``realist", in the sense that its mathematical
formulation resembles the mathematical formulation of classical
theory. It is in this sense that we often say that we want quantum
theory to `look like' classical theory.

In such a resemblance we wanted to include the way in which a
group acts on the state space. We know that in classical theory,
given a state space $S$, the action of the group is defined by \be
G\times S\rightarrow S \ee
i.e., the group maps the state space to itself.

Similarly, in the topos formalism, we were looking for a group
action which mapped the topos analogue of the state space to
itself, i.e., \be \underline{G}\times \us\rightarrow \us \ee
However, we have seen that such an action is not possible if the
topos we utilise to describe quantum theory is $Sh(\mv(\mh))$. In
fact, in such a topos, we obtain the twisted presheaves, i.e., for
each $g\in G$ we have \be \iota^{\hat{U}_g}:\us\rightarrow
\us^{\hat{U}_g} \ee
In order to solve such a problem and eliminate these twisted presheaves we had to change the topos we worked with. \\
We did this through the following steps:
\begin{enumerate}
\item [1)] First of all we introduced the category $\mv_f(\mh)$ of abelian von-Neumann sub-algebras which, however, was chosen to be invariant under group transformations, i.e., we assumed that the group does not act on it.
\item [2)] We then defined the sheaf $\ps{G/G_F}$
over $\mv_f(\mh)$. This sheaf associates to each context $V$ the
quotient space $G/G_{FV}$, where $G_{FV}$ is the fixed point group
of $V$. Such a sheaf was shown to be isomorphic to the sheaf which
associates to each context $V$ all possible other contexts which
are related to it via a group action, i.e., all possible faithful
representations of such a context:
\be
\ps{G/G_F}\simeq \ps{Hom}(\mv_f(\mh), \mv(\mh))
\ee
such that for all $V\in\mv_f(\mh)$ we obtain
\be
(\ps{G/G_F})_V:=G/G_{FV}\simeq Hom(\downarrow V,\mv(\mh))=:{\ps{Hom}(\mv_f(\mh), \mv(\mh))}_V
\ee
\item [3)] We then utilised the etal\'e space $\Lambda(\ps{G/G_F})$ as our new context category. Given such a category we were able to map all the sheaves in $Sh(\mv(\mh))$ to sheaves in $Sh(\Lambda(\ps{G/G_F}))$ via the functor
\be I:Sh(\mv(\mh))\rightarrow Sh(\Lambda(\ps{G/G_F})) \ee The
advantage of the topos $Sh(\Lambda(\ps{G/G_F}))$ is that it allows
a definition of the group action on individual sheaves in terms of
the group action on the elements of $\Lambda(\ps{G/G_F})$. Such an
action was shown to be continuous if $\Lambda(\ps{G/G_F})$ is
equipped with the bucket topology. However there are still
problems since a) $\Lambda(\ps{G/G_F})$ is a much bigger space
than the original category $\mv(\mh)$, thus we are over counting
information b) we still get twisted presehaves.
\item [4)] To solve the above mentioned problems we utilised the functor
\be p_{!}:Sh(\Lambda(\ps{G/G_F}))\rightarrow Sh(\mv_f(\mh)) \ee
to map all the sheaves over $\Lambda(\ps{G/G_F})$ to sheaves over $\mv_f(\mh)$. \\
Combining such a functor with the previously defined
$I$ allowed us to define a functor \be
F:Sh(\mv(\mh))\rightarrow Sh(\mv_f(\mh)) \ee which decomposes as
follows \be Sh(\mv(\mh))\xrightarrow{I}
Sh(\Lambda(\ps{G/G_F}))\xrightarrow{p_{!}} Sh(\mv_f(\mh)) \ee
We can now map all the important sheaves of the old formalism to sheaves in $Sh(\mv_f(\mh))$.

\noindent
For sheaves obtained in such a way it is possible to define a group action on them, even if $\mv_f(\mh)$ does not allow one. This is because the group action is defined at the level of the stalks, i.e., in terms of the elements in $\Lambda(\ps{G/G_F})$. Such an action does not induce twisted sheaves, thus solving our problem.

We also managed to the isomorphism of truth values \be \uom^{\mv_f(\mh)}\cong
F(\uom^{\mv(\mh)})/\ug \ee which is an expected result considering
the fact that the category $\mv_f(\mh)$ does not admit group
transformations.
\end{enumerate}
This work introduced a strategy for defining sheaves which we think will be very useful to eventually define the concept of quantisation in a topos. 

\bigskip
\bigskip

\textbf{Acknowledgements. }I would like to thank Professor C. J.
Isham and Dr A. D\"oring for useful and extensive discussions on
the initial stages of the project. I would also like to thank Professor Ieke Moerdijk for very useful discussion on sheaves and topology. A special thanks Sander Wolters for very useful comments on the manuscript.
A thanks also goes to my
mother Elena Romani and my father Luciano Flori for constant
support and encouragement. Thank you. This work was supported by
the Perimeter Institute of Theoretical Physics Waterloo Ontario.
\bigskip
\section{Appendix}
\begin{Lemma}
Let $f, h : X \rightarrow Y$ be continuous maps between
topological spaces $X, Y$, and let $Y$ be Hausdorff. Then $E(f, h)
:= \{x\in X | f(x) = h(x)\}$ is closed.
\end{Lemma}
\begin{Proof}
We have $E^c(f, h) := \{x\in X | f(x)\neq h(x)\}$. Let $x\in
E^c(f, h)$. Then since $Y$ is Hausdorff, there exists open
neighbourhoods $N_{x,f}$ of $f(x)$ and $N_{x,h}$ of $h(x)$ such
that $N_{x,f}\cap N_{x,h} = \emptyset$. Since $f, h$ are
continuous $f^{-1}(N_{x,f} )$ and $h^{-1}(N_{x,h})$ are open. Thus
$f^{-1}(N_{x,f} )\cap h^{-1}(N_{x,h})$ is open and non-empty
(since $x\in f^{-1}(N_{x,f} )\cap h^{-1}(N_{x,h})$). In fact, for
all $y\in f^{-1}(N_{x,f} )\cap h^{-1}(N_{x,h})$ we have that
$f(y)\neq h(y)$. It follows that $E^c(f, h)$ is open, and hence that
$E(f, h)$ is closed.
\end{Proof}
As a direct consequence we have the following corollary:
\begin{Corollary}
Let the topological group G act on a Hausdorff topological space
$X$ in a continuous way, i.e., the G-action map $G\times X
\rightarrow X$ is continuous. Then the stabiliser, $G_x$, of any
$x\in X$ is a closed subgroup of $G$.
\end{Corollary}
\begin{Proof}
For any given $x\in X$, consider the maps $f_x, h_x : G\rightarrow
X$ defined by $f_x(g) := gx$ and $h_x(g) := x$ for all $g\in G$.
The first is continuous since the $G$-action on $X$ is continuous,
and the second is continuous because constant maps are always
continuous. Now \be
E(f_x, h_x) = \{g \in G | f_x(g) = h_x(g)\}\\
= \{g\in G | gx = x\} = G_x \ee It follows from the Lemma that
$G_x$ is closed.
\end{Proof}
\begin{Theorem}
Given the etal\'e map $f : X\rightarrow Y$ the left adjoint functor $f!:Sh(X)\rightarrow Sh(Y)$ is defined as follows 
\be
f!(p_A : A\rightarrow X) = f\circ p_A : A\rightarrow Y 
\ee
for $p_A:A\rightarrow Y$ an etal\'e bundle
\end{Theorem}
\begin{Proof}
In the proof we will first define the functor $f!$ for general presheaf situation, then restrict our attention to the case of sheaves ($Sh(X)\subseteq \Sets^{X^{\op}}$) and $f$ etal\'e. 

Consider the map $f:X\rightarrow Y$, this gives rise to the functor $f!:\Sets^{X^{\op}}\rightarrow \Sets^{Y^{\op}}$. The standard definition of $f!$ is as follows:
\be
f!:=-\otimes_{X}(_{f}X^{\bullet})
\ee
which is defined on objects $A\in \Sets^{X^{\op}}$ as
\be
A\otimes_{X}(_{f}Y^{\bullet})
\ee
This is a presheaf in $\Sets^{Y^{\op}}$, thus for each element $y\in Y$ we obtain the set
\be
(A\otimes_{X}(_{f}Y^{\bullet}))y:=A\otimes_{X}(_{f}Y^{\bullet})(-, y)
\ee
where $(_{f}Y^{\bullet})$ is the presheaf 
\be
(_{f}Y^{\bullet}):X\times Y^{\op}\rightarrow \Sets
\ee
This presheaf derives from the composition of $f\times id_{Y^{\op}}:X\times Y^{op}\rightarrow Y\times Y^{op}$ ($(f\times id_{Y^{\op}})^*:\Sets^{Y\times Y^{op}}\rightarrow \Sets^{X\times Y^{op}}$) with $^{\bullet}Y^{\bullet}:Y\times Y^{\op}\rightarrow \Sets$, i.e.,
\be
(_{f}Y^{\bullet}):=(f\times id_{Y^{\op}})^*(^{\bullet}Y^{\bullet})=^{\bullet}Y^{\bullet}\circ (f\times id_{Y^{\op}})
\ee
where $^{\bullet}Y^{\bullet}$ is the bi-functor 
\ba
^{\bullet}Y^{\bullet}:Y\times Y^{\op}&\rightarrow&\Sets\\
(y,y^{'})&\mapsto&Hom_Y(y^{'},y)
\ea
Now coming back to our situation we then have the restricted functor
\ba
(_{f}Y^{\bullet})(-, y):(X, y)&\rightarrow& \Sets\\
(x, y)&\mapsto &(_{f}Y^{\bullet})(x,y)
\ea
which from the definition given above is
\ba
(_{f}Y^{\bullet})(x,y)=^{\bullet}Y^{\bullet}\circ (f\times id_{Y^{\op}})(x,y)=^{\bullet}Y^{\bullet}(f(x), y)=Hom_Y(y, f(x))
\ea
Therefore putting all the results together we have that for each $y\in Y$ we obtain $A\otimes_{X}(_{f}Y^{\bullet})(-, y)$, defined for each $x\in X$ as 
\be
A(-)\otimes_{X}(_{f}Y^{\bullet})(x,y):=A(x)\otimes_{X}Hom_Y(y, f(x))
\ee
This represents the presheaf $A$ defined over the element $x$, plus a collection of maps in $Y$ mapping the original $y$ to the image of $x$ via $f$. 

In particular $A(x)\otimes_{X}(_{f}X^{\bullet})=A(x)\otimes_{X}Hom_Y(y, f(-))$ represents the following equaliser: 
\[\xymatrix{
\coprod_{x, x^{'}}A(x)\times Hom_X(x^{'},x)\times Hom_Y(y, f(x^{'})\ar@<3pt>[rr]^{\;\;\;\;\;\;\; \tau} \ar@<-3pt>[rr]_{\;\;\;\;\;\;\; \theta}&&\coprod_x A(x)\times Hom_Y(y, f(x))\ar[dd]^{\sigma}\\
&&\\
&&A(-)\otimes_{X}Hom_Y(y, f(-)) \\
}\]
Such that given a triplet $(a, g, h)\in A(x)\times Hom_X(x^{'},x)\times Hom_Y(y, f(x^{'})$ we then obtain that 
\be\label{equ:equivalence}
\tau(a, g, h)=(ag, h)=\theta(a,g,h)=(a, gh)
\ee
Therefore $A(-)\otimes_{X}Hom_Y(y, f(-))$ is the quotient space of $\coprod_x A(x)\times Hom_Y(y, f(x))$ by the above equivalence conditions.

We now consider the situation in which $A$ is a sheaf on $X$, in particular it is an etal\'e bundle $p_A:A\rightarrow X$ and $f$ is an etal\'e map which means that it is a local homeomorphism, i.e. for each $x\in X$ there is an open set $V$ such that $x\in V$ and such that $f_{|V}:V\rightarrow f(V)$ is a homeomorphism. It follows that for each $x_i\in V$ there is a unique element $y_i$ such that $f_{|V}(x_i)=y_i$. In particular for each $V\subset X$ then $f_{|V}(V)=U$ for some $U\subset Y$.

\noindent
It can be the case that $f_{|V_i}(V_i)=f_{|V_j}(V_j)$ even if $V_i\neq V_j$, since the condition of being a homeomorphism is only local, however in these cases the restricted etal'e maps have to agree on the intersections, i.e. $f_{|V_i}(V_i\cap V_j)=f_{|V_j}(V_j\cap V_j)$

Let us now consider an open set $V$ with local homeomorphism $f_{|V}$. In this setting each element $y_i\in f_{|V}(V)$ will be of the form $f(x_i)$ for a unique $x_i$. Moreover, if we consider two open sets $V_1,V_2\subseteq V$, then to each map $V_1\rightarrow V_2$ in $X$, with associated bundle map $A(V_2)\rightarrow A(V_1)$, there corresponds a map $f_{|V}V_1\rightarrow f_{|V}(V_2)$ in $Y$. Therefore evaluating $A(-)\otimes_{X}Hom_Y(-, f(-))$ at the open set $f_{|V}(V)\subset Y$ we get, for each $V_i\subseteq V$ the equivalence classes $[A(V_i)\times_XHom_Y(f_{|V}(V), f_{|V}(V_i))]$ where $A(V_j)\times_XHom_Y(f_{|V}(V), f_{|V}(V_j))\simeq A(V_k)\times_XHom_Y(f_{|V}(V), f_{|V}(V_k))$ iff there exists a map $f_{|V}(V_j)\rightarrow f_{|V}(V_k)$ (which combines giving $f_{|V}(V)\rightarrow f_{|V}(V_k)$)  and corresponding bundle map $A(V_k)\rightarrow A(V_j)$ (which combine giving $A(V_k)\rightarrow A(V)$) given by the map $V_j\rightarrow V_k$ ( which combined gives $V\rightarrow V_k$) in $X$. A moment of thought reveals that such an equivalence class is nothing but $p_A^{-1}(V)$ (the fibre of $p_A$ at $V$) with associated fibre maps induced from the base maps.

We will now denote such an equivalence class by $[A(V)\times_XHom_Y(f_{|V}(V), f_{|V}(V))]$, since obviously in each equivalence class there will be the element $[A(V)\times_XHom_Y(f_{|V}(V), f_{|V}(V))]$

We apply the same procedure for each open set $V_i\subset X$. We can obtain two cases: \begin{enumerate}
\item [i)] $f_{|V_i}(V_i)=U\neq f_{V}(V)$. In that case we simply get an independent equivalence class for $U$.

\item [ii)] If $f_{|V_i}(V_i)=U=f_{V}(V)$ and there is no map $i:V\rightarrow V_i$ in $X$ then, in this case, we obtain for $U$ two distinct equivalence classes $[A(V_i)\times_XHom_Y(f_{|V_i}(V_i), f_{|V_i}(V_i))]$ and $[A(V)\times_XHom_Y(f_{|V}(V), f_{|V}(V))]$. 
\end{enumerate}
Thus the sheaf $A(-)\otimes_X(_fY^{\bullet})$ is defined for each open set $f_V(V)\subset Y$ as the set $[A(V)\times_XHom_Y(f_{|V}(V), f_{|V}(V))]\simeq A(V)))$, and for each map $f_{V^{'}}(V^{'})\rightarrow f_V(V)$ in $Y$ (with associated map $V^{'}\rightarrow V$ in $X$), the corresponding maps $[A(V)\times_XHom_Y(f_{|V}(V), f_{|V}(V))]\simeq A(V)\rightarrow [A(V^{'})\times_XHom_Y(f_{|V^{'}}(V)^{'}, f_{|V^{'}}(V^{'}))]\simeq A(V^{'})$

This is precisely what the etal\'e bundle $f\circ p_A:A\rightarrow Y$ is.

\end{Proof}

\end{document}